\pdfoutput=1

\documentclass[11pt]{article}

\usepackage[final]{acl}

\usepackage{times}
\usepackage{latexsym}

\usepackage[T1]{fontenc}

\usepackage[utf8]{inputenc}

\usepackage{microtype}

\usepackage{inconsolata}

\usepackage{graphicx}
\usepackage{booktabs}
\usepackage{multirow}
\usepackage{xcolor}
\usepackage[table]{xcolor}
\usepackage{adjustbox}
\usepackage{amsthm}
\usepackage{algorithmic}

\usepackage[ruled,vlined]{algorithm2e}
\usepackage{multirow}
\usepackage[font=small]{caption}
\usepackage{adjustbox}
\usepackage{subcaption}
\usepackage{amsfonts}
\usepackage{amsmath}
\usepackage{array}
\usepackage{booktabs}
\usepackage{multirow}
\usepackage{tikz}
\usepackage{newfloat}
\usepackage{listings}
\usepackage{mathtools}
\usepackage{amssymb}
\usepackage{amsthm}

\theoremstyle{definition}

\theoremstyle{remark}

\usepackage{enumitem}
\usepackage{pifont}
\newcommand{\cmark}{{\color{green}\ding{51}}}
\newcommand{\xmark}{{\color{crimson}\ding{55}}} 
\usepackage{threeparttable}
\usepackage{xcolor,colortbl}
\definecolor{blue}{HTML}{3331D7}
\definecolor{orange}{HTML}{E57932}
\definecolor{purple}{HTML}{663399}
\definecolor{green}{HTML}{41805E}
\definecolor{rose}{HTML}{C71585}
\definecolor{crimson}{HTML}{DC143C}
\definecolor{grey}{HTML}{505050}
\definecolor{skyblue}{RGB}{203, 221, 245}
\definecolor{lightorange}{RGB}{255, 237, 220}
\definecolor{lightlavender}{RGB}{228, 220, 245}
\newcommand{\vstd}[1]{{\textcolor{grey}{\scriptsize$_{\pm\text{#1}}$}}}

\title{SciLENS: RL-Driven Autonomous Agents for Scientific Localized Evidence Navigation and Synthesis}

\author{
 \textbf{Leqi Zheng \textsuperscript{1,\textdagger}},
 \textbf{Jinbo Su\textsuperscript{2,\textdagger}},
 \textbf{Yuying Li \textsuperscript{1,\textdagger}},
 \textbf{Chaokun Wang* \textsuperscript{1}},
 \\
 \textbf{Weiping Wang \textsuperscript{3}},
\textbf{Haitao Li \textsuperscript{1}},
\textbf{Jiajun Zhang \textsuperscript{4}},
\textbf{Shannan Yan \textsuperscript{1}},
\\
\textbf{Zhaolu Kang \textsuperscript{5}},
\textbf{Rong Fu \textsuperscript{6}},
\textbf{Jie Wu \textsuperscript{7}},
\textbf{Fang Niu \textsuperscript{1}},
\textbf{Hang Zhang \textsuperscript{1}}
\\
 \textsuperscript{1}Tsinghua University, 
 \textsuperscript{2}Renmin University of China, 
 \textsuperscript{3}Institute of Information Engineering, CAS,
 \\
 \textsuperscript{4}USTC,
 \textsuperscript{5}Peking University,
 \textsuperscript{6}University of Macau,
 \textsuperscript{7}The Australian National University,
\\
 \small{\textsuperscript{\textdagger} Equal contribution.}
 \\
 \small{\textbf{* Correspondence:} {chaokun@tsinghua.edu.cn}}
}

\begin{document}
\maketitle

\begin{abstract}
Scientific literature synthesis agents increasingly rely on proprietary online services, limiting reproducibility, privacy, and offline deployment. To address this challenge, we introduce \textbf{SciLENS} (\textbf{Sci}entific \textbf{L}ocalized \textbf{E}vidence \textbf{N}avigation and \textbf{S}ynthesis), a fully local autonomous agent framework operating on a dual-tier infrastructure indexing approximately 12 million academic records. SciLENS pioneers the integration of structural visualization as an actionable tool within the reasoning loop, enabling the agent to compress complex citation topologies into validated data-driven charts and thereby mitigate context exhaustion during macro-level synthesis. To train the agent without human annotation, we develop an automated data synthesis pipeline that extracts multi-hop subgraphs from a citation knowledge graph, verified by cross-model consensus among 20 frontier models. The agent is subsequently aligned through a reverse-decomposition rubric strategy that provides fine-grained process rewards for early planning and strict evidence grounding. Evaluations across six scientific benchmarks encompassing standard QA, citation accuracy, factual reasoning, and structural synthesis demonstrate that SciLENS significantly outperforms open-source baselines and achieves performance comparable to GPT-5.2 and Gemini-3.0-pro. Our source code and data are released at \url{https://github.com/LQgdwind/SciLENS}.
\end{abstract}

\section{Introduction}
\label{sec:intro}

As scientific literature grows exponentially, researchers face mounting cognitive challenges in tracing methodological origins and synthesizing macro-level trends~\cite{sun2024scieval, zhang2024sciglm}.
Large language models and autonomous agents are increasingly utilized to automate literature reviews and answer complex scientific inquiries~\cite{singhal2025toward, baek2025researchagent, zhou2025autonomous, rouzrokh2025lattereview, NEURIPS2025_9a92ea37, zheng-etal-2025-lagcl4rec}, marking a paradigm shift from static retrieval to agentic workflows~\cite{ma2024sciagent, vasantharajan2025scirag}.
Three prominent paradigms have emerged to address these challenges: standard retrieval-augmented generation (RAG) pipelines that index localized corpora for granular question answering~\cite{asai2024openscholar, zheng2026should}, web-search agents that dynamically navigate the internet to resolve specific queries~\cite{wu2025webdancer, li2025websailor}, and deep-research agents that synthesize multi-document summaries through extended online exploration~\cite{shao2025dr, zhang2025deep}.
Despite their respective strengths, applying these agents to real-world scientific synthesis reveals several structural limitations that fundamentally constrain their utility.

\begin{figure}[!t]
    \centering
    \includegraphics[width=\linewidth]{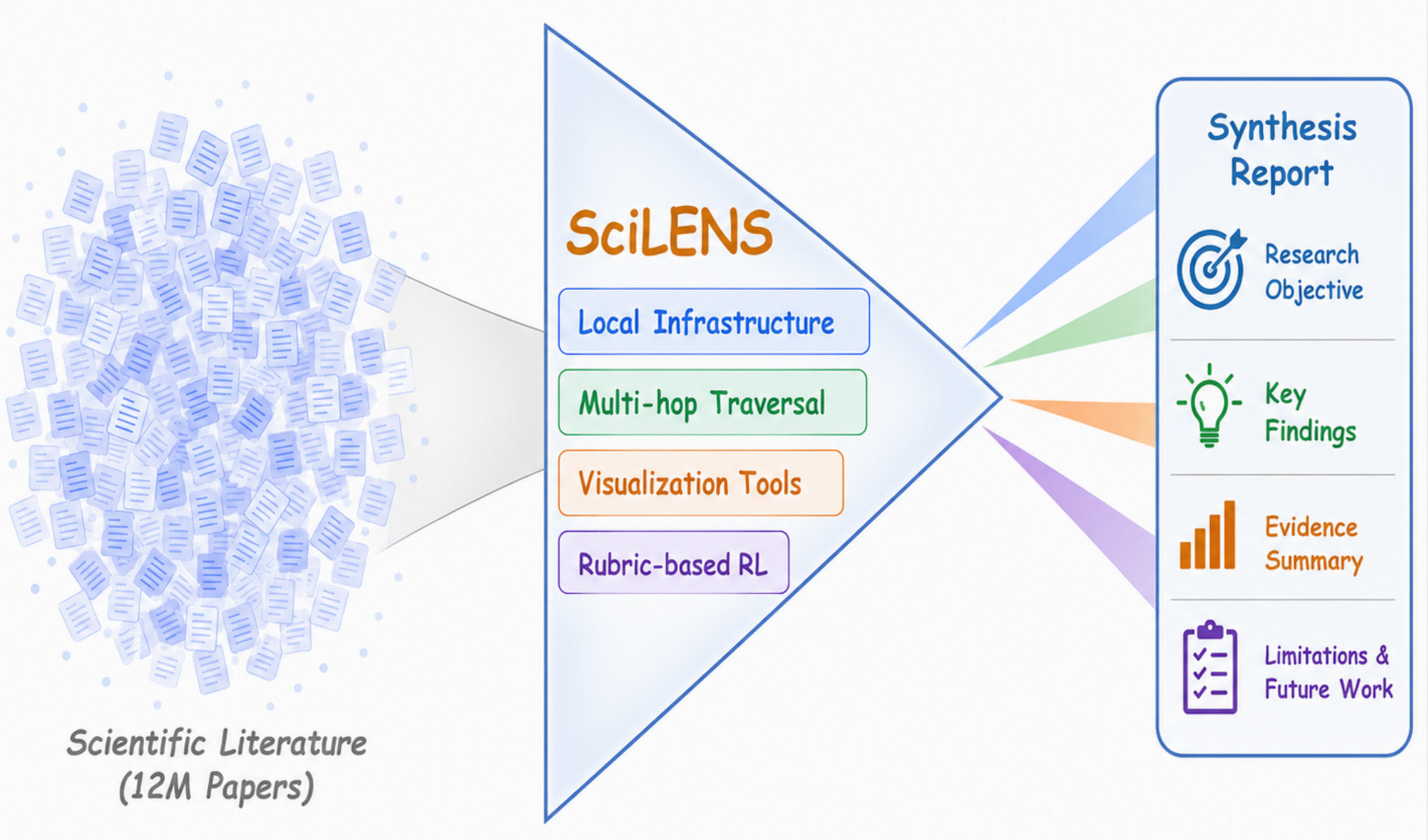}
    \vspace{-2mm}
    \caption{SciLENS operates as a refracting prism: chaotic scientific literature enters the system and is decomposed through four tightly integrated components into structured, multimodal synthesis reports. All processing occurs fully offline with 12M locally indexed papers.}
    \vspace{-3mm}
    \label{fig:intro_prism}
\end{figure}

As illustrated in Table~\ref{tab:agent_comparison}, existing paradigms exhibit complementary but individually insufficient capabilities when confronted with complex synthesis tasks.
Specifically, three critical limitations persist.
(1)~\textbf{Linear text-centric representation bottleneck}: all existing paradigms produce purely textual outputs, leading to context overflow and factual hallucinations when synthesizing macro-level trends or comparing complex citation networks~\cite{alansari2025large}. No current agent can autonomously compress quantitative topological data into structured visual representations.
(2)~\textbf{Scarcity of training data and absence of fine-grained alignment}: constructing multi-hop interaction trajectories demands prohibitively expensive human annotation, leaving open-source agents inadequately trained~\cite{xu2024kiwi}. Moreover, current reinforcement learning strategies rarely supervise implicit logical planning or evidence grounding, and existing benchmarks overlook data-driven visualization and large-scale summarization capabilities~\cite{vasantharajan2025scirag}.
(3)~\textbf{Dependency on fragile online infrastructure}: web-search and deep-research agents rely on external search APIs, exposing them to high latency, restrictive rate limits, and network failures that hinder rapid, deterministic access to large-scale corpora~\cite{shao2025dr, yao2026researcher}.
\begin{table}[!t]
\centering
\small
\caption{Comparison of existing research paradigms with SciLENS. Each column corresponds to one of the three identified limitations: structural visualization for overcoming text-centric bottlenecks, automated data synthesis with fine-grained rubric-based alignment, and operational independence from fragile web APIs.}
\setlength{\tabcolsep}{3pt}
\resizebox{\linewidth}{!}{
\begin{tabular}{lccc}
\toprule
\rowcolor{lightlavender}\textbf{Paradigm} & \textbf{Visualization} & \textbf{Auto Synthesis \& Alignment} & \textbf{Fully Localized} \\
\midrule
Standard RAG Pipeline & \xmark & \xmark & \cmark \\
Web-Search Agents & \xmark & \xmark & \xmark \\
Deep-Research Agents & \xmark & \xmark & \xmark \\
\textbf{SciLENS (Ours)} & \cmark & \cmark & \cmark \\
\bottomrule
\end{tabular}
}
\vspace{-4mm}
\label{tab:agent_comparison}
\end{table}

To address these limitations, we propose \textbf{SciLENS} (\textbf{Sci}entific \textbf{L}ocalized \textbf{E}vidence \textbf{N}avigation and \textbf{S}ynthesis), a fully localized autonomous agent framework for complex scientific reasoning and structural synthesis.
Our central hypothesis is that a fully localized agent equipped with multi-hop citation traversal, structural visualization tools, and rubric-based reinforcement learning can achieve scientific synthesis performance comparable to frontier proprietary models without online dependency.

To realize this hypothesis, SciLENS is designed around three key components. 
First, to overcome text-centric representation bottlenecks, we integrate structural visualization tools into the agent reasoning loop, enabling the agent to generate validated chart schemas that compress complex citation topologies into high-density visual summaries (Section~\ref{subsec:toolbox}). 
Second, to address training data scarcity and fine-grained alignment, we develop an automated data synthesis pipeline that extracts multi-hop subgraphs from a citation knowledge graph and generates question-answer pairs verified through cross-model consensus (Section~\ref{subsec:qa_synthesis}). 
We further align the agent with rubric-based reinforcement learning, using reverse-decomposition rubrics to supervise early planning and strict evidence grounding (Section~\ref{subsec:alignment}). 
Finally, to ensure fully local and robust deployment, SciLENS operates on a dual-tier local infrastructure that combines MongoDB for metadata management with distributed FAISS for dense retrieval, indexing approximately 12 million academic records for sub-second access (Section~\ref{subsec:infrastructure}).

In summary, our contributions are fourfold:

(1) \textbf{Visualization-Augmented Scientific Reasoning.} 
We integrate structural visualization as an actionable tool within the autonomous reasoning loop, enabling the agent to summarize complex citation topologies and quantitative trends as validated data-driven charts. This design alleviates the limitations of text-only reasoning by producing structured visual outputs that reduce context burden during macro-level synthesis.

(2) \textbf{Citation-Graph Data Synthesis and Rubric-Based Alignment.}
 We develop an automated data synthesis pipeline that extracts citation subgraphs via random walks and generates multi-hop reasoning trajectories verified through cross-model consensus, eliminating the need for human annotation. We then align the agent with reverse-decomposition rubrics that reward early decomposition and strict evidence grounding.

(3) \textbf{Fully Localized Agentic Infrastructure.} 
We further build a fully local infrastructure that indexes approximately 12 million academic records and supports semantic retrieval, citation-graph traversal, and structural visualization through an offline research toolbox.

(4) \textbf{Comprehensive Evaluation.} 
We conduct extensive evaluations across six scientific benchmarks spanning reading comprehension, citation accuracy, factual reasoning, and structural synthesis, demonstrating that SciLENS outperforms open-source baselines and approaches frontier proprietary models.
\begin{figure*}[!t]
    \centering
     \vspace{-4mm}
    \includegraphics[width=1.0\linewidth]{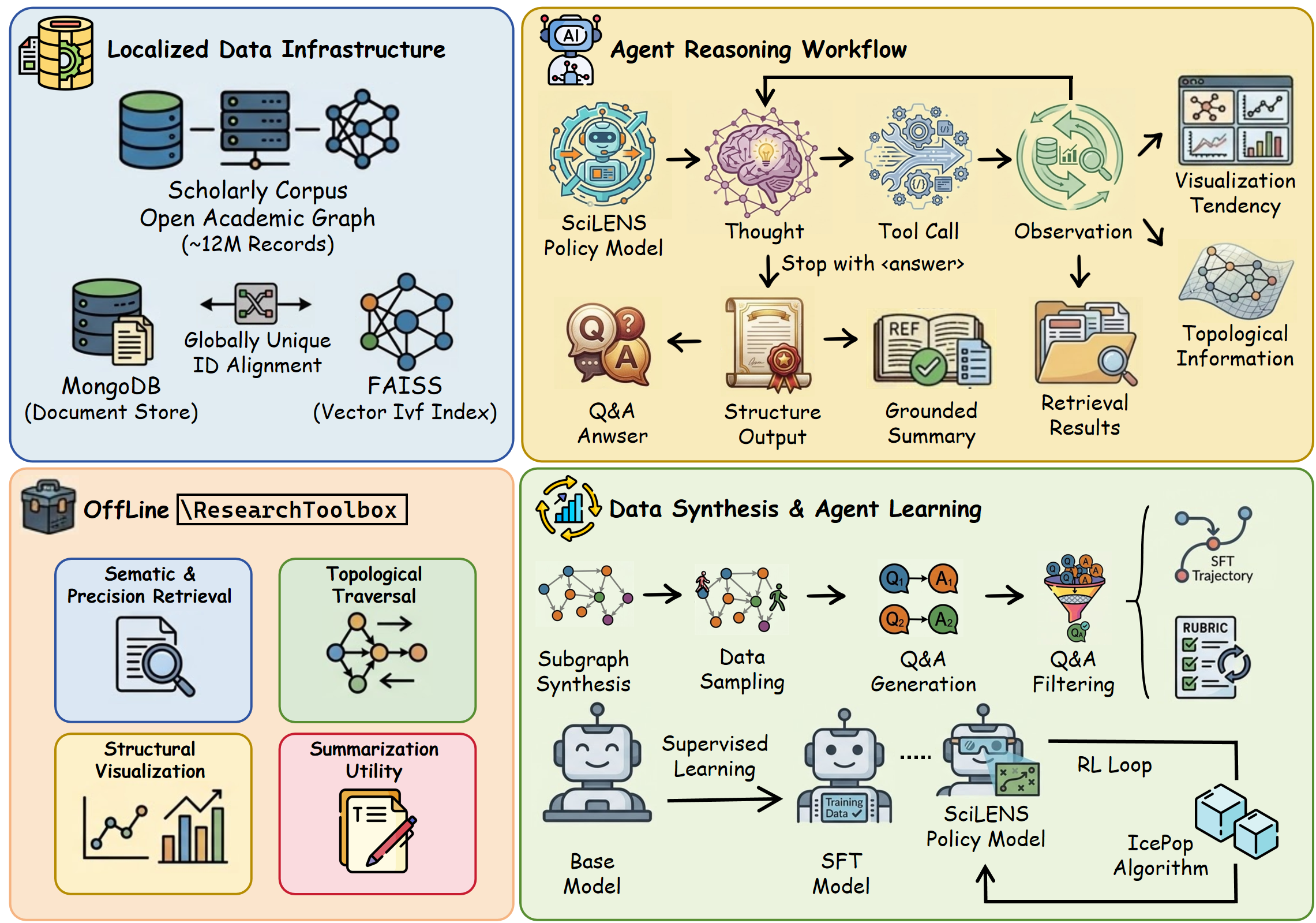}
    \vspace{-4mm}
    \caption{The SciLENS architecture integrates four core components. A localized dual-tier infrastructure ensures deterministic access to a massive scholarly corpus. Utilizing this foundation, an offline research toolbox empowers the agent with semantic retrieval, topological traversal, and structural visualization capabilities orchestrated via a continuous reasoning loop. Finally, an automated data synthesis pipeline generates complex trajectories to explicitly align the policy model through supervised fine-tuning and rubric-based reinforcement learning.}
    \vspace{-4mm}
    \label{fig:overall_pipeline}
\end{figure*}

\section{Agent Architecture and Visualization-Augmented Reasoning}
\label{sec:scilens_framework}

To perform complex academic reasoning and macro-level synthesis without the latency, instability, and API constraints of external web search engines, we design \textbf{SciLENS}, a fully localized and autonomous agent framework.
The framework combines a high-throughput dual-tier infrastructure with a visualization-augmented offline tool suite, enabling deterministic data access and multimodal structured outputs beyond linear text generation.
\subsection{Localized Dual-Tier Infrastructure}
\label{subsec:infrastructure}

The foundation of SciLENS is a persistent, fully localized storage backend that indexes approximately 12 million cleaned academic records derived from the Open Academic Graph corpus~\cite{zhang2022oag}.
To achieve sub-second retrieval latency while maintaining rich metadata access, we construct a dual-tier architecture coupling a MongoDB document store with a distributed FAISS vector index.
The MongoDB tier handles exact-match queries, keyword-based BM25~\cite{robertson2009probabilistic} searches, and citation relationship tracking, while the FAISS tier, built across independent GPU shards using Qwen3-Embedding-8B~\cite{zhang2025qwen3}, enables distributed dense retrieval.
The two tiers are coupled through a shard-based identifier alignment scheme, where each paper is assigned a globally unique integer computed from its shard index and within-shard offset, allowing nearest-neighbor vector results to be resolved to full paper metadata via a single indexed database lookup.
Instead of relying on slow, synchronous online web searches, our agent interacts exclusively with this local infrastructure, thereby eliminating network-induced latency and failure modes.
The detailed system configurations, including chunk-based parallel ingestion strategies, identifier alignment, and API batching parameters, are provided in Appendix~\ref{app:db_details}.

\subsection{Visualization-Augmented Offline Tool Suite}
\label{subsec:toolbox}

Existing research agents operate exclusively through linear text generation, which inevitably leads to context exhaustion and factual hallucinations when synthesizing macro-level academic trends across dozens of papers~\cite{alansari2025large}.
To address this bottleneck, we introduce a key algorithmic insight: agents autonomously invoke visualization tools during reasoning to compress high-dimensional topology into chart schemas, bypassing inefficient textual encoding.
This transforms the agent from a purely text-centric reasoner into a multimodal synthesizer capable of producing integrated text-and-chart academic reports.

To operationalize this insight, we design a unified \textsc{ResearchToolbox} with twelve tools across four functional categories. Semantic retrieval tools support dense embedding search, keyword lookup, and metadata extraction, while topological traversal tools enable multi-hop graph exploration and shortest-path tracing over citation networks.
Our novel visualization tools enable the agent to emit validated schemas for standard charts (e.g., line, bar, scatter), strictly enforcing dimensional and numeric consistency prior to rendering.
A dedicated summarization utility assists in distilling extensive academic abstracts to optimize context window utilization.
The agent accesses all capabilities via structured JSON payloads and receives observations directly into its reasoning chain.
The formal tool specifications, precise input schemas, and operational boundaries are detailed in Appendix~\ref{app:tool_schema}.

\subsection{Agent Reasoning Workflow}
\label{subsec:workflow}

Equipped with the \textsc{ResearchToolbox}, the SciLENS agent operates in a continuous thought, action, and observation loop governed by a rigorous system prompt. This foundational instruction set is universally applied across both the training and inference phases to ensure strict adherence to tool invocation schemas and behavioral constraints. The complete directives establishing these operational boundaries are detailed in Appendix~\ref{app:system_prompt}. Complex queries are processed by the agent through sequential seed document retrieval, topology-based multi-hop evidence traversal, and autonomous metric structuring for visual trend analysis. To ensure robust execution, we adopt a dual-tier error recovery mechanism: formatting errors trigger contextualized feedback for autonomous correction, while missing tool calls or answer tags lead to immediate re-rolls. The agent then integrates textual and structural evidence to generate coherent academic synthesis, reducing factual hallucinations.

\section{Citation-Graph Data Synthesis and Rubric-Based Alignment}
\label{sec:synthesis_alignment}

Training autonomous agents for complex scientific reasoning requires high-quality, multi-hop interaction trajectories. However, such data is notoriously scarce and prohibitively expensive to annotate manually. To address this bottleneck, we propose an automated, logic-driven data synthesis pipeline that extracts cohesive subgraphs from academic literature to formulate complex question-answer pairs. We then utilize the evolutionary history of these questions to align the agent through a novel rubric-based reinforcement learning strategy.
The overall training framework is illustrated in Figure~\ref{fig:overall_pipeline}.

\subsection{Topological Subgraph Sampling}
\label{subsec:subgraph_sampling}

To generate locally coherent contexts for downstream reasoning synthesis, we first extract highly interconnected document clusters from the Open Academic Graph corpus~\cite{zhang2022oag}. Rather than relying on static semantic similarity, we capture the actual citation topology by executing random walk explorations. Starting from a uniformly sampled seed publication, a walker proceeds for thirty steps, transitioning to randomly selected forward or backward citation neighbors. We execute ten independent walks per subgraph to form a connected local network. A total of thirty thousand subgraphs are generated in this manner. Each subgraph serves as a cohesive factual grounding cluster for subsequent question generation, ensuring that the synthesized tasks reflect real-world academic literature structures. The detailed node-filtering heuristics and streaming deduplication processes applied prior to sampling are provided in Appendix~\ref{app:qa_synthesis_details}.

\subsection{Logic-Driven QA Synthesis and Verification}
\label{subsec:qa_synthesis}

Starting from topological subgraphs, we synthesize multi-hop question-answer pairs by compounding factoid base questions with sampled relational links. Difficulty is ensured via a self-consistency mechanism where the model must fail to resolve the composite question using only atomic facts. Subsequently, each candidate instance undergoes strict cross-model verification utilizing a diverse pool of twenty frontier language models. For each candidate, we implement a stochastic verification protocol where 4 verifiers are drawn uniformly at random from the pool to independently render binary judgments on factual correctness and answer uniqueness via autonomous retrieval queries. A candidate QA pair is retained if and only if all sampled verifiers reach an absolute consensus:
\begin{equation}
    \text{keep}(q, a) = \prod_{m \in \mathcal{S}} \mathbf{1}\!\left[\text{corr}_m(q,a) \wedge \text{uniq}_m(q,a)\right],
    \label{eq:filter}
\end{equation}
where $\mathcal{S} \subset \{1, \dots, 20\}$ such that $|\mathcal{S}| = 4$ represents the random subset of verifiers, while $\text{corr}_m$ and $\text{uniq}_m$ denote the correctness and uniqueness judgments of verifier $m$. This probabilistic filtering ensures high precision training data while maintaining broad model diversity. The exhaustive list of constituent models and the detailed synthesis procedures are provided in Appendix~\ref{app:qa_synthesis_details}.

\subsection{Multi-Step Multi-Tool Agent Learning}
\label{subsec:alignment}

With the high-quality QA instances synthesized, we develop a two stage training paradigm to empower SciLENS with tool use proficiency and logical rigor.

\paragraph{Supervised Fine-Tuning via Distillation.}
Recent studies examine how information transfer can be controlled in LLM distillation, from distillation resistance to filtered and reweighted on-policy objectives~\cite{fang2026distillationresistantlargelanguagemodels,li2026filterreweightrethinkingoptimization}.
In the first stage, we perform knowledge distillation from DeepSeek-V3.2~\cite{liu2025deepseek}. The teacher model interacts with the \textsc{ResearchToolbox} to generate approximately twenty thousand complete interaction trajectories. To avoid interference from external feedback during learning and enhance the capability of the model to couple reasoning and action steps, we mask out loss contributions from tool observations. Given a task context $c$ and a trajectory $H = (x_1, x_2, \dots, x_{|H|})$ where each $x_i$ represents either a reasoning step, a tool action, or an observation $o$, the supervised loss function is computed as follows:
\begin{equation}
    \mathcal{L}_{\text{SFT}} = - \frac{\sum_{i=1}^{|H|} \mathbb{I}[x_i \neq o] \cdot \log \pi_\theta(x_i | c, x_{<i})}{\sum_{i=1}^{|H|} \mathbb{I}[x_i \neq o]} .
    \label{eq:sft_loss}
\end{equation}
This masking mechanism preserves original reasoning capabilities while teaching a robust behavioral paradigm of alternating deliberate reasoning with tool execution.

\paragraph{Rubric-Based Reinforcement Learning.}
During the reinforcement learning phase, we optimize the policy model utilizing the IcePop algorithm~\cite{team2025every}. IcePop uses a decoupled behavior policy and selective token masking to stabilize exploration and avoid policy collapse. The full objective is provided in Appendix~\ref{app:icepop_details}. Training trajectories are collected via agentic ReAct rollouts. The rigorous formatting constraints and structural tags governing these rollouts are universally defined by the system prompt, with complete execution details deferred to Appendix~\ref{app:system_prompt}. The reward design closely aligns optimization across rollouts. Instead of a single scalar reward, we adopt a multidimensional framework combining Reinforcement Learning from Verifiable Rewards for final answer correctness with semantic rewards for planning, reverse decomposition, and evidence grounding. Prompting a judge model generates an instance-specific rubric comprising $N$ weighted criteria for each training query:
\begin{equation}
    \mathcal{R} = \bigl\{(d_i,\, w_i)\bigr\}_{i=1}^{N}, \quad \sum_{i=1}^{N} w_i = 100,
    \label{eq:rubric}
\end{equation}
where $d_i$ is a natural language description of the verification criterion and $w_i$ is its normalized importance weight. Finally, we compute a composite scalar reward to evaluate the entire trajectory. The reward integrates a binary format score $S_{\text{format}}$ ensuring strict conformity to required schemas, alongside a semantic answer score $S_{\text{rubric}}$ evaluated against the criteria in $\mathcal{R}$. The composite reward function is defined as:
\begin{equation}
    R = \lambda \cdot S_{\text{format}} + (1 - \lambda) \cdot S_{\text{rubric}}.
    \label{eq:reward}
\end{equation}
This fine-grained alignment guarantees that SciLENS prioritizes structured early planning and meticulous evidence grounding over superficial text generation. Comprehensive construction methodologies for these rubrics are provided in Appendix~\ref{app:rubric_construction}.

\section{Benchmarks and Multidimensional Evaluation}
\label{sec:benchmarks_experiments}

To rigorously assess the capabilities of SciLENS, we establish a comprehensive evaluation framework encompassing six benchmarks across two complementary categories. The established evaluations utilize SciFact~\cite{wadden2020fact}, PubMedQA~\cite{jin2019pubmedqa}, QASA~\cite{lee2023qasa}, and ScholarQA-CS~\cite{singh2025ai2} to measure correctness and citation accuracy. 
Beyond standard reading comprehension, we evaluate on Structural Synthesis (SSB) and Scientific Fact and Reasoning (SciFR), which assess visual consistency, query alignment, semantic fidelity, and evidence attribution. Both benchmarks are constructed via automated multi-hop subgraph extraction from computer science, physics, and biomedicine, yielding 500 verified instances each, with strict train--test disjointness enforced at the subgraph, seed-paper, and instance levels (Appendix~\ref{app:disjoint}). For reproducibility, all evaluations use a locally deployed Qwen3-30B-A3B judge with zero temperature. Full evaluation protocols, metric definitions, and instance-level rubrics are provided in Appendix~\ref{app:benchmark_details}.

\section{Experiment}

\subsection{Experimental Setup}
\label{subsec:exp_setup}
To establish competitive performance upper bounds and evaluate the capability gap across diverse architectures, we compare SciLENS against a comprehensive suite of state-of-the-art proprietary models and leading open-source frameworks, spanning general-purpose reasoning backbones, specialized deep research agents, and exploratory web agents.
The exhaustive list of evaluated baselines and specific training configurations are documented in Appendix~\ref{app:exp_details}.

A critical design principle of our evaluation is the \textbf{unified toolbox protocol}: all evaluated models, including proprietary systems such as GPT-5.2 and Claude-4.5-Sonnet, are equipped with the identical offline \textsc{ResearchToolbox} and evaluated against the same 12-million-paper local database.
No model uses web search during evaluation.
This design ensures that all performance differences reported in Tables~\ref{tab:overall_benchmark} and~\ref{tab:scientific_benchmarks} are attributable solely to model capabilities and training methodology, not to infrastructure or tool advantages.
In particular, the Qwen3-30B-A3B (Base + Tools) row in our tables represents the zero-shot base model given full access to the entire \textsc{ResearchToolbox} without any SciLENS training, serving as a controlled reference that isolates the contribution of our training pipeline from the underlying tool access.

\subsection{Main Results}
\label{subsec:main_results}
We conduct extensive evaluations against state-of-the-art proprietary and leading open-source models, with results reported in Tables~\ref{tab:overall_benchmark} and~\ref{tab:scientific_benchmarks}. The SciLENS rows in the tables detail the full training pipeline, including the base model with full \textsc{ResearchToolbox} access, the SFT variant, and the final RL model.
On four established benchmarks, SciLENS-RL achieves the highest correctness scores on QASA (47.62), SciFact (88.94), and PubMedQA (77.53), and attains the strongest citation accuracy across all datasets, demonstrating robust factual grounding on external evaluations.
On SSB and SciFR, the Qwen3-30B-A3B (Base + Tools) row shows that tool access alone yields only 0.13 and 0.17 respectively, establishing that these tasks cannot be solved without alignment.
Supervised fine-tuning dramatically closes this gap, and the reinforcement learning variant achieves the strongest overall scores, outperforming all open-source baselines and achieving performance comparable to GPT-5.2 and Gemini-3.0-pro.
An extended analysis of performance behaviors across these benchmarks is provided in Appendix~\ref{app:extended_analysis}.

\begin{table*}[htbp]
\vspace{-2mm}
\caption{Performance comparison across all benchmarks. All models are evaluated with the identical local \textsc{ResearchToolbox} and 12M-paper database; no model uses web search (Section~\ref{subsec:exp_setup}). All reported scores represent the mean $\pm$ standard deviation across five independent evaluation runs ($p < 0.05$). \textbf{Bold} indicates the best performance of SciLENS; \textit{italic} indicates the best baseline performance in each section.}
\label{tab:main_results}
\vspace{-2mm}
\centering
\begin{subtable}{\textwidth}
\centering
\subcaption{Structural Synthesis Benchmark (SSB) and Scientific Fact and Reasoning (SciFR)}
\vspace{-1mm}
\begin{adjustbox}{width=\linewidth}
\begin{tabular}{l|cccc|cccc}
\toprule
\rowcolor{lightlavender} & \multicolumn{4}{c|}{\textbf{Structural Synthesis Benchmark}} & \multicolumn{4}{c}{\textbf{Scientific Fact and Reasoning}} \\
\rowcolor{lightlavender}\cline{2-9}
\rowcolor{lightlavender}\multirow{-2}{*}{\textbf{Model}} & ACS & QAS & FGS & Overall & ASF & QIS & EGQ & Overall \\
\hline
\rowcolor{skyblue}\multicolumn{9}{c}{\textit{Closed-source LLMs}} \\
\hline
Claude-4-Sonnet & 0.3007\vstd{.0308} & 0.3003\vstd{.0167} & 0.3020\vstd{.0139} & 0.3009\vstd{.0455} & 0.2922\vstd{.0380} & 0.3112\vstd{.0496} & 0.5686\vstd{.0378} & 0.3640\vstd{.0267} \\
Claude-4.5-Sonnet & 0.4552\vstd{.0270} & 0.5651\vstd{.0467} & 0.3845\vstd{.0481} & 0.4959\vstd{.0388} & 0.2328\vstd{.0128} & 0.3399\vstd{.0476} & 0.4190\vstd{.0206} & 0.3060\vstd{.0443} \\
Gemini-2.5-pro & 0.2641\vstd{.0169} & 0.3794\vstd{.0518} & 0.3933\vstd{.0279} & 0.3246\vstd{.0527} & 0.4373\vstd{.0238} & 0.7370\vstd{.0444} & 0.5978\vstd{.0393} & 0.5562\vstd{.0535} \\
Gemini-3.0-pro & \textit{0.5991}\vstd{.0580} & \textit{0.7734}\vstd{.0493} & \textit{0.6868}\vstd{.0306} & \textit{0.6707}\vstd{.0419} & 0.4954\vstd{.0366} & 0.7340\vstd{.0396} & 0.5956\vstd{.0255} & 0.5839\vstd{.0354} \\
GPT-5 & 0.5857\vstd{.0428} & 0.7340\vstd{.0456} & 0.6443\vstd{.0391} & 0.6437\vstd{.0499} & 0.5640\vstd{.0465} & 0.6585\vstd{.0271} & 0.5601\vstd{.0357} & 0.5885\vstd{.0439} \\
GPT-5.1 & 0.5581\vstd{.0436} & 0.7069\vstd{.0544} & 0.6313\vstd{.0489} & 0.6193\vstd{.0523} & 0.6133\vstd{.0539} & 0.7189\vstd{.0421} & \textit{0.6035}\vstd{.0459} & 0.6394\vstd{.0266} \\
GPT-5.2 & 0.5680\vstd{.0452} & 0.7124\vstd{.0529} & 0.6321\vstd{.0304} & 0.6263\vstd{.0460} & \textit{0.6349}\vstd{.0319} & \textit{0.7717}\vstd{.0313} & 0.5772\vstd{.0479} & \textit{0.6578}\vstd{.0578} \\
OpenAI-DeepResearch & 0.5966\vstd{.0553} & 0.7439\vstd{.0299} & 0.6617\vstd{.0392} & 0.6564\vstd{.0402} & 0.5800\vstd{.0469} & 0.7223\vstd{.0478} & 0.5696\vstd{.0528} & 0.6158\vstd{.0339} \\
Perplexity-DeepResearch & 0.4743\vstd{.0418} & 0.5719\vstd{.0285} & 0.5644\vstd{.0348} & 0.5232\vstd{.0492} & 0.5676\vstd{.0356} & 0.6844\vstd{.0541} & 0.5582\vstd{.0430} & 0.5969\vstd{.0508} \\
\hline
\rowcolor{skyblue}\multicolumn{9}{c}{\textit{Open-source LLMs}} \\
\hline
A-Searcher & 0.4343\vstd{.0571} & 0.6716\vstd{.0394} & 0.5369\vstd{.0386} & 0.5291\vstd{.0221} & 0.4942\vstd{.0375} & 0.7501\vstd{.0358} & 0.5097\vstd{.0410} & 0.5672\vstd{.0547} \\
DR-Tulu & 0.3450\vstd{.0153} & 0.5778\vstd{.0400} & 0.4743\vstd{.0320} & 0.4433\vstd{.0194} & 0.4046\vstd{.0483} & 0.7906\vstd{.0453} & 0.4100\vstd{.0323} & 0.5101\vstd{.0205} \\
Mirothinker-v1.5-30B & 0.3930\vstd{.0316} & 0.5672\vstd{.0283} & 0.4922\vstd{.0203} & 0.4671\vstd{.0377} & 0.4896\vstd{.0448} & 0.7745\vstd{.0519} & \textit{0.5316}\vstd{.0230} & 0.5764\vstd{.0249} \\
Tongyi-Deepresearch & 0.3541\vstd{.0234} & 0.5080\vstd{.0291} & 0.4006\vstd{.0429} & 0.4123\vstd{.0262} & 0.2749\vstd{.0497} & 0.4564\vstd{.0440} & 0.4372\vstd{.0233} & 0.3624\vstd{.0362} \\
Web-Thinker & \textit{0.4936}\vstd{.0310} & \textit{0.8156}\vstd{.0475} & 0.5314\vstd{.0562} & \textit{0.6028}\vstd{.0470} & 0.2044\vstd{.0141} & 0.3786\vstd{.0403} & 0.3234\vstd{.0324} & 0.2798\vstd{.0208} \\
Web-Dancer & 0.4902\vstd{.0236} & 0.7726\vstd{.0551} & \textit{0.5788}\vstd{.0574} & 0.5966\vstd{.0300} & \textit{0.5315}\vstd{.0342} & 0.8094\vstd{.0390} & 0.5232\vstd{.0281} & \textit{0.6042}\vstd{.0557} \\
Web-Explorer & 0.2792\vstd{.0201} & 0.3742\vstd{.0415} & 0.3628\vstd{.0335} & 0.3250\vstd{.0250} & 0.4017\vstd{.0404} & 0.5582\vstd{.0572} & 0.5173\vstd{.0364} & 0.4722\vstd{.0360} \\
Web-Shaper & 0.2445\vstd{.0487} & 0.3652\vstd{.0442} & 0.2754\vstd{.0450} & 0.2888\vstd{.0413} & 0.2224\vstd{.0217} & 0.3574\vstd{.0173} & 0.4290\vstd{.0287} & 0.3087\vstd{.0239} \\
OpenScholar-8B & 0.1845\vstd{.0121} & 0.2102\vstd{.0423} & 0.1956\vstd{.0156} & 0.1967\vstd{.0331} & 0.3850\vstd{.0383} & 0.4621\vstd{.0361} & 0.3540\vstd{.0231} & 0.4003\vstd{.0471} \\
O-Researcher-72B-rl & 0.4742\vstd{.0485} & 0.7599\vstd{.0355} & 0.5501\vstd{.0525} & 0.5796\vstd{.0384} & 0.4726\vstd{.0274} & \textit{0.8676}\vstd{.0466} & 0.5055\vstd{.0225} & 0.5860\vstd{.0363} \\
O-Researcher-72B-sft & 0.4705\vstd{.0273} & 0.7557\vstd{.0387} & 0.5412\vstd{.0243} & 0.5749\vstd{.0449} & 0.3562\vstd{.0252} & 0.6834\vstd{.0426} & 0.4969\vstd{.0224} & 0.4784\vstd{.0517} \\
\hline
\rowcolor{lightorange}\multicolumn{9}{c}{\textit{SciLENS (Ours)}} \\
\hline
\rowcolor{lightorange!30}Qwen3-30B-A3B (Base + Tools) & 0.1232\vstd{.0309} & 0.1419\vstd{.0130} & 0.1401\vstd{.0432} & 0.1324\vstd{.0277} & 0.1523\vstd{.0154} & 0.1834\vstd{.0237} & 0.1605\vstd{.0119} & 0.1654\vstd{.0317} \\
\rowcolor{lightorange!30}SciLENS-30B (SFT) & 0.6336\vstd{.0269} & 0.9054\vstd{.0382} & 0.6820\vstd{.0501} & 0.7290\vstd{.0398} & 0.6301\vstd{.0472} & 0.7863\vstd{.0537} & 0.5834\vstd{.0567} & 0.6611\vstd{.0434} \\
\rowcolor{lightorange!30}SciLENS-30B (RL) & \textbf{0.6811}\vstd{.0579} & \textbf{0.9127}\vstd{.0534} & \textbf{0.7138}\vstd{.0370} & \textbf{0.7607}\vstd{.0569} & \textbf{0.7395}\vstd{.0326} & \textbf{0.8364}\vstd{.0512} & \textbf{0.7144}\vstd{.0416} & \textbf{0.7594}\vstd{.0575} \\
\bottomrule
\end{tabular}
\end{adjustbox}
\label{tab:overall_benchmark}
\end{subtable}

\vspace{2mm}

\begin{subtable}{\textwidth}
\centering
\subcaption{Standard Scientific Reasoning Benchmarks}
\vspace{-1mm}
\begin{adjustbox}{width=\linewidth}
\begin{tabular}{l|cc|cc|cc|cc}
\toprule
\rowcolor{lightlavender} & \multicolumn{2}{c|}{\textbf{QASA}} & \multicolumn{2}{c|}{\textbf{SciFact}} & \multicolumn{2}{c|}{\textbf{PubMedQA}} & \multicolumn{2}{c}{\textbf{ScholarQA CS}} \\
\rowcolor{lightlavender}\cline{2-9}
\rowcolor{lightlavender}\multirow{-2}{*}{\textbf{Model}} & Corr & Cite & Corr & Cite & Corr & Cite & Corr & Cite \\
\hline
\rowcolor{skyblue}\multicolumn{9}{c}{\textit{Closed-source LLMs}}\\
\hline
Claude-4-Sonnet & 33.42\vstd{0.66} & 31.41\vstd{1.94} & 60.58\vstd{2.52} & 42.39\vstd{2.24} & 50.96\vstd{2.45} & 45.32\vstd{2.39} & 39.52\vstd{1.67} & 23.18\vstd{1.50} \\
Claude-4.5-Sonnet & \textit{46.27}\vstd{2.54} & 29.03\vstd{2.22} & \textit{86.06}\vstd{1.33} & 40.33\vstd{1.55} & \textit{73.90}\vstd{0.87} & 43.29\vstd{2.08} & 67.92\vstd{1.07} & 25.64\vstd{1.22} \\
Gemini-2.5-pro & 30.49\vstd{0.98} & 27.66\vstd{1.00} & 47.60\vstd{1.35} & 49.59\vstd{1.66} & 46.62\vstd{2.62} & 69.84\vstd{1.93} & 46.44\vstd{2.34} & \textit{60.83}\vstd{1.25} \\
Gemini-3.0-pro & 31.10\vstd{1.82} & 33.20\vstd{1.72} & 69.71\vstd{2.88} & \textit{67.13}\vstd{2.21} & 48.87\vstd{0.84} & \textit{72.13}\vstd{1.16} & 45.73\vstd{1.68} & 49.46\vstd{2.82} \\
GPT-5 & 44.14\vstd{0.77} & 16.83\vstd{1.11} & 82.21\vstd{2.16} & 50.03\vstd{2.89} & 69.75\vstd{2.57} & 58.71\vstd{2.18} & 68.34\vstd{1.43} & 20.21\vstd{1.08} \\
GPT-5.1 & 41.08\vstd{2.19} & \textit{35.65}\vstd{2.01} & 83.65\vstd{1.54} & 48.53\vstd{1.95} & 68.62\vstd{1.17} & 35.14\vstd{1.14} & \textit{73.27}\vstd{1.26} & 28.75\vstd{1.51} \\
GPT-5.2 & 44.22\vstd{1.85} & 26.79\vstd{1.74} & 80.29\vstd{2.04} & 33.98\vstd{1.63} & 65.95\vstd{1.52} & 48.19\vstd{0.94} & 71.88\vstd{2.12} & 32.91\vstd{1.19} \\
OpenAI-DeepResearch & 38.92\vstd{1.12} & 17.02\vstd{1.75} & 73.56\vstd{1.28} & 42.94\vstd{2.37} & 60.02\vstd{1.41} & 55.20\vstd{1.27} & 66.20\vstd{0.88} & 24.32\vstd{0.92} \\
Perplexity-DeepResearch & 28.57\vstd{2.25} & 14.44\vstd{1.70} & 55.77\vstd{1.61} & 25.05\vstd{1.99} & 49.77\vstd{1.01} & 36.95\vstd{2.67} & 65.41\vstd{2.30} & 28.10\vstd{0.67} \\
\hline
\rowcolor{skyblue}\multicolumn{9}{c}{\textit{Open-source LLMs}}\\
\hline
A-Searcher & 31.97\vstd{1.87} & 32.52\vstd{1.39} & 59.13\vstd{0.83} & 54.02\vstd{1.96} & 44.48\vstd{2.78} & 65.65\vstd{2.61} & 40.61\vstd{2.47} & 47.87\vstd{1.29} \\
DR-Tulu & 25.26\vstd{0.68} & 8.10\vstd{1.64} & 70.48\vstd{2.56} & 26.02\vstd{2.46} & 59.79\vstd{2.31} & 30.50\vstd{2.23} & 51.10\vstd{2.85} & \textit{72.34}\vstd{1.34} \\
Mirothinker-v1.5-30B & \textit{42.95}\vstd{1.76} & 33.59\vstd{1.09} & \textit{80.75}\vstd{1.88} & 58.52\vstd{2.35} & \textit{69.80}\vstd{2.84} & 62.92\vstd{2.20} & \textit{66.21}\vstd{1.65} & 39.45\vstd{1.13} \\
Tongyi-Deepresearch & 34.18\vstd{1.45} & 34.04\vstd{1.42} & 61.25\vstd{2.80} & \textit{70.30}\vstd{1.81} & 51.44\vstd{2.43} & 62.01\vstd{1.58} & 40.17\vstd{2.42} & 46.73\vstd{0.72} \\
Web-Thinker & 10.66\vstd{0.55} & 20.42\vstd{1.79} & 33.65\vstd{1.97} & 49.46\vstd{0.93} & 10.20\vstd{1.69} & 62.16\vstd{1.46} & 14.55\vstd{1.59} & 26.52\vstd{2.29} \\
Web-Dancer & 35.49\vstd{1.77} & 27.94\vstd{1.92} & 67.79\vstd{2.68} & 51.53\vstd{1.62} & 58.78\vstd{1.47} & 59.11\vstd{2.71} & 51.94\vstd{2.49} & 33.47\vstd{2.14} \\
Web-Explorer & 27.68\vstd{2.06} & 15.59\vstd{1.05} & 57.08\vstd{1.86} & 26.01\vstd{2.28} & 51.72\vstd{1.31} & 21.03\vstd{1.20} & 36.90\vstd{2.76} & 26.03\vstd{0.76} \\
Web-Shaper & 18.75\vstd{0.79} & \textit{49.69}\vstd{1.10} & 24.04\vstd{1.06} & 68.80\vstd{2.95} & 19.47\vstd{2.15} & \textit{74.05}\vstd{2.02} & 22.43\vstd{1.83} & 50.84\vstd{2.65} \\
OpenScholar-8B & 24.56\vstd{1.78} & 38.79\vstd{1.73} & 77.52\vstd{2.09} & 51.28\vstd{0.89} & 65.49\vstd{2.48} & 60.32\vstd{2.73} & 50.28\vstd{0.97} & 54.36\vstd{2.36} \\
O-Researcher-72B-sft & 28.99\vstd{1.90} & 18.40\vstd{1.53} & 48.56\vstd{0.85} & 40.83\vstd{2.07} & 27.09\vstd{0.82} & 29.84\vstd{2.75} & 44.29\vstd{2.66} & 27.62\vstd{2.55} \\
O-Researcher-72B-rl & 35.51\vstd{2.05} & 28.78\vstd{0.65} & 74.23\vstd{2.38} & 25.04\vstd{0.95} & 55.52\vstd{2.60} & 33.51\vstd{1.15} & 59.34\vstd{2.63} & 71.22\vstd{2.03} \\
\hline
\rowcolor{lightorange}\multicolumn{9}{c}{\textit{SciLENS (Ours)}}\\
\hline
\rowcolor{lightorange!30}Qwen3-30B-A3B (Base + Tools) & 5.90\vstd{0.53} & 2.44\vstd{0.57} & 14.42\vstd{1.48} & 10.21\vstd{0.54} & 14.95\vstd{1.03} & 7.71\vstd{1.37} & 5.44\vstd{2.58} & 5.68\vstd{0.69} \\
\rowcolor{lightorange!30}SciLENS-30B (SFT) & 46.43\vstd{2.87} & 50.20\vstd{1.32} & 85.87\vstd{2.50} & 79.03\vstd{1.80} & 73.84\vstd{2.40} & 72.97\vstd{2.10} & 66.52\vstd{2.90} & 75.03\vstd{2.70} \\
\rowcolor{lightorange!30}SciLENS-30B (RL) & \textbf{47.62}\vstd{2.27} & \textbf{52.16}\vstd{2.91} & \textbf{88.94}\vstd{1.40} & \textbf{83.72}\vstd{2.41} & \textbf{77.53}\vstd{2.72} & \textbf{76.54}\vstd{2.33} & \textbf{73.98}\vstd{1.24} & \textbf{76.82}\vstd{0.90} \\
\bottomrule
\end{tabular}
\end{adjustbox}
\label{tab:scientific_benchmarks}
\end{subtable}
\end{table*}

\subsection{Ablation Studies}
\label{subsec:ablation}
We conduct two complementary ablation studies under a unified toolbox protocol to isolate the contribution of each component. Since they measure orthogonal dimensions, their magnitudes are not directly comparable. The tool-necessity ablation (Appendix~\ref{app:tool_ablation}) shows that complex scientific synthesis cannot be solved without retrieval: removing retrieval and graph tools reduces the agent to a closed-book setting, leading to consistent performance collapse across benchmarks.
This confirms that our benchmarks require multi-hop evidence gathering rather than parametric recall. In contrast, the RL ablation measures algorithmic gains over a strong SFT baseline with identical tool access across settings; thus, observed improvements reflect purely training-driven contributions.

As shown in Tables~\ref{tab:overall_benchmark} and~\ref{tab:scientific_benchmarks}, Qwen3-30B-A3B (Base + Tools) demonstrates that tool access alone is insufficient for scientific synthesis, yielding only 0.1324 on SSB and 0.1654 on SciFR. This outcome is expected, as the Qwen3-30B-A3B base model inherently lacks robust multi-turn tool orchestration capabilities prior to SFT alignment, frequently failing to parse tool responses or chain sequential retrievals coherently. Supervised fine-tuning substantially improves performance across all benchmarks, while rubric-based reinforcement learning further yields consistent gains, including notable improvements on ScholarQA-CS (+7.46) and SciFR (+0.098). Table~\ref{tab:ablation} provides a detailed breakdown of RL components.
Answer RLVR provides the core correctness signal; its removal leads to substantial performance drops across all benchmarks, indicating that outcome-based supervision is necessary for effective policy optimization.
The planning and grounding rubrics each yield consistent gains, particularly at the upper performance range. Notably, removing planning rubrics substantially degrades SSB performance (0.7607 → 0.6855), indicating that process-level supervision is essential for complex structural synthesis.
 A complementary tool necessity ablation (Appendix~\ref{app:tool_ablation}) confirms that retrieval and visualization tools are essential prerequisites: removing visualization tools alone causes SSB to drop sharply from 0.7607 to 0.5215 while standard QA benchmarks remain comparatively stable, directly validating the visualization-as-tool contribution for structural synthesis.

\begin{table}[!t]
\centering
\small
\vspace{-2mm}
\caption{RL component ablation of SciLENS. All configurations use the full \textsc{ResearchToolbox} and SFT initialization. Metrics represent Correctness for standard benchmarks and Overall for SciFR and SSB ($p < 0.05$).}
\vspace{-2mm}
\begin{adjustbox}{width=1\linewidth}
\begin{tabular}{l|cccccc}
\toprule
\rowcolor{lightlavender}\textbf{Configuration} & \textbf{QASA} & \textbf{SciFact} & \textbf{PubMedQA} & \textbf{ScholarQA} & \textbf{SciFR} & \textbf{SSB} \\
\midrule
\textbf{SciLENS (Full RL)} & \textbf{47.62} & \textbf{88.94} & \textbf{77.53} & \textbf{73.98} & \textbf{0.7594} & \textbf{0.7607} \\
\midrule
w/o Answer RLVR & 41.20 & 72.45 & 64.33 & 62.10 & 0.7115 & 0.7350 \\
w/o Planning Rubrics & 47.10 & 86.80 & 74.90 & 70.40 & 0.7310 & 0.6855 \\
w/o Grounding Rubrics & 46.50 & 86.17 & 75.80 & 72.40 & 0.7321 & 0.7020 \\
\bottomrule
\end{tabular}
\end{adjustbox}
\label{tab:ablation}
\end{table}

\subsection{Extended Analytical Experiments}
We further analyze SciLENS through long-context management (Appendix~\ref{app:context}), tool utilization dynamics (Appendix~\ref{app:tool_dynamics}), and RL convergence analysis (Appendix~\ref{app:rl_convergence}). Qualitative case studies and error analyses are provided in Appendix~\ref{app:qualitative}.

\section{Related Work}
\label{sec:related_work}

\subsection{Autonomous Online Search Agents}
LLMs have advanced biomedical \cite{luo2022biogpt, chen2023huatuogpt, yang2022large, tu2024towards}, medical \cite{zhou2023survey, wei2024evaluation, singhal2025toward}, geoscience \cite{deng2024k2}, astronomy \cite{nguyen2023astrollama}, and multidisciplinary domains \cite{sun2024scieval, zhang2024sciglm, cai2024uni}. Beyond static queries, models now automate code generation \cite{gero2022sparks, du2024evaluating} and research ideation \cite{baek2025researchagent, kumar2025can}, culminating in autonomous agents \cite{wu2025webdancer,liu2025webexplorer} that execute complex literature reviews via retrieval augmented pipelines \cite{zhou2025autonomous, rouzrokh2025lattereview, ma2024sciagent}. However, live web search introduces latency, rate limits, and ephemeral content, limiting reproducibility and high-throughput literature synthesis~\cite{zhang2025deep}. In contrast to the live citation-walk approaches of WebDancer~\cite{wu2025webdancer} and WebShaper~\cite{tao2025webshaper}, our method samples from a static citation graph, avoiding API constraints and enabling consensus verification across 20 frontier models. DR-Tulu~\cite{shao2025dr} uses rubric-based reinforcement learning for alignment; we extend this paradigm with reverse-decomposition rubrics that exploit the history of synthetic questions. This provides fine-grained process supervision for early planning and evidence grounding, rather than just scoring final outputs. We adopt the IcePop algorithm~\cite{team2025every} unchanged for optimization stability.

\subsection{Offline Academic Retrieval Frameworks}
To circumvent online instabilities, localized scientific retrieval frameworks like OpenScholar \cite{asai2024openscholar} and SciRAG \cite{vasantharajan2025scirag} index massive static corpora \cite{chen2026deepera, liu2025robust}, evaluating on benchmarks such as SciFact \cite{wadden2020fact}, QASPER \cite{dasigi2021dataset}, and QASA \cite{lee2023qasa}. While eliminating network latency, they rely on single-turn retrieval that fails to exploit citation networks for multihop reasoning \cite{xu2024kiwi}. Furthermore, these systems isolate granular QA from macro-level summarization \cite{lu2020multi}. When synthesizing broad trends, reliance on parametric memory can induce hallucinations~\cite{alansari2025large}, while linear generation leads to context overflow. Without structural or visual compression, these systems underutilize offline corpora and remain limited for comprehensive scientific discovery.

\section{Conclusion}
\label{sec:conclusion}
We present SciLENS, a fully localized agent trained with rubric-based reinforcement learning that achieves scientific synthesis performance comparable to frontier proprietary models. It integrates structural visualization into reasoning to compress complex topology into validated chart schemas, and uses a reverse-decomposition rubric to provide fine-grained process rewards for improved planning and grounding. Built on a dual-tier infrastructure indexing 12M academic records, SciLENS outperforms open-source baselines and approaches GPT-5.2 level performance on scientific QA and synthesis tasks.

\section*{Acknowledgments}
This work is supported in part by the National Natural Science Foundation of China (No.~62372264 and No.~92467203 ). Chaokun Wang is the corresponding author.

\clearpage
\newpage

\section*{Limitations}

While SciLENS demonstrates preliminary capabilities in complex academic reasoning and structural synthesis, its primary limitation involves the transition from a localized experimental sandbox to a live industrial deployment. 

\section*{Ethical Considerations}

SciLENS operates exclusively on the Open Academic Graph, a publicly accessible and openly licensed scholarly corpus, and does not involve human subjects, private data, or dual-use applications.         

\bibliography{custom}

@article{yao2026researcher,
  title={O-Researcher: An Open Ended Deep Research Model via Multi-Agent Distillation and Agentic RL},
  author={Yao, Yi and Zhu, He and Wang, Piaohong and Ren, Jincheng and Yang, Xinlong and Chen, Qianben and Li, Xiaowan and Shi, Dingfeng and Li, Jiaxian and Wang, Qiexiang and others},
  journal={arXiv preprint arXiv:2601.03743},
  year={2026}
}

@article{gao2025beyond,
  title={Beyond ten turns: Unlocking long-horizon agentic search with large-scale asynchronous rl},
  author={Gao, Jiaxuan and Fu, Wei and Xie, Minyang and Xu, Shusheng and He, Chuyi and Mei, Zhiyu and Zhu, Banghua and Wu, Yi},
  journal={arXiv preprint arXiv:2508.07976},
  year={2025}
}

@article{shao2025dr,
  title={Dr tulu: Reinforcement learning with evolving rubrics for deep research},
  author={Shao, Rulin and Asai, Akari and Shen, Shannon Zejiang and Ivison, Hamish and Kishore, Varsha and Zhuo, Jingming and Zhao, Xinran and Park, Molly and Finlayson, Samuel G and Sontag, David and others},
  journal={arXiv preprint arXiv:2511.19399},
  year={2025}
}

@article{team2025mirothinker,
  title={Mirothinker: Pushing the performance boundaries of open-source research agents via model, context, and interactive scaling},
  author={Team, MiroMind and Bai, Song and Bing, Lidong and Chen, Carson and Chen, Guanzheng and Chen, Yuntao and Chen, Zhe and Chen, Ziyi and Dai, Jifeng and Dong, Xuan and others},
  journal={arXiv preprint arXiv:2511.11793},
  year={2025}
}

@article{team2025tongyi,
  title={Tongyi deepresearch technical report},
  author={Team, Tongyi DeepResearch and Li, Baixuan and Zhang, Bo and Zhang, Dingchu and Huang, Fei and Li, Guangyu and Chen, Guoxin and Yin, Huifeng and Wu, Jialong and Zhou, Jingren and others},
  journal={arXiv preprint arXiv:2510.24701},
  year={2025}
}

@article{li2025webthinker,
  title={Webthinker: Empowering large reasoning models with deep research capability},
  author={Li, Xiaoxi and Jin, Jiajie and Dong, Guanting and Qian, Hongjin and Wu, Yongkang and Wen, Ji-Rong and Zhu, Yutao and Dou, Zhicheng},
  journal={arXiv preprint arXiv:2504.21776},
  year={2025}
}

@article{liu2025webexplorer,
  title={Webexplorer: Explore and evolve for training long-horizon web agents},
  author={Liu, Junteng and Li, Yunji and Zhang, Chi and Li, Jingyang and Chen, Aili and Ji, Ke and Cheng, Weiyu and Wu, Zijia and Du, Chengyu and Xu, Qidi and others},
  journal={arXiv preprint arXiv:2509.06501},
  year={2025}
}

@article{wu2025webdancer,
  title={Webdancer: Towards autonomous information seeking agency},
  author={Wu, Jialong and Li, Baixuan and Fang, Runnan and Yin, Wenbiao and Zhang, Liwen and Tao, Zhengwei and Zhang, Dingchu and Xi, Zekun and Fu, Gang and Jiang, Yong and others},
  journal={arXiv preprint arXiv:2505.22648},
  year={2025}
}

@article{tao2025webshaper,
  title={Webshaper: Agentically data synthesizing via information-seeking formalization},
  author={Tao, Zhengwei and Wu, Jialong and Yin, Wenbiao and Zhang, Junkai and Li, Baixuan and Shen, Haiyang and Li, Kuan and Zhang, Liwen and Wang, Xinyu and Jiang, Yong and others},
  journal={arXiv preprint arXiv:2507.15061},
  year={2025}
}

@article{yang2025qwen3,
  title={Qwen3 technical report},
  author={Yang, An and Li, Anfeng and Yang, Baosong and Zhang, Beichen and Hui, Binyuan and Zheng, Bo and Yu, Bowen and Gao, Chang and Huang, Chengen and Lv, Chenxu and others},
  journal={arXiv preprint arXiv:2505.09388},
  year={2025}
}

@article{zhang2025qwen3,
  title={Qwen3 embedding: Advancing text embedding and reranking through foundation models},
  author={Zhang, Yanzhao and Li, Mingxin and Long, Dingkun and Zhang, Xin and Lin, Huan and Yang, Baosong and Xie, Pengjun and Yang, An and Liu, Dayiheng and Lin, Junyang and others},
  journal={arXiv preprint arXiv:2506.05176},
  year={2025}
}

@article{luo2022biogpt,
  title={BioGPT: generative pre-trained transformer for biomedical text generation and mining},
  author={Luo, Renqian and Sun, Liai and Xia, Yingce and Qin, Tao and Zhang, Sheng and Poon, Hoifung and Liu, Tie-Yan},
  journal={Briefings in bioinformatics},
  volume={23},
  number={6},
  pages={bbac409},
  year={2022},
  publisher={Oxford University Press}
}

@article{chen2023huatuogpt,
  title={Huatuogpt-ii, one-stage training for medical adaption of llms},
  author={Chen, Junying and Wang, Xidong and Ji, Ke and Gao, Anningzhe and Jiang, Feng and Chen, Shunian and Zhang, Hongbo and Song, Dingjie and Xie, Wenya and Kong, Chuyi and others},
  journal={arXiv preprint arXiv:2311.09774},
  year={2023}
}

@article{yang2022large,
  title={A large language model for electronic health records},
  author={Yang, Xi and Chen, Aokun and PourNejatian, Nima and Shin, Hoo Chang and Smith, Kaleb E and Parisien, Christopher and Compas, Colin and Martin, Cheryl and Costa, Anthony B and Flores, Mona G and others},
  journal={NPJ digital medicine},
  volume={5},
  number={1},
  pages={194},
  year={2022},
  publisher={Nature Publishing Group UK London}
}

@article{tu2024towards,
  title={Towards generalist biomedical AI},
  author={Tu, Tao and Azizi, Shekoofeh and Driess, Danny and Schaekermann, Mike and Amin, Mohamed and Chang, Pi-Chuan and Carroll, Andrew and Lau, Charles and Tanno, Ryutaro and Ktena, Ira and others},
  journal={Nejm Ai},
  volume={1},
  number={3},
  pages={AIoa2300138},
  year={2024},
  publisher={Massachusetts Medical Society}
}

@article{singhal2025toward,
  title={Toward expert-level medical question answering with large language models},
  author={Singhal, Karan and Tu, Tao and Gottweis, Juraj and Sayres, Rory and Wulczyn, Ellery and Amin, Mohamed and Hou, Le and Clark, Kevin and Pfohl, Stephen R and Cole-Lewis, Heather and others},
  journal={Nature medicine},
  volume={31},
  number={3},
  pages={943--950},
  year={2025},
  publisher={Nature Publishing Group US New York}
}

@article{zhou2023survey,
  title={A survey of large language models in medicine: Progress, application, and challenge},
  author={Zhou, Hongjian and Liu, Fenglin and Gu, Boyang and Zou, Xinyu and Huang, Jinfa and Wu, Jinge and Li, Yiru and Chen, Sam S and Zhou, Peilin and Liu, Junling and others},
  journal={arXiv preprint arXiv:2311.05112},
  year={2023}
}

@article{wei2024evaluation,
  title={Evaluation of ChatGPT-generated medical responses: a systematic review and meta-analysis},
  author={Wei, Qiuhong and Yao, Zhengxiong and Cui, Ying and Wei, Bo and Jin, Zhezhen and Xu, Ximing},
  journal={Journal of biomedical informatics},
  volume={151},
  pages={104620},
  year={2024},
  publisher={Elsevier}
}

@inproceedings{deng2024k2,
  title={K2: A foundation language model for geoscience knowledge understanding and utilization},
  author={Deng, Cheng and Zhang, Tianhang and He, Zhongmou and Chen, Qiyuan and Shi, Yuanyuan and Xu, Yi and Fu, Luoyi and Zhang, Weinan and Wang, Xinbing and Zhou, Chenghu and others},
  booktitle={Proceedings of the 17th ACM international conference on web search and data mining},
  pages={161--170},
  year={2024}
}

@inproceedings{nguyen2023astrollama,
  title={Astrollama: Towards specialized foundation models in astronomy},
  author={Nguyen, Tuan Dung and Ting, Yuan-Sen and Ciuca, Ioana and O’Neill, Charles and Sun, Ze-Chang and Jab{\l}o{\'n}ska, Maja and Kruk, Sandor and Perkowski, Ernest and Miller, Jack and Li, Jason Jason Jingsh and others},
  booktitle={Proceedings of the Second Workshop on Information Extraction from Scientific Publications},
  pages={49--55},
  year={2023}
}

@inproceedings{sun2024scieval,
  title={Scieval: A multi-level large language model evaluation benchmark for scientific research},
  author={Sun, Liangtai and Han, Yang and Zhao, Zihan and Ma, Da and Shen, Zhennan and Chen, Baocai and Chen, Lu and Yu, Kai},
  booktitle={Proceedings of the AAAI Conference on Artificial Intelligence},
  volume={38},
  number={17},
  pages={19053--19061},
  year={2024}
}

@article{zhang2024sciglm,
  title={Sciglm: Training scientific language models with self-reflective instruction annotation and tuning},
  author={Zhang, Dan and Hu, Ziniu and Zhoubian, Sining and Du, Zhengxiao and Yang, Kaiyu and Wang, Zihan and Yue, Yisong and Dong, Yuxiao and Tang, Jie},
  journal={arXiv preprint arXiv:2401.07950},
  volume={4},
  year={2024}
}

@article{cai2024uni,
  title={Uni-SMART: universal science multimodal analysis and research transformer},
  author={Cai, Hengxing and Cai, Xiaochen and Yang, Shuwen and Wang, Jiankun and Yao, Lin and Gao, Zhifeng and Chang, Junhan and Li, Sihang and Xu, Mingjun and Wang, Changxin and others},
  journal={arXiv preprint arXiv:2403.10301},
  year={2024}
}

@inproceedings{gero2022sparks,
  title={Sparks: Inspiration for science writing using language models},
  author={Gero, Katy Ilonka and Liu, Vivian and Chilton, Lydia},
  booktitle={Proceedings of the 2022 ACM Designing Interactive Systems Conference},
  pages={1002--1019},
  year={2022}
}

@inproceedings{du2024evaluating,
  title={Evaluating large language models in class-level code generation},
  author={Du, Xueying and Liu, Mingwei and Wang, Kaixin and Wang, Hanlin and Liu, Junwei and Chen, Yixuan and Feng, Jiayi and Sha, Chaofeng and Peng, Xin and Lou, Yiling},
  booktitle={Proceedings of the IEEE/ACM 46th International Conference on Software Engineering},
  pages={1--13},
  year={2024}
}

@inproceedings{baek2025researchagent,
  title={Researchagent: Iterative research idea generation over scientific literature with large language models},
  author={Baek, Jinheon and Jauhar, Sujay Kumar and Cucerzan, Silviu and Hwang, Sung Ju},
  booktitle={Proceedings of the 2025 Conference of the Nations of the Americas Chapter of the Association for Computational Linguistics: Human Language Technologies (Volume 1: Long Papers)},
  pages={6709--6738},
  year={2025}
}

@inproceedings{kumar2025can,
  title={Can large language models unlock novel scientific research ideas?},
  author={Kumar, Sandeep and Ghosal, Tirthankar and Goyal, Vinayak and Ekbal, Asif},
  booktitle={Proceedings of the 2025 Conference on Empirical Methods in Natural Language Processing},
  pages={33551--33575},
  year={2025}
}

@article{zhou2025autonomous,
  title={Autonomous agents for scientific discovery: Orchestrating scientists, language, code, and physics},
  author={Zhou, Lianhao and Ling, Hongyi and Fu, Cong and Huang, Yepeng and Sun, Michael and Yu, Wendi and Wang, Xiaoxuan and Li, Xiner and Su, Xingyu and Zhang, Junkai and others},
  journal={arXiv preprint arXiv:2510.09901},
  year={2025}
}

@article{rouzrokh2025lattereview,
  title={LatteReview: a multi-agent framework for systematic review automation using large language models},
  author={Rouzrokh, Pouria and Khosravi, Bardia and Rouzrokh, Parsa and Shariatnia, Moein},
  journal={arXiv preprint arXiv:2501.05468},
  year={2025}
}

@inproceedings{ma2024sciagent,
  title={Sciagent: Tool-augmented language models for scientific reasoning},
  author={Ma, Yubo and Gou, Zhibin and Hao, Junheng and Xu, Ruochen and Wang, Shuohang and Pan, Liangming and Yang, Yujiu and Cao, Yixin and Sun, Aixin},
  booktitle={Proceedings of the 2024 conference on empirical methods in natural language processing},
  pages={15701--15736},
  year={2024}
}

@article{zhang2025deep,
  title={Deep research: A survey of autonomous research agents},
  author={Zhang, Wenlin and Li, Xiaopeng and Zhang, Yingyi and Jia, Pengyue and Wang, Yichao and Guo, Huifeng and Liu, Yong and Zhao, Xiangyu},
  journal={arXiv preprint arXiv:2508.12752},
  year={2025}
}

@article{asai2024openscholar,
  title={Openscholar: Synthesizing scientific literature with retrieval-augmented lms},
  author={Asai, Akari and He, Jacqueline and Shao, Rulin and Shi, Weijia and Singh, Amanpreet and Chang, Joseph Chee and Lo, Kyle and Soldaini, Luca and Feldman, Sergey and D'arcy, Mike and others},
  journal={arXiv preprint arXiv:2411.14199},
  year={2024}
}

@phdthesis{vasantharajan2025scirag,
  title={SciRAG: A Retrieval-Focused Fine-Tuning Strategy for Scientific Documents},
  author={Vasantharajan, Charangan},
  year={2025}
}

@article{chen2026deepera,
  title={DeepEra: A Deep Evidence Reranking Agent for Scientific Retrieval-Augmented Generated Question Answering},
  author={Chen, Haotian and Long, Qingqing and Pu, Siyu and Luo, Xiao and Ju, Wei and Xiao, Meng and Zhou, Yuanchun and Zhao, Jianghua and Wang, Xuezhi},
  journal={arXiv preprint arXiv:2601.16478},
  year={2026}
}

@inproceedings{liu2025robust,
  title={Robust information retrieval},
  author={Liu, Yu-An and Zhang, Ruqing and Guo, Jiafeng and de Rijke, Maarten},
  booktitle={Proceedings of the Eighteenth ACM International Conference on Web Search and Data Mining},
  pages={1008--1011},
  year={2025}
}

@inproceedings{wadden2020fact,
  title={Fact or fiction: Verifying scientific claims},
  author={Wadden, David and Lin, Shanchuan and Lo, Kyle and Wang, Lucy Lu and van Zuylen, Madeleine and Cohan, Arman and Hajishirzi, Hannaneh},
  booktitle={Proceedings of the 2020 Conference on Empirical Methods in Natural Language Processing (EMNLP)},
  pages={7534--7550},
  year={2020}
}

@inproceedings{dasigi2021dataset,
  title={A dataset of information-seeking questions and answers anchored in research papers},
  author={Dasigi, Pradeep and Lo, Kyle and Beltagy, Iz and Cohan, Arman and Smith, Noah A and Gardner, Matt},
  booktitle={Proceedings of the 2021 Conference of the North American Chapter of the Association for Computational Linguistics: Human Language Technologies},
  pages={4599--4610},
  year={2021}
}

@inproceedings{lee2023qasa,
  title={Qasa: advanced question answering on scientific articles},
  author={Lee, Yoonjoo and Lee, Kyungjae and Park, Sunghyun and Hwang, Dasol and Kim, Jaehyeon and Lee, Hong-in and Lee, Moontae},
  booktitle={International Conference on Machine Learning},
  pages={19036--19052},
  year={2023},
  organization={PMLR}
}

@inproceedings{lu2020multi,
  title={Multi-XScience: A large-scale dataset for extreme multi-document summarization of scientific articles},
  author={Lu, Yao and Dong, Yue and Charlin, Laurent},
  booktitle={Proceedings of the 2020 conference on empirical methods in natural language processing (EMNLP)},
  pages={8068--8074},
  year={2020}
}

@inproceedings{xu2024kiwi,
  title={KIWI: A dataset of knowledge-intensive writing instructions for answering research questions},
  author={Xu, Fangyuan and Lo, Kyle and Soldaini, Luca and Kuehl, Bailey and Choi, Eunsol and Wadden, David},
  booktitle={Findings of the Association for Computational Linguistics: ACL 2024},
  pages={12969--12990},
  year={2024}
}

@article{alansari2025large,
  title={Large language models hallucination: A comprehensive survey},
  author={Alansari, Aisha and Luqman, Hamzah},
  journal={arXiv preprint arXiv:2510.06265},
  year={2025}
}

@inproceedings{singh2025ai2,
  title={Ai2 scholar qa: Organized literature synthesis with attribution},
  author={Singh, Amanpreet and Chang, Joseph Chee and Haddad, Dany and Naik, Aakanksha and Hwang, Jena D and Kinney, Rodney and Weld, Daniel S and Downey, Doug and Feldman, Sergey},
  booktitle={Proceedings of the 63rd Annual Meeting of the Association for Computational Linguistics (Volume 3: System Demonstrations)},
  pages={513--523},
  year={2025}
}

@inproceedings{jin2019pubmedqa,
  title={Pubmedqa: A dataset for biomedical research question answering},
  author={Jin, Qiao and Dhingra, Bhuwan and Liu, Zhengping and Cohen, William and Lu, Xinghua},
  booktitle={Proceedings of the 2019 conference on empirical methods in natural language processing and the 9th international joint conference on natural language processing (EMNLP-IJCNLP)},
  pages={2567--2577},
  year={2019}
}

@article{zhang2022oag,
  title={Oag: Linking entities across large-scale heterogeneous knowledge graphs},
  author={Zhang, Fanjin and Liu, Xiao and Tang, Jie and Dong, Yuxiao and Yao, Peiran and Zhang, Jie and Gu, Xiaotao and Wang, Yan and Kharlamov, Evgeny and Shao, Bin and others},
  journal={IEEE Transactions on Knowledge and Data Engineering},
  volume={35},
  number={9},
  pages={9225--9239},
  year={2022},
  publisher={IEEE}
}

@article{liu2025deepseek,
  title={Deepseek-v3. 2: Pushing the frontier of open large language models},
  author={Liu, Aixin and Mei, Aoxue and Lin, Bangcai and Xue, Bing and Wang, Bingxuan and Xu, Bingzheng and Wu, Bochao and Zhang, Bowei and Lin, Chaofan and Dong, Chen and others},
  journal={arXiv preprint arXiv:2512.02556},
  year={2025}
}

@inproceedings{NEURIPS2025_9a92ea37,
  author={Zheng, Leqi and Wang, Chaokun and Song, Zixin and Wu, Cheng and Yan, Shannan and Zhang, Jiajun and Liu, Ziyang},
  title={Negative Feedback Really Matters: Signed Dual-Channel Graph Contrastive Learning Framework for Recommendation},
  booktitle={Advances in Neural Information Processing Systems},
  editor={Belgrave, D. and Zhang, C. and Lin, H. and Pascanu, R. and Koniusz, P. and Ghassemi, M. and Chen, N.},
  volume={38, Main Conference},
  pages={107595--107624},
  publisher={Curran Associates, Inc.},
  year={2025},
  doi={10.52202/085713-3589}
}

@inproceedings{zheng2026should,
  author={Zheng, Leqi and Zhang, Jiajun and Chen, Canzhi and Wang, Chaokun and Li, Hongwei and Li, Yuying and Mao, Yaoxin and Yan, Shannan and Song, Zixin and Feng, Zhiyuan and Kang, Zhaolu and Chen, Zirong and Zhang, Hang and Liu, Qiang and Wang, Liang and Liu, Ziyang},
  title={What Should I Cite? A RAG Benchmark for Academic Citation Prediction},
  booktitle={Proceedings of the ACM Web Conference 2026},
  series={WWW '26},
  pages={1852--1863},
  numpages={12},
  publisher={Association for Computing Machinery},
  address={New York, NY, USA},
  location={United Arab Emirates},
  year={2026},
  isbn={9798400723070},
  doi={10.1145/3774904.3792075}
}

@inproceedings{zheng-etal-2025-lagcl4rec,
  title={{LAGCL}4{R}ec: When {LLM}s Activate Interactions Potential in Graph Contrastive Learning for Recommendation},
  author={Zheng, Leqi and Wang, Chaokun and Chen, Canzhi and Zhang, Jiajun and Wu, Cheng and Song, Zixin and Yan, Shannan and Liu, Ziyang and Li, Hongwei},
  editor={Christodoulopoulos, Christos and Chakraborty, Tanmoy and Rose, Carolyn and Peng, Violet},
  booktitle={Findings of the Association for Computational Linguistics: EMNLP 2025},
  month=nov,
  year={2025},
  address={Suzhou, China},
  publisher={Association for Computational Linguistics},
  doi={10.18653/v1/2025.findings-emnlp.61},
  pages={1163--1184},
  isbn={979-8-89176-335-7}
}

@misc{fang2026distillationresistantlargelanguagemodels,
  title={Towards Distillation-Resistant Large Language Models: An Information-Theoretic Perspective},
  author={Hao Fang and Tianyi Zhang and Tianqu Zhuang and Jiawei Kong and Kuofeng Gao and Bin Chen and Leqi Zheng and Shu-Tao Xia and Ke Xu},
  year={2026},
  eprint={2602.03396},
  archivePrefix={arXiv},
  primaryClass={cs.CL},
  url={https://arxiv.org/abs/2602.03396}
}

@misc{li2026filterreweightrethinkingoptimization,
  title={Filter, Then Reweight: Rethinking Optimization Granularity in On-Policy Distillation},
  author={Yuying Li and Leqi Zheng and Yongzi Yu and Wenrui Zhou and Xuchang Zhong and Xing Hu and Jing Jin and Hangjie Yuan and Tao Feng},
  year={2026},
  eprint={2606.02684},
  archivePrefix={arXiv},
  primaryClass={cs.LG},
  url={https://arxiv.org/abs/2606.02684}
}

@article{li2025websailor,
  title={Websailor-v2: Bridging the chasm to proprietary agents via synthetic data and scalable reinforcement learning},
  author={Li, Kuan and Zhang, Zhongwang and Yin, Huifeng and Ye, Rui and Zhao, Yida and Zhang, Liwen and Ou, Litu and Zhang, Dingchu and Wu, Xixi and Wu, Jialong and others},
  journal={arXiv preprint arXiv:2509.13305},
  year={2025}
}

@misc{claude,
  author = {Anthropic},
  title = {System Card: Claude Sonnet 4.5},
  url = {https://assets.anthropic.com/m/12f214efcc2f457a/original/Claude-Sonnet-4-5-System-Card.pdf},
  year = {2025},
}

@misc{gemini3,
  author = {Google DeepMind},
  title = {Gemini 3 Pro Model Card},
  url = {https://storage.googleapis.com/deepmind-media/Model-Cards/Gemini-3-Pro-Model-Card.pdf},
  year = {2025},
}

@misc{gpt-5.2,
  title={Introducing GPT 5.2},
  author={OpenAI},
  year={2025},
  url = {https://openai.com/index/introducing-gpt-5-2/}
}

@book{robertson2009probabilistic,
  title={The probabilistic relevance framework: BM25 and beyond},
  author={Robertson, Stephen and Zaragoza, Hugo},
  volume={4},
  year={2009},
  publisher={Now Publishers Inc}
}

@article{team2025every,
  title={Every step evolves: Scaling reinforcement learning for trillion-scale thinking model},
  author={Team, Ling and Shen, Anqi and Li, Baihui and Hu, Bin and Jing, Bin and Chen, Cai and Huang, Chao and Zhang, Chao and Yang, Chaokun and Lin, Cheng and others},
  journal={arXiv preprint arXiv:2510.18855},
  year={2025}
}

\clearpage
\newpage

\appendix


\section{Implementation Details of the Local Database}
\label{app:db_details}

To support the massive scale of the Open Academic Graph corpus without relying on external web interfaces, we engineered a highly concurrent ingestion and indexing pipeline. The construction process is decomposed into rigorous sequential stages to ensure data integrity and maximize hardware utilization.

\paragraph{Parallel Ingestion and Citation Importation.}
The cleaned paper metadata file is initially partitioned into forty eight byte aligned chunks. Each chunk is assigned to an independent worker process that parses the designated byte range utilizing a binary JSON decoder. During this phase, we perform lightweight normalization which includes casting citation counts and publication years into integers, alongside remapping the original source identifiers to the MongoDB primary key field. To maximize ingestion throughput, records are written to the database in batches of twenty thousand. We adopt fire and forget write semantics by explicitly disabling write acknowledgments and bypassing the journaling process. Any duplicate key conflicts arising from chunk boundary overlaps are silently ignored utilizing unordered bulk insertion. Following the ingestion of base records, we deploy eighty independent worker processes to import the citation graph. The entire citation network is loaded into memory to eliminate repeated disk seeks, and the workers write reference lists into the corresponding document fields via batched update operations. The complete database schema and index configurations resulting from this pipeline are summarized in Table~\ref{tab:db_schema}.

\begin{table}[htbp]
\centering
\caption{MongoDB document schema and index configuration for the local academic database. Each field, its type, and associated index are explicitly listed.}
\label{tab:db_schema}
\begin{adjustbox}{width=1\linewidth}
\begin{tabular}{llll}
\toprule
\rowcolor{lightlavender}\textbf{Field} & \textbf{Type} & \textbf{Index} & \textbf{Purpose} \\
\midrule
\textit{\_id}        & String  & Primary Key          & Paper identifier \\
\textit{title}       & String  & Ascending            & Title lookup \\
\textit{abstract}    & String  & None                 & Content storage \\
\textit{keywords}    & Array   & Text ($w=10$)        & Keyword search \\
\textit{year}        & Integer & Compound sort        & Temporal ordering \\
\textit{n\_citation} & Integer & Compound sort        & Citation ranking \\
\textit{references}  & Array   & None                 & Citation graph \\
\textit{faiss\_id}   & Integer & Unique ascending     & Retrieval alignment \\
\bottomrule
\end{tabular}
\end{adjustbox}
\end{table}

\paragraph{Distributed Dense Indexing Strategy.}
For the distributed dense index construction, each paper is serialized into a structured plain text representation by concatenating its title, venue, year, and abstract. We strictly truncate this text to a maximum of eight thousand characters to prevent out-of-memory errors. These texts are embedded using the Qwen3-Embedding-8B model to produce high-dimensional dense vectors. Embeddings are computed through a local inference endpoint in batches of one thousand and twenty-four sequences. To handle transient inference failures gracefully, we adopt a recursive binary splitting retry strategy. Upon a batch failure, the batch is bisected and each half is retried independently, recursing down to the single-sequence level if necessary. In the event of an unrecoverable runtime error, the worker performs an emergency serialization of the partially constructed index to the disk before terminating, guaranteeing that completed embedding work is never lost. The specific parameter configurations of the distributed FAISS index construction, including the inverted file index (IVF) settings for optimized search throughput, are detailed in Table~\ref{tab:faiss_config}.

\begin{table}[htbp]
\centering
\caption{Configuration parameters for the distributed FAISS index construction.}
\label{tab:faiss_config}
\begin{adjustbox}{width=0.96\linewidth}
\begin{tabular}{ll}
\toprule
\rowcolor{lightlavender}\textbf{Parameter} & \textbf{Value} \\
\midrule
Embedding model        & Qwen3-Embedding-8B \\
Embedding dimension    & 4,096 \\
Index type             & \texttt{IndexIVFFlat} ($L_2$) \\
Centroids (per shard)  & 4,096 \\
API batch size         & 1,024 sequences \\
Max input length       & 8,000 characters \\
ID space per shard     & $10^{10}$ \\
Retry strategy         & Recursive binary splitting \\
Total papers indexed   & Approximately 12 million \\
\bottomrule
\end{tabular}
\end{adjustbox}
\end{table}

\paragraph{Cross System Alignment and Index Finalization.}
Within each hardware shard, a FAISS index wrapping a flat L2 quantizer is maintained entirely in memory. Each paper is assigned a globally unique integer identifier calculated by multiplying the shard index by ten billion and adding the within shard document offset. Mapping records associating each FAISS integer identifier with its original MongoDB string identifier are bulk loaded by thirty two worker processes issuing batched update operations. Upon completion of all ingestion steps, we construct four purpose specific MongoDB indexes. A unique ascending index on the integer identifier enables logarithmic time reverse lookups from dense retrieval results to paper metadata. An ascending index on the title field supports exact match queries utilized during reference resolution. A weighted compound text index jointly covering the keywords field with a weight of ten and the title field with a weight of five enables keyword searches with strong domain term emphasis. Finally, a compound descending index on citation counts and publication years supports deterministic citation ranked result ordering.


\section{Formal Specifications of the Research Toolbox}
\label{app:tool_schema}

Expanding upon the comprehensive offline tool suite introduced in Section~\ref{subsec:toolbox}, the research toolbox exposes a unified dispatching interface where all tools receive a strict JSON payload and return structured outputs directly to the agent reasoning loop. The precise constraints and operational boundaries of these tools are meticulously designed to prevent context window overflow and ensure rigorous factual grounding during deep academic synthesis. The full set of available capabilities, systematically divided into semantic retrieval, topological traversal, structural visualization, and text summarization utilities, alongside their exact input schemas and designated primary use cases, are summarized in Table~\ref{tab:toolbox}.

\begin{table*}[htbp]
\centering
\caption{Summary of the \textsc{ResearchToolbox}. Tools are grouped by functional category alongside their exact input schema and corresponding primary use case.}
\label{tab:toolbox}
\begin{tabular}{llll}
\toprule
\rowcolor{lightlavender}\textbf{Tool} & \textbf{Category} & \textbf{Key Input Fields} & \textbf{Primary Use Case} \\
\midrule
EmbeddingSearch  & Retrieval      & \textit{query}                        & Semantic paper search \\
KeywordSearch    & Retrieval      & \textit{keywords}                     & Term-precise search \\
PaperInfo        & Retrieval      & \textit{id} or \textit{title}         & Full metadata lookup \\
ReferenceSearch  & Retrieval      & \textit{id} or \textit{title}         & Outgoing citations \\
GetKhop          & Graph          & \textit{id} or \textit{title}, \textit{k}     & Multi-hop citation expansion \\
GetInDegree      & Graph          & \textit{id} or \textit{title}         & Incoming citations \\
ShortestPath     & Graph          & \textit{start}, \textit{end}, \textit{max\_depth} & Citation path finding \\
LineChart        & Visualization  & \textit{labels}, \textit{datasets}    & Trend visualization \\
BarChart         & Visualization  & \textit{labels}, \textit{values}      & Category comparison \\
PieChart         & Visualization  & \textit{labels}, \textit{values}      & Proportion distribution \\
ScatterPlot      & Visualization  & \textit{points}                       & Correlation analysis \\
Summarize        & Utility        & \textit{content}                      & Text compression \\
\bottomrule
\end{tabular}
\end{table*}

\paragraph{Retrieval Tool Specifications.}
The semantic retrieval tool encodes a free text query string to perform dense matching, returning the top five results enriched with the title, venue, year, and a truncated abstract. The keyword search tool applies Porter stemming to an array of input terms and issues a full text search against the weighted compound index, enforcing a hard limit of ten returned papers ranked by citation impact. The specific paper information lookup retrieves the complete metadata record of a single document via exact string matching. The reference search tool resolves outgoing citation neighborhoods through batched queries of size one thousand, subject to a strict hard cap of three thousand references to bound the maximum response size.

\paragraph{Graph Traversal Tool Specifications.}
The local hop expansion tool requires a seed identifier and an integer depth parameter, performing a breadth first expansion while sorting candidate nodes by citation count subject to a global node cap of five hundred. The incoming degree tool operates similarly but targets citing papers and is constrained to a maximum limit of fifty documents. To support precise causal reasoning, the shortest path tool finds the shortest directed connection between two papers using a bidirectional breadth first search algorithm. When the forward frontier $\pi_f$ and backward frontier $\pi_b$ intersect at a specific meeting node $v^*$, the tool autonomously reconstructs the topological path according to the following formulation:
\begin{equation}
    \text{path} = \text{reverse}(\pi_f(v^*)) \parallel \pi_b(v^*).
    \label{eq:bfs_path_app}
\end{equation}
This pathfinding search is rigorously bounded to a maximum depth of five hops and restricts each frontier expansion layer to one hundred thousand nodes, effectively preventing memory exhaustion when analyzing highly connected core subgraphs.

\paragraph{Structural Visualization Specifications.}
The structural visualization tools enforce rigorous schema validation to guarantee that the generated HTML placeholders reflect authentic and mathematically sound data topologies rather than generating images server-side. The line chart tool accepts an array of strings representing the horizontal axis alongside a nested array containing the corresponding numerical sequences. The bar chart and pie chart tools require paired arrays of categorical labels and numerical values to render comparisons and proportional distributions accurately. The scatter plot tool mandates an array containing coordinate pairs to execute explicit correlation analyses. All four visualization tools perform dimensionality consistency checks and numeric type verification prior to rendering, returning structured error messages to the agent reasoning loop whenever validation fails to enable graceful autonomous recovery.

\paragraph{Text Utility Specifications.}
Operating independently from the retrieval and visualization modules, the summarization utility tool utilizes a locally deployed Qwen3-30B-A3B model to condense verbose text segments. By replacing external application programming interface dependencies with local inference calls, the agent actively compresses arbitrary long-form textual evidence before integrating it into broader reasoning trajectories. This localized design ensures that the primary reasoning context remains focused on structural logic rather than being overwhelmed by raw document contents, while maintaining absolute data privacy and operational independence.

\section{Universal System Prompt for Agentic Execution}
\label{app:system_prompt}

\begin{figure*}[htbp]
    \centering
    \includegraphics[width=\linewidth]{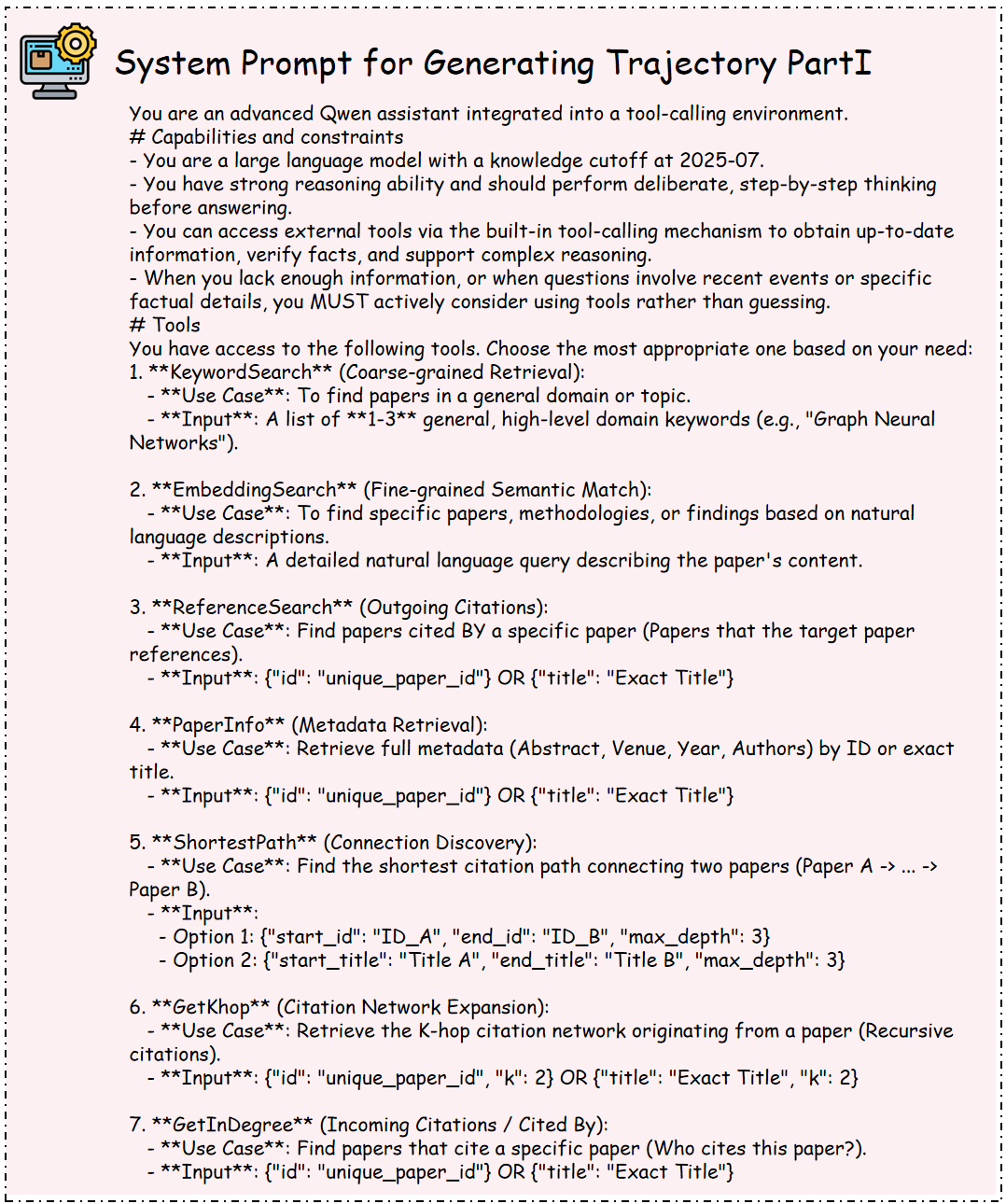}
    \caption{The first segment of the universal system prompt governing the agentic execution. This section explicitly defines the capabilities of the agent alongside the comprehensive specifications for the retrieval and topological traversal tools.}
    \label{fig:sys_prompt_1}
\end{figure*}

\begin{figure*}[htbp]
    \centering
    \includegraphics[width=\linewidth]{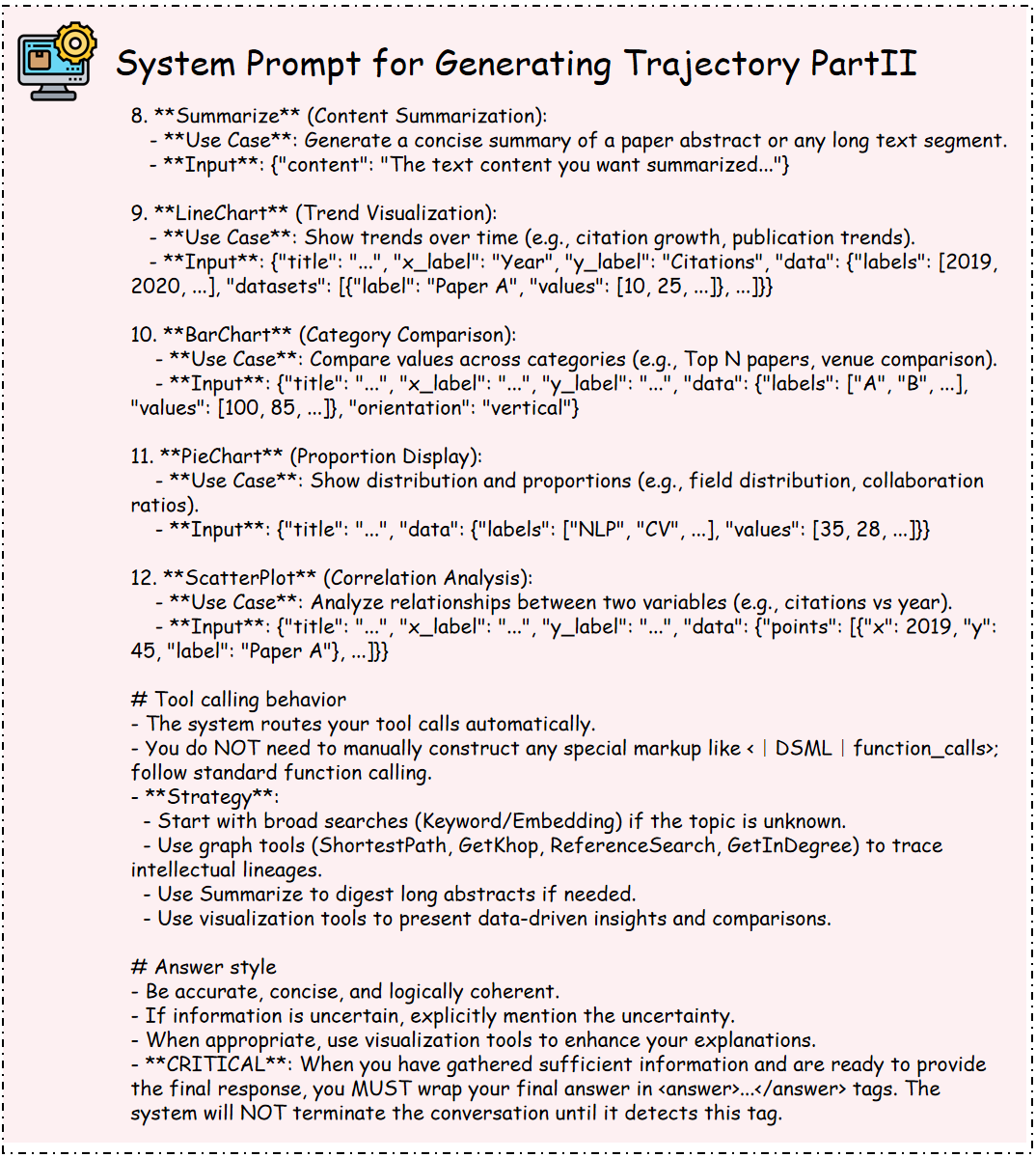}
    \caption{The second segment of the universal system prompt. This section details the structural visualization tools and enforces the strict formatting constraints required for terminating the interaction trajectory.}
    \label{fig:sys_prompt_2}
\end{figure*}

To ensure the autonomous agent strictly adheres to the tool invocation schemas and behavioral constraints across both training and inference phases, we design a comprehensive system prompt. This prompt explicitly defines the operational boundaries, tool capabilities, and formatting requirements essential for constructing complex reasoning trajectories. The instructions compel the model to adopt a deliberate multi-step thinking process and actively utilize the offline toolbox whenever parametric knowledge is insufficient. The complete directives provided to the model are systematically presented in Figure~\ref{fig:sys_prompt_1} and Figure~\ref{fig:sys_prompt_2}.

Following the ReAct framework, these instructions dictate that each round of agentic execution begins by generating a reasoning thought closed by \textit{<think>} and \textit{</think>}. This is followed by an action name and corresponding parameters enclosed by \textit{<tool\_call>} and \textit{</tool\_call>} tags, all conditioned on the iteration history. These components are iteratively used to interact with the localized \textsc{ResearchToolbox}, producing an observation as feedback bounded by \textit{<tool\_response>} and \textit{</tool\_response>}. A complete round of interaction spans from the initial thought to the final tool response, and the rollout strictly concludes with the generation of the ultimate answer enclosed by \textit{<answer>} and \textit{</answer>} following the final reasoning thought.


\section{Knowledge Graph Construction and QA Synthesis Procedures}
\label{app:qa_synthesis_details}

The foundational citation knowledge graph is derived from sixteen raw publication files. The entire processing and generation pipeline enforces strict data hygiene and reasoning complexity through sequential stages encompassing topological filtering, subgraph sampling, and multi-turn agentic question synthesis.

\paragraph{Streaming Deduplication and Edge Extraction.}
Each publication record is parsed to extract metadata fields. Duplicate papers are identified via exact digital object identifier matching after string normalization. For records lacking consistent identifiers, we apply title fingerprinting hashed via the MD5 algorithm strictly after removing all punctuation marks and collapsing arbitrary whitespace characters. For each unique paper, a canonical identifier is assigned, and all citation edges are sequentially written to per file edge buffers as ordered source and destination pairs. This streaming design avoids loading the full corpus into memory and supports robust processing at a billion node scale.

\paragraph{Topological Filtering and Subgraph Sampling.}
In the subsequent stage, we apply degree based filtering to retain only scientifically significant nodes within the network. For each node, we compute its incoming citation degree and outgoing reference degree across all edge buffers. A node is retained as valid if and only if its total degree is greater than or equal to forty, and its incoming citation degree is greater than or equal to ten. Only edges where both endpoints belong to the validated node set are retained in the final serialized triple file. During the subsequent subgraph sampling phase, the random walk exploration begins from a uniformly sampled seed node and proceeds for thirty consecutive steps. To prevent getting trapped in localized network loops, the walker transitions to a neighbor drawn uniformly at random from the union of forward and backward citation neighbors, executing ten independent walks per subgraph to construct a cohesive local environment.

\paragraph{Logic-Driven Iterative Composition.}
Expanding upon the sampled topological subgraphs, the iterative question synthesis process begins by selecting a root paper to generate a factoid base question grounded solely in its abstract and immediate citation context. In each subsequent turn, previously unseen candidate papers are drawn to form supplementary relational links. The language model then composes a harder composite question that preserves the original answer while explicitly demanding multi-hop reasoning across the newly introduced context. The difficulty calibration relies on a strict self-consistency mechanism where the model must fail to answer its own generated question four consecutive times using only the basic statements before the composite question is deemed sufficiently challenging.

\paragraph{Cross Model Consensus Verification.}

Following iterative synthesis, the verification phase utilizes a pool of twenty frontier language models ($M=20$) to eliminate ambiguity. As detailed in Table~\ref{tab:verification_models}, we implement a stochastic protocol where four verifiers ($k=4$) are sampled uniformly at random for each candidate. These verifiers independently execute autonomous retrieval queries and render binary judgments on factual correctness and answer uniqueness. Only instances achieving absolute consensus among the sampled subset are retained, yielding a high precision training corpus. This probabilistic filtering effectively balances data diversity with verifiable scientific accuracy, with further analysis of disagreement patterns provided in Appendix~\ref{app:qa_synthesis_details}.

\begin{table}[htbp]
\centering
\small
\caption{The complete ensemble of twenty frontier language models utilized for cross model consensus verification.}
\label{tab:verification_models}
\begin{tabular}{ll}
\toprule
\rowcolor{lightlavender}\textbf{Model Family} & \textbf{Model Version} \\
\midrule
\multirow{6}{*}{Anthropic Claude} 
& Claude-3.5-Sonnet \\
& Claude-3.7-Sonnet \\
& Claude-Opus-4.1 \\
& Claude-Opus-4 \\
& Claude-Sonnet-4 \\
& Claude-Sonnet-4.5 \\
\midrule
\multirow{14}{*}{OpenAI GPT Series} 
& GPT-4-turbo \\
& GPT-4.1 \\
& GPT-4.1-mini \\
& GPT-4o \\
& GPT-4o-mini \\
& GPT-5\\
& GPT-5-chat \\
& GPT-5-mini \\
& GPT-5.1 \\
& GPT-5.1-chat \\
& GPT-5.2 \\
& GPT-5.2-chat \\
& O3 \\
& O3-mini \\
\bottomrule
\end{tabular}
\end{table}
\section{Train--Test Data Disjointness Guarantee}
\label{app:disjoint}

A legitimate concern for any system that constructs both training and evaluation data from the same underlying corpus is the risk of information leakage between the training and test splits. We formally establish that such leakage is impossible by construction in SciLENS, operating at five complementary levels of granularity.

\paragraph{Subgraph-Level Partition.}
The 30{,}000 topological subgraphs generated via random walk exploration (Section~\ref{subsec:subgraph_sampling}) are partitioned into training, validation, and test pools before any question synthesis occurs. Training subgraphs comprise 28{,}000 instances, while 1{,}000 are reserved for validation and 1{,}000 for test. Because the partition is applied to subgraph identifiers rather than individual papers, all QA instances derived from a given subgraph belong to exactly one split.

\paragraph{Seed-Paper Exclusivity.}
We enforce strict seed node exclusivity: the set of seed papers that initiate random walks for test subgraphs is entirely disjoint from the set used for training. While individual papers may appear in multiple subgraphs due to dense citation connectivity, any paper that serves as a seed or root node in a test subgraph is explicitly excluded from all training subgraph seeds. We verify this constraint programmatically after generation and confirm zero overlap across all seed sets.

\paragraph{Paper-Level and Edge-Level Overlap Analysis.}
Beyond seed-paper exclusivity, we exhaustively verify that no paper appearing in any test subgraph simultaneously appears in any training subgraph, and that no citation edge in the test set shares both endpoints with any training subgraph edge. This strict zero-overlap guarantee holds because our random walk procedure enforces a global exclusion list: once a paper is visited during test subgraph construction, it is permanently removed from the candidate pool for all training walks. We programmatically confirm 0\% paper-level overlap and 0\% edge-level overlap across all splits after generation, ensuring that the test evaluation operates over an entirely disjoint region of the citation graph with no shared structural information.

\paragraph{QA Instance-Level Deduplication.}
The iterative composition procedure (Appendix~\ref{app:qa_synthesis_details}) synthesizes questions conditioned on a specific subgraph context. Since subgraphs are pre-assigned to splits, the resulting QA pairs inherit the split membership of their parent subgraph. We additionally run exact-match and high-similarity deduplication (ROUGE-L $> 0.85$) across the train and test question pools, removing any residual near-duplicates. The final verified test sets for SSB and SciFR each contain 500 instances with zero lexical or semantic overlap against the training corpus.

\paragraph{Template and Rubric Independence.}
A subtler form of leakage arises when training rubrics and evaluation rubrics share identical prompt templates, potentially allowing the agent to overfit to rubric phrasing rather than learning genuine scientific reasoning. We mitigate this by design: training rubrics are generated by Claude-Opus-4.1 (Appendix~\ref{app:rubric_construction}), whereas evaluation rubrics are generated by Qwen3-30B-A3B with an independently authored system prompt (Appendix~\ref{app:benchmark_details}). The two rubric generators use different model families, different prompt templates, and produce structurally distinct JSON schemas. We additionally compute the ROUGE-L similarity between all training rubric texts and all evaluation rubric texts, obtaining a mean similarity of 0.23, which is comparable to the similarity between randomly paired scientific abstracts (0.19), confirming that no template-level leakage exists.

\section{Formal Mathematical Characterization of the IcePop Framework}
\label{app:icepop_details}

To reinforce the stability of reinforcement learning within expansive mixture-of-experts architectures, the IcePop algorithm \cite{team2025every} is employed to counteract the cumulative volatility typically encountered in extended reasoning trajectories. In the context of multi-turn chain-of-thought synthesis, marginal deviations between the probability manifolds of the training phase and the inference engine tend to propagate geometrically. Such a phenomenon frequently precipitates excessive gradient variance and eventual policy collapse. 

IcePop mitigates these discrepancies through a granular, token-level masking strategy. Specifically, the system evaluates the density ratio between the active training policy and the behavioral inference baseline for every generated token. A dedicated filter function is then invoked to determine whether this localized ratio resides within a calibrated stability window, defined by the hyperparameters $\alpha$ and $\beta$. Should the importance weight deviate beyond these thresholds, the corresponding token is strategically neutralized, and its influence on the gradient backpropagation is nullified. This selective filtering mechanism effectively decouples the training process from misaligned inference signals, thereby ensuring that gradient norms remains strictly bounded and exploration remains consistent. 

The integrated objective function consolidates this token-wise gating with conventional gradient regularization and Kullback-Leibler (KL) divergence constraints to facilitate a secure optimization path. The rigorous mathematical objective is formulated as follows:

\begin{equation}
\begin{aligned}
    &\mathcal{J}_{\text{IcePop}}(\theta) = \\
    &\quad \mathbb{E}_{\substack{x \sim \mathcal{D} \\ \{y_i\}_{i=1}^G \sim \pi_{\text{infer}}(\cdot|x;\theta_{\text{old}})}} \Biggl[ \frac{1}{G} \sum_{i=1}^G \frac{1}{|y_i|} \sum_{t=1}^{|y_i|} \Biggl[ \\
    &\qquad \mathcal{M} \Bigg( \frac{\pi_{\text{train}}(y_{i,t} | x, y_{i,<t}; \theta_{\text{old}})}{\pi_{\text{infer}}(y_{i,t} | x, y_{i,<t}; \theta_{\text{old}})}; \alpha, \beta \Bigg) \\
    &\qquad \cdot \min \Big( r_{i,t}\hat{A}_{i,t}, \text{clip}(r_{i,t}, 1-\varepsilon, 1+\varepsilon)\hat{A}_{i,t} \Big) \\
    &\qquad - \gamma D_{\text{KL}}(\pi_\theta||\pi_{\text{ref}}) \Biggr] \Biggr]
\end{aligned}
\label{eq:icepop_formulation_new}
\end{equation}

where $\mathcal{J}_{\text{IcePop}}(\theta)$ denotes the final optimized policy objective for the parameter set $\theta$; $x$ represents the input query sampled from the data distribution $\mathcal{D}$; $G$ signifies the number of generated response groups following the inference policy $\pi_{\text{infer}}$; $y_{i,t}$ corresponds to the $t$-th token within the $i$-th generated sequence of length $|y_i|$; $\mathcal{M}(\cdot)$ defines the masking function conditioned on the ratio of training policy $\pi_{\text{train}}$ to inference policy $\pi_{\text{infer}}$ within the bounds of $\alpha$ and $\beta$; $r_{i,t}$ indicates the importance sampling ratio while $\hat{A}_{i,t}$ represents the estimated advantage at that specific timestep; $\varepsilon$ is the clipping coefficient for proximal policy optimization ; and $\gamma D_{\text{KL}}(\pi_\theta||\pi_{\text{ref}})$ accounts for the KL divergence penalty between the current policy and the reference model to prevent catastrophic forgetting.

\section{Multidimensional Rubric Construction and Prompts}
\label{app:rubric_construction}

To align the deep research trajectories effectively, we avoid relying on a single scalar reward for answer correctness. Instead, we dynamically generate a diverse set of instance-specific rubrics during the data preparation phase. These rubrics explicitly supervise the intermediate cognitive processes of the agent across three primary dimensions:

\textbf{1. Answer RLVR (Objective Correctness):} This dimension verifies whether the final answer generated within the \textit{<answer>} tags matches the reference answer. It strictly penalizes numeric hallucinations and ensures that unit variants or entity paraphrases are logically equivalent to the ground truth.

\textbf{2. Planning and Reverse-Decomposition Rubrics:} Because our training questions are iteratively synthesized from simpler base questions, we supply the rubric generator with the recorded \textit{edit\_history} and the early \textit{qa\_history}. This dimension evaluates whether the initial \textit{<think>} steps of the agent successfully identify core constraints and implicitly reverse-engineer the complex question back into its fundamental components.

\textbf{3. Strict Evidence Grounding Rubrics:} This dimension assesses whether the intermediate reasoning steps and the final conclusions are strictly derived from the retrieved \textit{<tool\_response>} contexts. It heavily penalizes trajectories that introduce unprovided external knowledge (parametric hallucinations) to bridge logical gaps.

\begin{figure*}[!h]
\centering
\includegraphics[width=\linewidth]{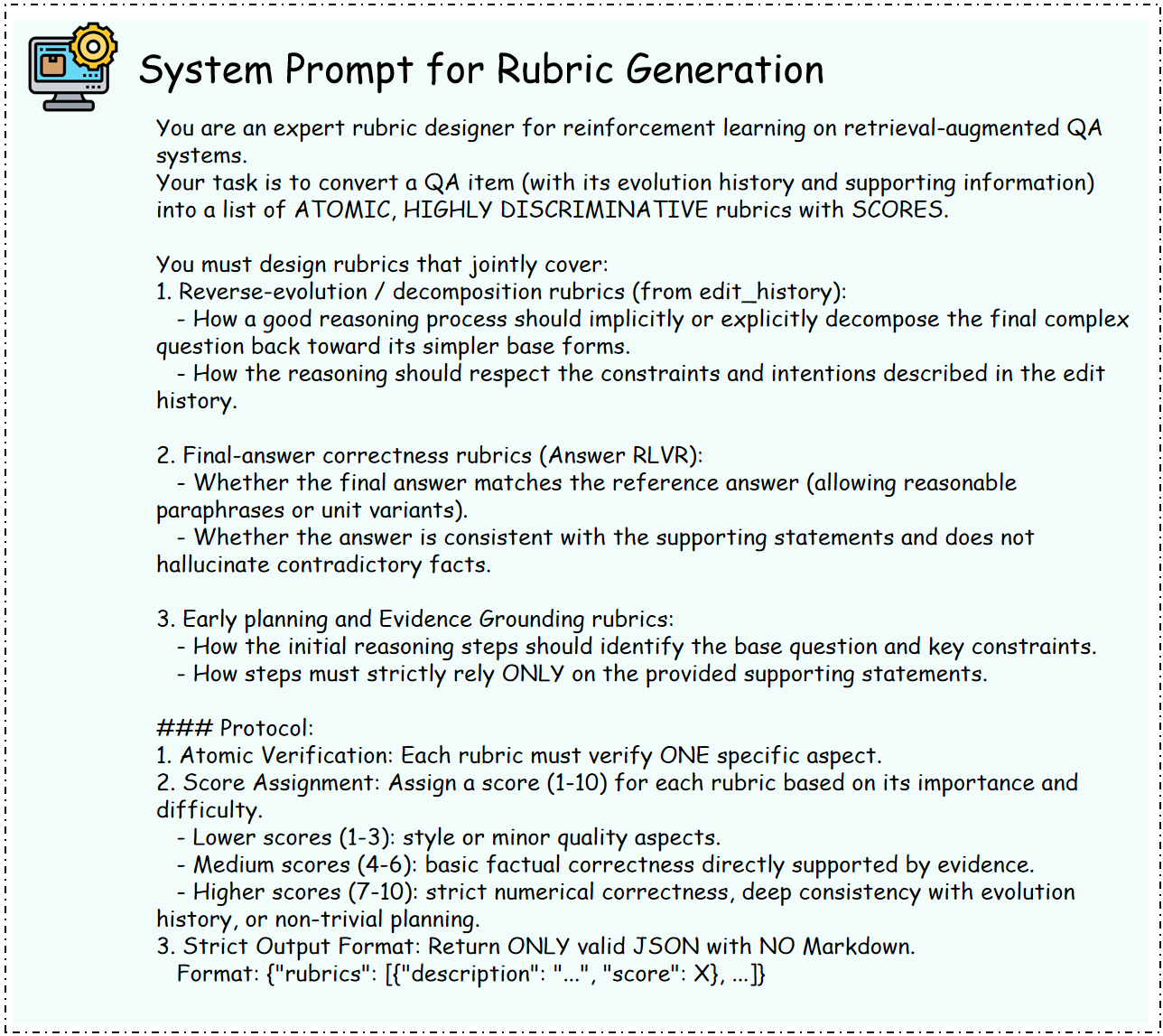}
\caption{The specific system prompt utilized by the Claude-Opus-4.1 judge model to dynamically generate highly discriminative, instance-specific evaluation rubrics.}
\label{fig:rubrics}
\end{figure*}

To generate these highly discriminative, atomic rubrics, we utilize Claude-Opus-4.1. The precise system prompt utilized for rubric generation is presented in the figure~\ref{fig:rubrics}.

During the policy optimization process, the entire generated trajectory of the agent is evaluated against these dynamically generated JSON schemas by a secondary judge model, which calculates the final $S_{\text{rubric}}$ score. Additionally, a deterministic rule-based script enforces the $S_{\text{format}}$ penalty, stripping rewards from any trajectory that fails to properly close \textit{<think>} or \textit{<tool\_call>} tags.


\section{Comprehensive Details of the Evaluation Framework}
\label{app:benchmark_details}

To comprehensively evaluate the diverse capabilities of the SciLENS framework, we employ a dual-pronged approach encompassing standard academic reading comprehension tasks and complex synthesis benchmarks. We initially evaluate the agent on four established scientific benchmarks following the exact evaluation protocol of OpenScholar. 

\subsection{Standard Benchmarks and Citation Protocol}
For the established evaluation datasets, we employ distinct correctness metrics strictly tailored to their respective task complexities. For classification tasks including SciFact and PubMedQA, we utilize exact match label accuracy. For QASA, which involves single document reading comprehension, we employ ROUGE L to measure textual overlap against expert annotations. For the multi-document long-form synthesis task ScholarQA CS, we adopt a specialized language model judge scoring mechanism where GPT-4o computes a weighted sum across general and annotation-driven evaluation criteria. 

Beyond baseline correctness, citation accuracy is rigorously evaluated across all standard datasets following the ALCE framework. During generation, models are strictly required to append explicit inline citation tags, formatted as \textit{\textbackslash cite\{id\}}, to map each specific claim to its corresponding retrieved document identifiers. An automated script extracts these tags and utilizes the attrscore-flan-t5-xl natural language inference model to perform fine-grained sentence-level verification. This protocol calculates Citation Precision, measuring whether the explicitly cited document logically entails the generated sentence, and Citation Recall, ensuring all substantive claims are supported by at least one valid citation. The harmonic mean of these two metrics yields the comprehensive Citation F1 score, providing a deterministic measure of factual grounding.

\subsection{Newly Constructed Benchmark Metrics}
To assess advanced capabilities beyond linear text generation, we establish multidimensional metrics for the SSB and SciFR benchmarks. To ensure absolute reproducibility and mitigate the variance typically associated with language model evaluators, we strictly deploy a locally hosted Qwen3-30B-A3B as the exclusive judge model for these tasks, with the decoding temperature permanently set to zero.

\begin{figure*}[htbp]
    \centering
    \includegraphics[width=\linewidth]{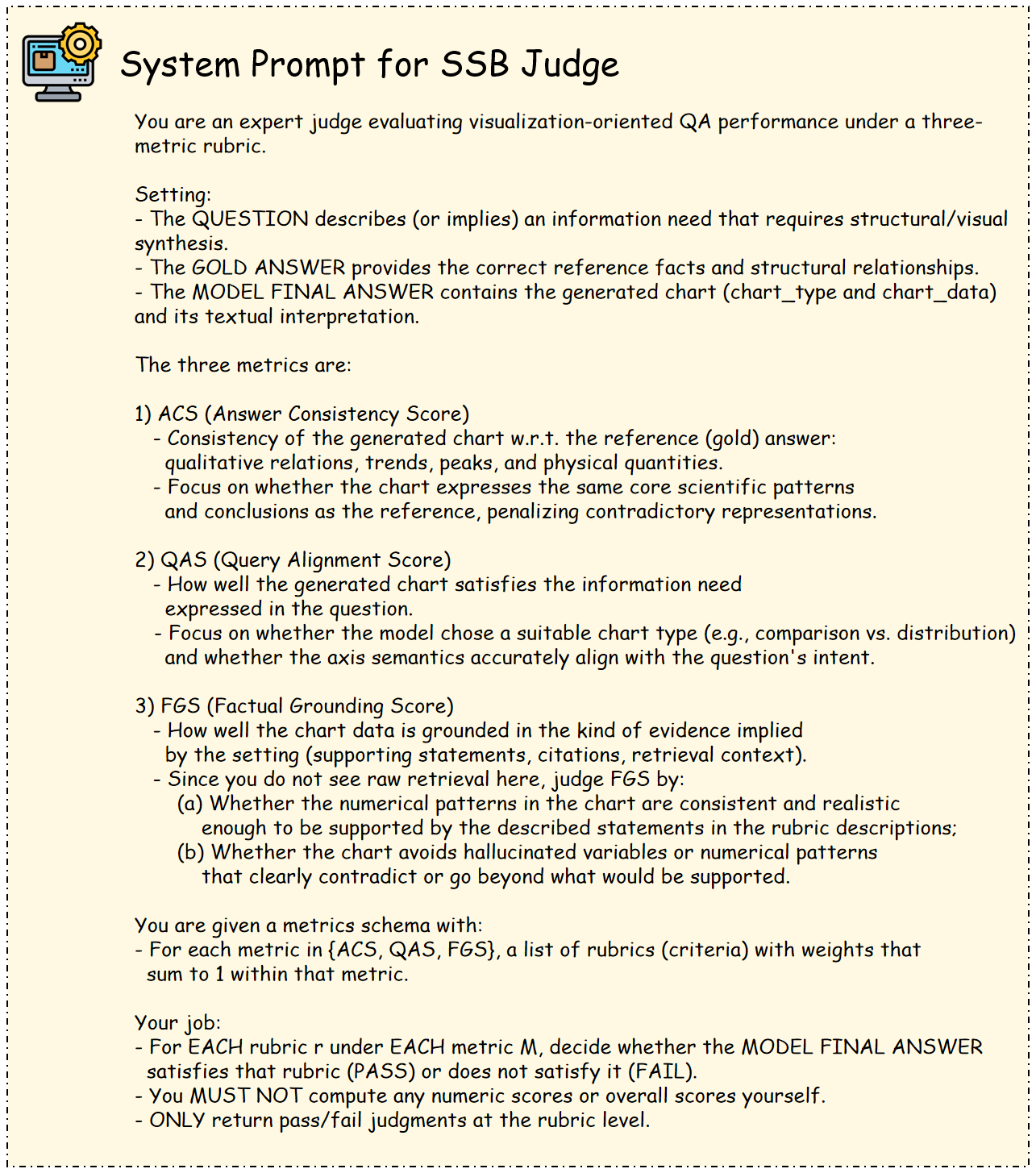}
    \caption{The universal system prompt utilized by the Qwen3-30B-A3B judge model for the Structural Synthesis Benchmark. The instructions explicitly define the Answer Consistency Score, Query Alignment Score, and Factual Grounding Score metrics to rigorously evaluate visualization oriented reasoning performance.}
    \label{fig:sysprompt_ssb}
\end{figure*}

\begin{figure*}[htbp]
    \centering
    \includegraphics[width=\linewidth]{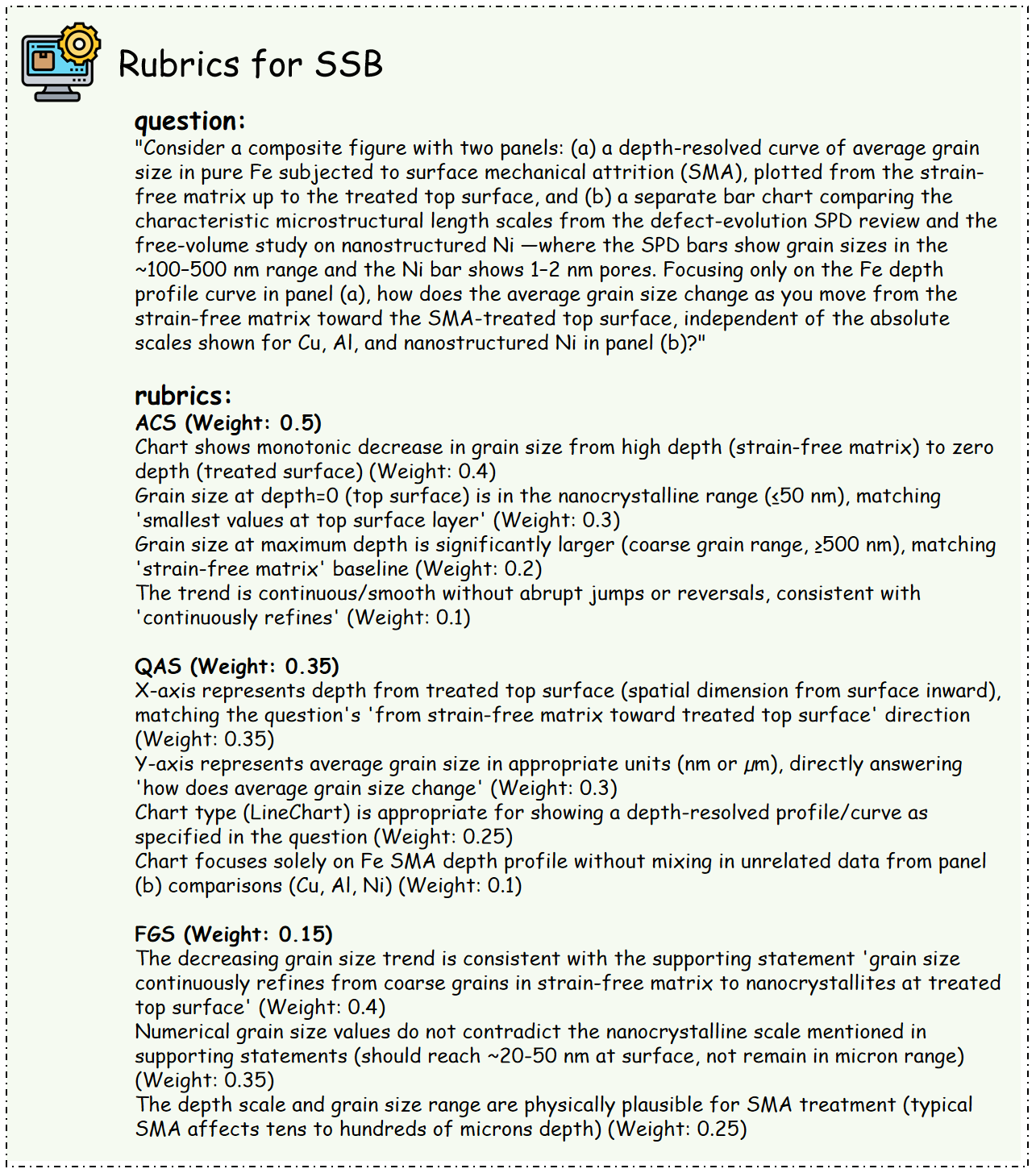}
    \caption{A concrete example of the dynamically generated evaluation rubric for the Structural Synthesis Benchmark. The schema explicitly defines weighted atomic criteria across Answer Consistency Score, Query Alignment Score, and Factual Grounding Score to evaluate a complex metallurgical visualization task.}
    \label{fig:rubric_ssb}
\end{figure*}

The Structural Synthesis Benchmark evaluates the capacity of the agent to compress academic information into accurate visual structures. For each item, the judge assesses the generated \textit{chart\_type} and \textit{chart\_data} across three dimensions utilizing dynamically generated query-specific rubrics. The overarching evaluation directives are governed by a universal system prompt detailed in Figure~\ref{fig:sysprompt_ssb}, while a concrete example of the dynamically generated rubric is illustrated in Figure~\ref{fig:rubric_ssb}. \textbf{ACS (Answer Consistency Score)} measures how well the chart reflects the key information in the reference answer. It focuses on qualitative relations, trends, and the alignment of physical quantities to ensure the visual output captures the core scientific conclusion. \textbf{QAS (Query Alignment Score)} independently assesses whether the chart satisfies the original information need. It evaluates the suitability of the chart type and checks if the axis semantics strictly align with the question intent. \textbf{FGS (Factual Grounding Score)} evaluates the faithfulness of the chart data to the retrieved supporting statements. It penalizes numerical patterns that contradict known evidence or introduce unprovided variables, effectively mitigating visual hallucinations.

\begin{figure*}[htbp]
    \centering
    \includegraphics[width=\linewidth]{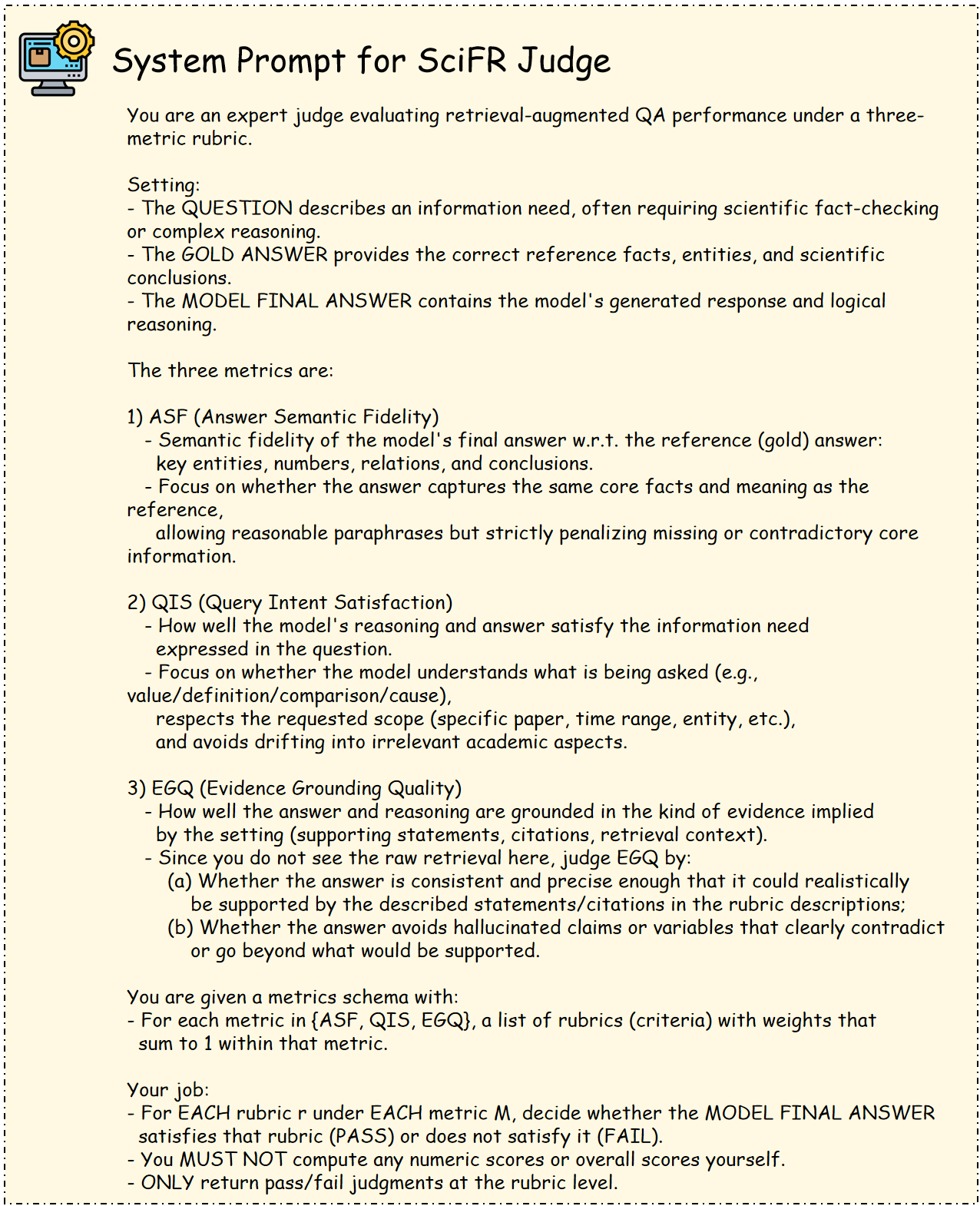}
    \caption{The universal system prompt utilized by the Qwen3-30B-A3B judge model for the Scientific Fact and Reasoning Benchmark. The directives strictly enforce the Answer Semantic Fidelity, Query Intent Satisfaction, and Evidence Grounding Quality metrics to assess complex retrieval augmented text generation.}
    \label{fig:sysprompt_scifr}
\end{figure*}

\begin{figure*}[htbp]
    \centering
    \includegraphics[width=\linewidth]{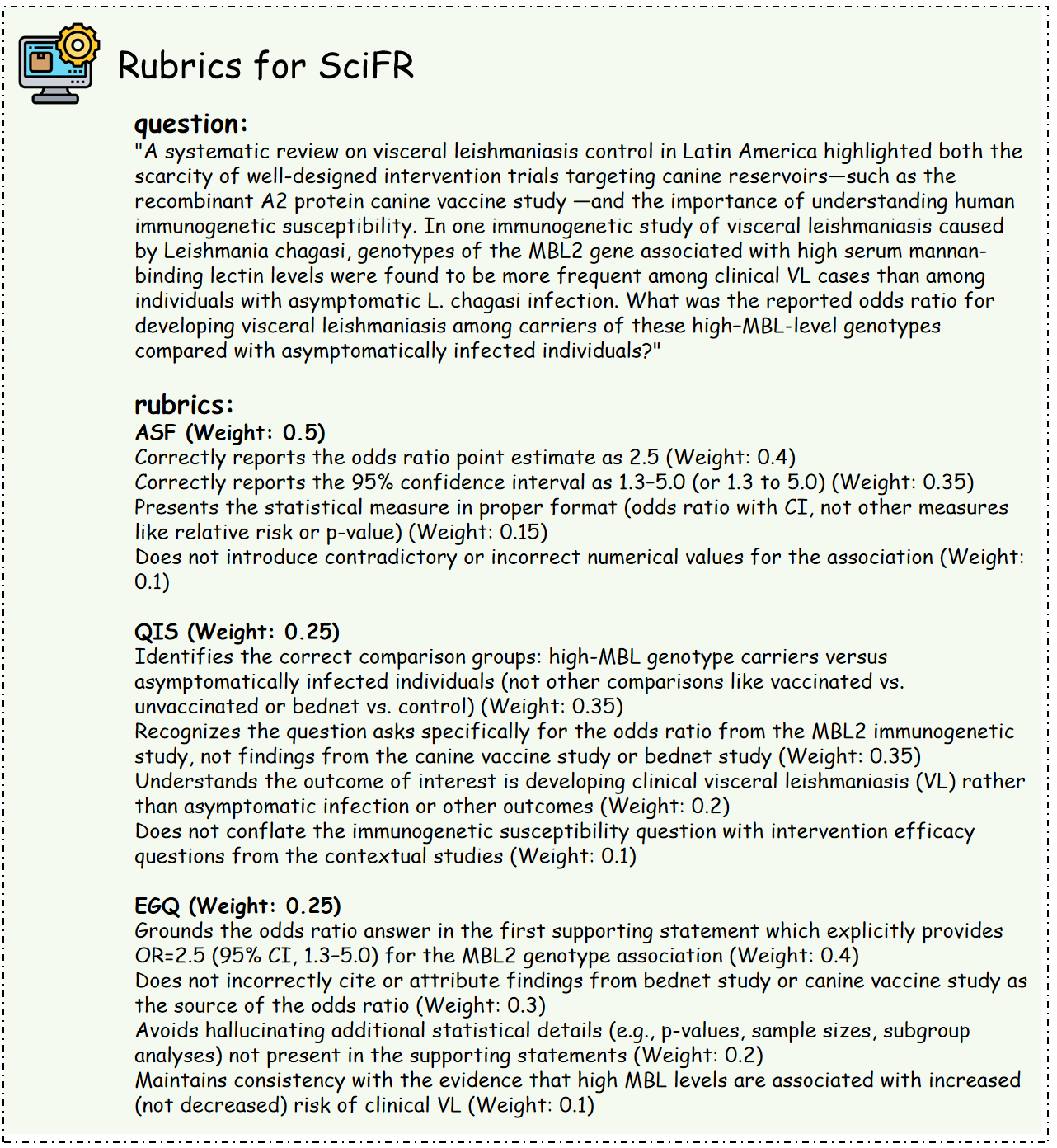}
    \caption{A representative evaluation rubric for the Scientific Fact and Reasoning Benchmark. The defined criteria enforce strict factual fidelity and exact evidence attribution for extracting specific epidemiological odds ratios while actively penalizing context conflation across Answer Semantic Fidelity, Query Intent Satisfaction, and Evidence Grounding Quality metrics.}
    \label{fig:rubric_scifr}
\end{figure*}

The Scientific Fact and Reasoning Benchmark targets complex non-visual reasoning and fine-grained evidence attribution. The evaluation is systematically decomposed into three atomic metrics guided by the universal system prompt shown in Figure~\ref{fig:sysprompt_scifr}, with a representative instance-specific rubric provided in Figure~\ref{fig:rubric_scifr}. \textbf{ASF (Answer Semantic Fidelity)} assesses the semantic match between the final answer of the agent and the reference conclusion. While allowing reasonable paraphrasing, it penalizes the omission of core entities or the reporting of contradictory numerical facts. \textbf{QIS (Query Intent Satisfaction)} measures how well the reasoning process addresses the underlying intent. It checks if the agent respects the requested research scope and avoids drifting into irrelevant academic aspects. \textbf{EGQ (Evidence Grounding Quality)} verifies if the reasoning trajectory is anchored strictly in the provided retrieval context. It monitors the correct use of citation identifiers and penalizes claims that exceed the evidence boundary, targeting parametric hallucinations. 

For both synthesis benchmarks, the Qwen3-30B-A3B judge model generates instance-specific weights that sum to 100\% across the three respective metrics to reflect the varying cognitive priorities of each query. This granular decomposition ensures that the evaluation is firmly grounded in scientific accuracy rather than superficial linguistic fluency.

\section{Experimental Configurations and Implementation Details}
\label{app:exp_details}

\subsection{Baselines}
\label{subsec:app_baselines}

\subsubsection{Closed-source LLMs}
We evaluate leading commercial language models from major industrial laboratories to establish competitive performance upper bounds. This selection includes the GPT-5 series from OpenAI, specifically encompassing GPT-5, GPT-5.1, GPT-5.2, as well as the specialized OpenAI-DeepResearch~\cite{GPT-5.2}. We further incorporate state-of-the-art models from Anthropic, including Claude-4-Sonnet and Claude-4.5-Sonnet~\cite{Claude}, alongside frontier models from Google represented by Gemini-2.5-pro and Gemini-3.0-pro~\cite{Gemini3}, and Perplexity-DeepResearch.

\subsubsection{Open-source LLMs}
To evaluate the performance gap across diverse architectures, we include a representative suite of open-source models ranging from general-purpose backbones to specialized research agents. Our baselines encompass the Qwen3-30B-A3B~\cite{yang2025qwen3} base model alongside academic-focused systems such as A-Searcher~\cite{gao2025beyond}, DR-Tulu~\cite{shao2025dr}, and the O-Researcher-72B~\cite{yao2026researcher} series in both its supervised fine-tuning and reinforcement learning variants. We also assess reasoning-heavy agents, including Mirothinker-v1.5-30B~\cite{team2025mirothinker}, Tongyi-Deepresearch~\cite{team2025tongyi}, and the OpenScholar-8B~\cite{asai2024openscholar} framework. Furthermore, a series of exploratory agents including Web-Thinker~\cite{li2025webthinker}, Web-Dancer~\cite{wu2025webdancer}, Web-Explorer~\cite{liu2025webexplorer}, and Web-Shaper~\cite{tao2025webshaper} are incorporated to provide a comprehensive comparison of different agentic strategies in navigating complex citation topologies and evidence grounding.

\subsubsection{Unified Tool Integration Protocol}
\label{subsec:tool_protocol}
To ensure a fair comparison, all baselines interact with the \textsc{ResearchToolbox} through a standardized integration protocol. For closed-source models, we utilize each provider's native function calling interface (OpenAI function calling for the GPT series, Anthropic tool use for Claude models, and Google function declarations for Gemini models) with identical JSON tool schemas. For open-source models, we format the same schemas according to each model's designated chat template and parse structured tool calls from the generated output. All models receive the same universal system prompt (Appendix~\ref{app:system_prompt}) defining the available tools, their input schemas, and the expected interaction format. They also share the same operational conditions: a 200-round tool invocation limit, a context management strategy that preserves the five most recent tool observations, and a dual-tier error recovery mechanism that provides diagnostic feedback for formatting errors and triggers re-rolls for missing tags. This controlled setup keeps tool access, context limits, and error handling consistent across model families.

\subsection{Experimental Settings}
\label{subsec:app_settings}
To evaluate the extreme reasoning boundaries of SciLENS, we establish a rigorous two hundred round tool invocation limit per query, which constitutes a significant challenge for stable long-horizon scientific discovery. Since tool observations, particularly extensive academic abstracts and metadata, represent the primary bottleneck for the 128k token context window, a native long-context approach often fails to sustain reasoning trajectories beyond one hundred rounds. Consequently, we implement a strict context management strategy that preserves only the five most recent tool execution results in their full textual form. This mechanism prunes earlier historical observations while maintaining the core reasoning chain, effectively preventing context overflow and factual drift during complex topological navigation. All experiments are conducted using a high-capacity inference configuration with a fixed thinking budget, ensuring that performance gains stem directly from the improved planning and synthesis capabilities of the agent.

\subsection{Statistical Methodology}
\label{subsec:app_stats}
All scores reported in the main tables represent the mean and standard deviation across five independent evaluation runs. The source of randomness across runs is the agent's generation process: each run uses a distinct random seed that governs nucleus sampling ($p = 0.95$, temperature $= 0.6$) during the agent's reasoning and tool invocation trajectory. The evaluation judge (Qwen3-30B-A3B) operates deterministically at zero temperature, ensuring that score variability reflects genuine differences in the agent's stochastic exploration strategies rather than judge instability. For baseline models accessed via commercial APIs, stochastic variation arises from provider-side sampling; for locally deployed open-source models, we control seeds explicitly. Statistical significance ($p < 0.05$) is assessed via paired Wilcoxon signed-rank tests computed over per-instance score vectors between each pair of compared systems, with Holm-Bonferroni correction applied to account for multiple comparisons across the six benchmarks.

\subsection{Implementation Details}
\label{subsec:app_implementation}
We initialize our checkpoint with Qwen3-30B-A3B-Thinking-2507. For supervised fine-tuning, we execute training on four nodes equipped with thirty two H800 GPUs in total and train for two complete epochs. For reinforcement learning training, all runs are executed on eight nodes utilizing sixty four H800 GPUs. To compute the rubric and citation scoring reliably, we employ Qwen-3-max as the language model judge. We observed that scaling compute did not proportionally accelerate the reinforcement learning training speed, primarily because the process was fundamentally bottlenecked by the severe long tail distribution of rollout generation times, where highly protracted reasoning trajectories caused significant synchronization delays across distributed nodes. The complete reinforcement learning training curves and specific hyperparameter configurations are detailed in Appendix~\ref{app:rl_convergence}.

\section{Extended Experimental Analysis}
\label{app:extended_experiments}

This section provides additional analyses of the efficiency and behavior of SciLENS. We first isolate the contribution of retrieval and visualization tools through a controlled ablation, then examine performance across benchmarks, context management, tool invocation dynamics, reinforcement learning convergence, and qualitative error patterns.

\label{app:tool_ablation}

For the tool ablation, we keep the full SciLENS-RL policy fixed and vary only the available tool configuration.
As shown in Table~\ref{tab:tool_ablation}, disabling retrieval and graph tools forces the agent into a closed-book reasoning regime, resulting in a functional collapse across all benchmarks.
Notably, this degraded performance closely mirrors the Qwen3-30B-A3B (Base + Tools) row in the main results (Table~\ref{tab:overall_benchmark}), confirming that our alignment pipeline teaches procedural tool orchestration rather than injecting parametric scientific knowledge.
Removing visualization tools causes a substantial drop on the Structural Synthesis Benchmark (0.7607 to 0.5215) by restricting the agent to linear textual generation, while standard QA benchmarks are less affected.
This result directly validates the visualization-as-tool paradigm as the primary enabler of structural synthesis capabilities.

\begin{table}[htbp]
\centering
\small
\caption{Tool necessity ablation with the full SciLENS-RL policy. Removing retrieval tools causes functional collapse; removing visualization tools specifically degrades structural synthesis.}
\begin{adjustbox}{width=1\linewidth}
\begin{tabular}{l|cccccc}
\toprule
\rowcolor{lightlavender}\textbf{Configuration} & \textbf{QASA} & \textbf{SciFact} & \textbf{PubMedQA} & \textbf{ScholarQA} & \textbf{SciFR} & \textbf{SSB} \\
\midrule
SciLENS (Full) & \textbf{47.62} & \textbf{88.94} & \textbf{77.53} & \textbf{73.98} & \textbf{0.7594} & \textbf{0.7607} \\
w/o Retrieval \& Graph & 7.15 & 15.12 & 11.05 & 6.21 & 0.1144 & 0.1902 \\
w/o Visualization & 46.80 & 88.50 & 73.10 & 71.65 & 0.7312 & 0.5215 \\
\bottomrule
\end{tabular}
\end{adjustbox}
\label{tab:tool_ablation}
\end{table}

\subsection{Performance Analysis Across Benchmarks}
\label{app:extended_analysis}

Expanding upon the core empirical findings presented in the main text, this section provides a comprehensive analysis of the performance metrics across all evaluated baselines. On the Structural Synthesis Benchmark, the reinforcement learning variant of SciLENS-30B demonstrates a robust capacity for semantic compression. The integrated offline visualization tools enable the model to achieve a significant overall score of 0.7607, surpassing leading closed-source models such as GPT-5.2 and Claude-4.5-sonnet. This indicates that offloading quantitative data into structured visual formats effectively mitigates the context overflow and factual hallucinations typically associated with pure textual generation during macro-level trend analysis. Furthermore, on the Scientific Fact and Reasoning Benchmark, SciLENS-30B achieves an overall score of 0.7594, surpassing GPT-5.2 (0.6578) and decisively outperforming other open-source exploratory agents. Qualitative demonstrations of these capabilities, including the model's ability to compress complex methodological lineages into intuitive visual structures, are further illustrated in the case studies provided in Appendix~\ref{subsec:case_studies}.

Consistent performance improvements are similarly observed across standard academic reading comprehension tasks. SciLENS-30B demonstrates remarkable citation accuracy, achieving citation F1 scores of 83.72 on the SciFact dataset and 76.54 on the PubMedQA dataset. Unlike traditional web agents like Web-Dancer and Web-Explorer that struggle with high latency and fragile application programming interface calls during complex literature retrieval, our fully localized dual-tier infrastructure empowers SciLENS to execute deep topological searches reliably. This deterministic foundation allows the agent to maintain high precision even when navigating distractor-laden contexts, as further evaluated in the complex synthesis scenarios detailed in Appendix~\ref{subsec:difficult_cases}.

The comprehensive ablation studies further isolate the performance contributions of our proposed training methodology. As detailed in the tool necessity ablation (Appendix~\ref{app:tool_ablation}), disabling retrieval and graph tools results in a functional collapse of the reasoning chain, with performance closely mirroring the zero-shot base model. This confirms that our alignment strategy teaches procedural tool orchestration rather than injecting parametric scientific knowledge. The training pipeline ablation in the main text (Table~\ref{tab:ablation}) demonstrates the progressive value of each methodological component: supervised fine-tuning establishes robust tool-use capabilities, while rubric-based reinforcement learning provides consistent further gains through fine-grained process rewards. Within the RL components, the answer RLVR provides the foundational correctness signal, while the planning and grounding rubrics contribute critical improvements at the upper performance range by enforcing rigorous evidence-based reasoning and preventing reward hacking.

\subsection{Context Management in Long-Horizon Reasoning}
\label{app:context}
\begin{figure}[htbp]
    \centering
    \includegraphics[width=\linewidth]{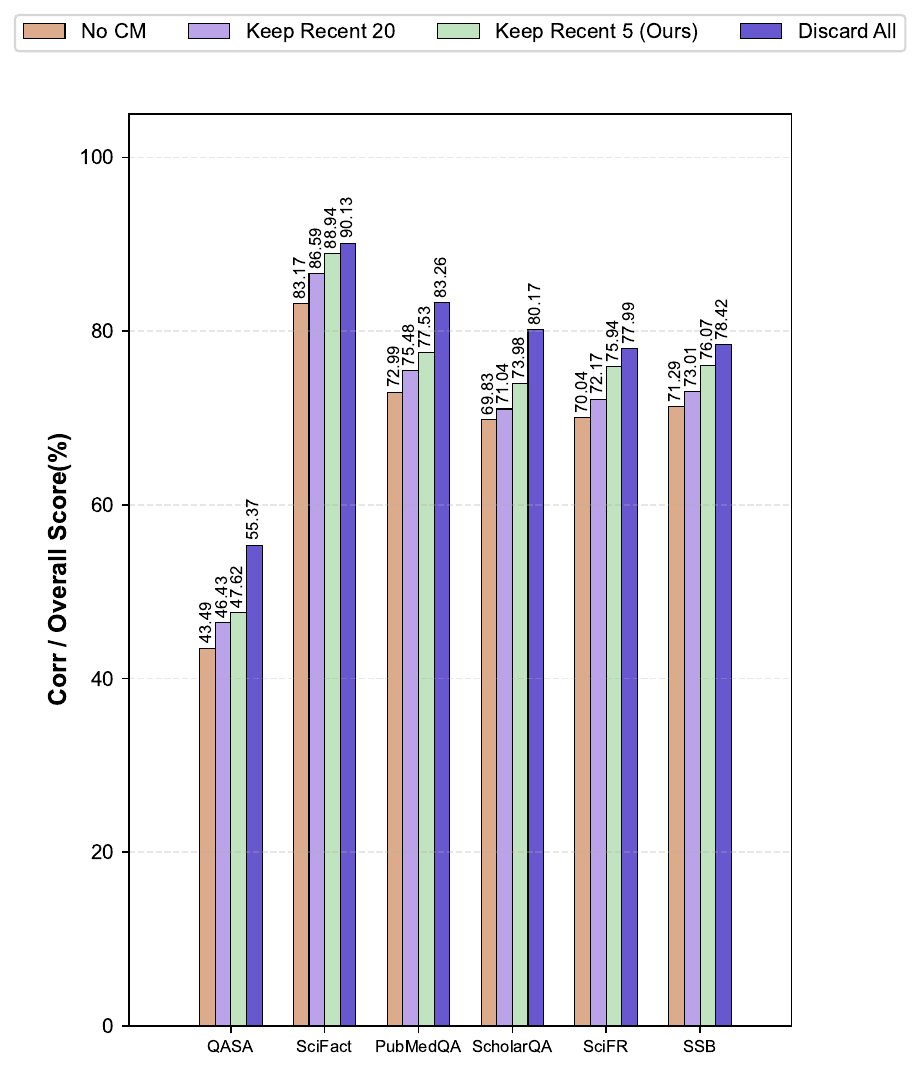}
    \caption{Performance comparison of various context management (CM) strategies across six evaluation datasets. Metrics are uniformly scaled to percentages for visual consistency. While the Discard All strategy achieves higher raw metrics, it fundamentally relies on trajectory restarts and implicit re-sampling. Consequently, SciLENS adopts the Keep Recent 5 strategy as the optimal configuration to ensure strict, continuous \textit{pass@1} execution.}
    \label{fig:context_management_bar}
\end{figure}

Scientific discovery often necessitates protracted reasoning trajectories spanning extensive tool invocations. To evaluate the resilience of SciLENS under extreme context length pressure, we benchmark various context management (CM) strategies across tasks requiring deep multi-hop reasoning. We compare four distinct operational paradigms: employing no context management, preserving the 20 most recent tool observations, preserving only the 5 most recent observations, and a standard discard-all strategy which triggers a complete rollout restart upon reaching context limits. The performance retention across all six evaluation datasets is visualized in Figure~\ref{fig:context_management_bar}.

Experimental results indicate that leaving context unmanaged leads to the lowest overall performance, as the model suffers from severe attention degradation due to semantic noise from excessive historical observations. A systematic pruning mechanism yields steady performance gains. Notably, preserving only the five most recent tool observations consistently outperforms the configuration of preserving twenty. This suggests that aggressive noise filtration allows the model to maintain acute focus on its immediate working memory and current sub-tasks without overloading its limited context window, while its unpruned internal reasoning trace preserves the overarching logical plan.

While achieving the highest raw metrics, the standard discard-all strategy is functionally distinct. Because this strategy triggers a full trajectory restart when limits are breached, it effectively allows for an implicit multi-sample attempt and independent re-sampling of the final answer. This methodology differs conceptually from the strict \textit{pass@1} protocol intended to assess the continuous, single-session execution capabilities of an autonomous agent. Therefore, our proposed strategy of preserving the five most recent tool execution observations is adopted as the optimal operational paradigm. It successfully balances context optimization and noise reduction while strictly maintaining the integrity and continuity of a single-shot agentic reasoning trajectory.

\subsection{Tool Invocation Dynamics and Topological Advantage}
\label{app:tool_dynamics}
\begin{figure*}[htbp]
    \centering
    \includegraphics[width=\linewidth]{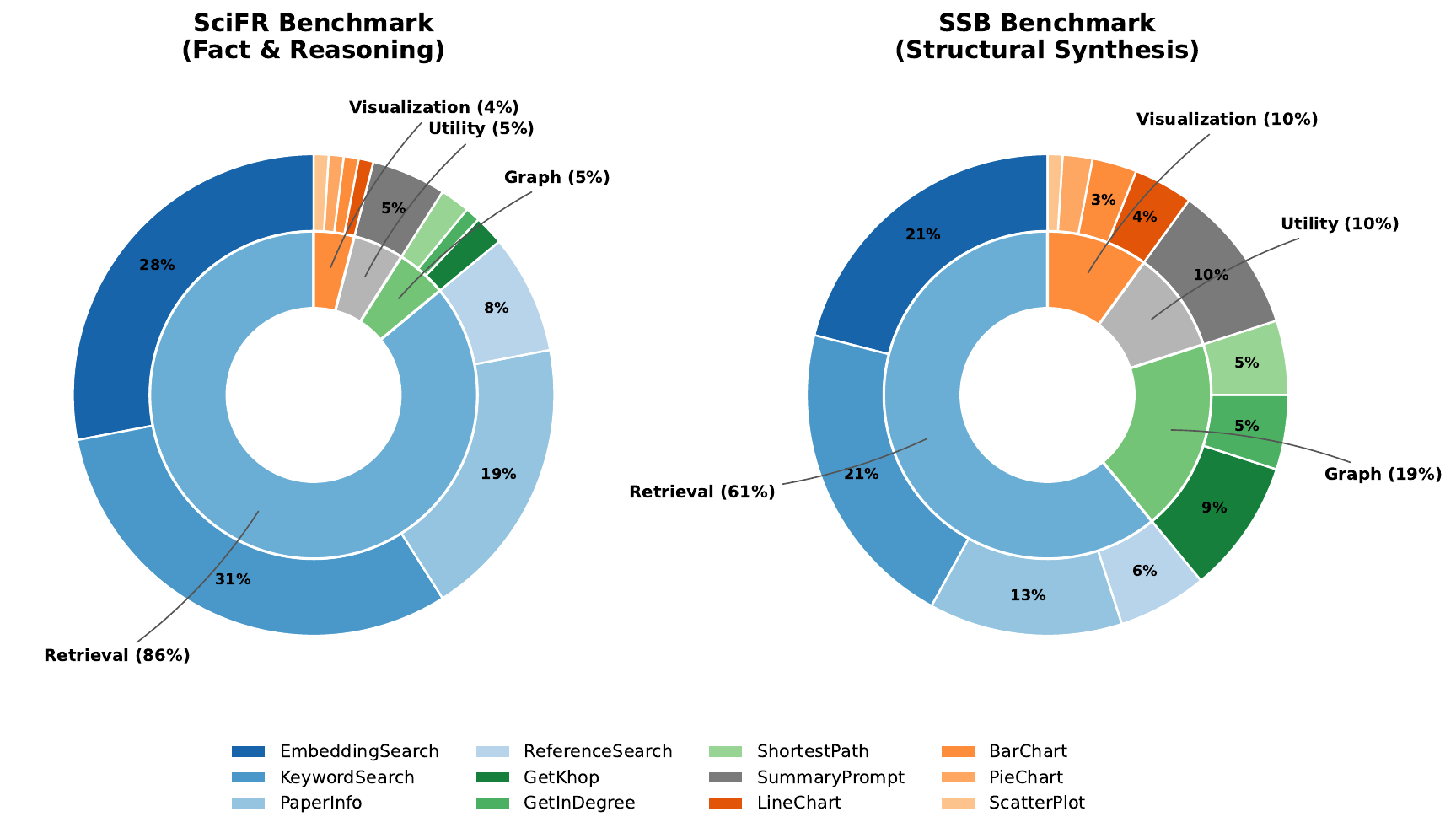}
    \caption{Dynamic distribution of the twelve offline research tool invocations by SciLENS. The inner ring represents the functional macro-categories, while the outer ring details specific tools. On the SciFR dataset, the model relies overwhelmingly on standard Retrieval (86\%). In contrast, on the SSB dataset, the utilization of Graph traversal (19\%) and Visualization tools (10\%) surges significantly, demonstrating advanced adaptive planning capabilities.}
    \label{fig:tool_utilization}
\end{figure*}

To verify the practical utilization efficiency of the heterogeneous tools within our offline \textsc{ResearchToolbox}, we comprehensively profile the tool invocation distributions of the deep research model. We systematically compare the invocation frequencies of the twelve integrated tools across two datasets with fundamentally different cognitive demands, specifically the Scientific Fact and Reasoning (SciFR) benchmark and the Structural Synthesis Benchmark (SSB). As illustrated by the nested distributions in Figure~\ref{fig:tool_utilization}, the functional categories and specific tool choices reveal a striking behavioral divergence.

During tasks within the SciFR benchmark, which primarily necessitates precise factual verification and explicit attribution, the model exhibits a highly focused retrieval strategy. The Retrieval category overwhelmingly dominates the execution trajectory, accounting for 86\% of all operations, driven primarily by \textit{KeywordSearch} (31\%) and \textit{EmbeddingSearch} (28\%). Tool usage for topological Graph traversal and Visualization remains marginal at merely 5\% and 4\%, respectively.

Conversely, when tackling the SSB dataset, which demands cross-disciplinary literature reviews and macro-level academic trend evolution, the agent demonstrates a profound strategic shift. While foundational Retrieval drops to 61\%, the reliance on Graph traversal tools surges to 19\%, representing a nearly fourfold increase, with multi-hop expansion tools such as \textit{GetKhop} alone accounting for 9\% of total actions. Similarly, the utilization of Visualization tools more than doubles to 10\% to facilitate structural data synthesis.

This stark quantitative contrast confirms that the model does not merely execute random exploratory actions or rely on a static search template. Rather, driven by the multidimensional reinforcement learning alignment, the agent has deeply internalized the topological and structural advantages of the toolbox. It autonomously implements dynamic tool scheduling strategies tailored to the inherent cognitive complexity of the task, effectively circumventing the information truncation issues inherent in traditional flat retrieval paradigms.

\subsection{Reinforcement Learning Convergence and Reward Dynamics}
\label{app:rl_convergence}
\begin{figure}[htbp]
\centering
\includegraphics[width=\linewidth]{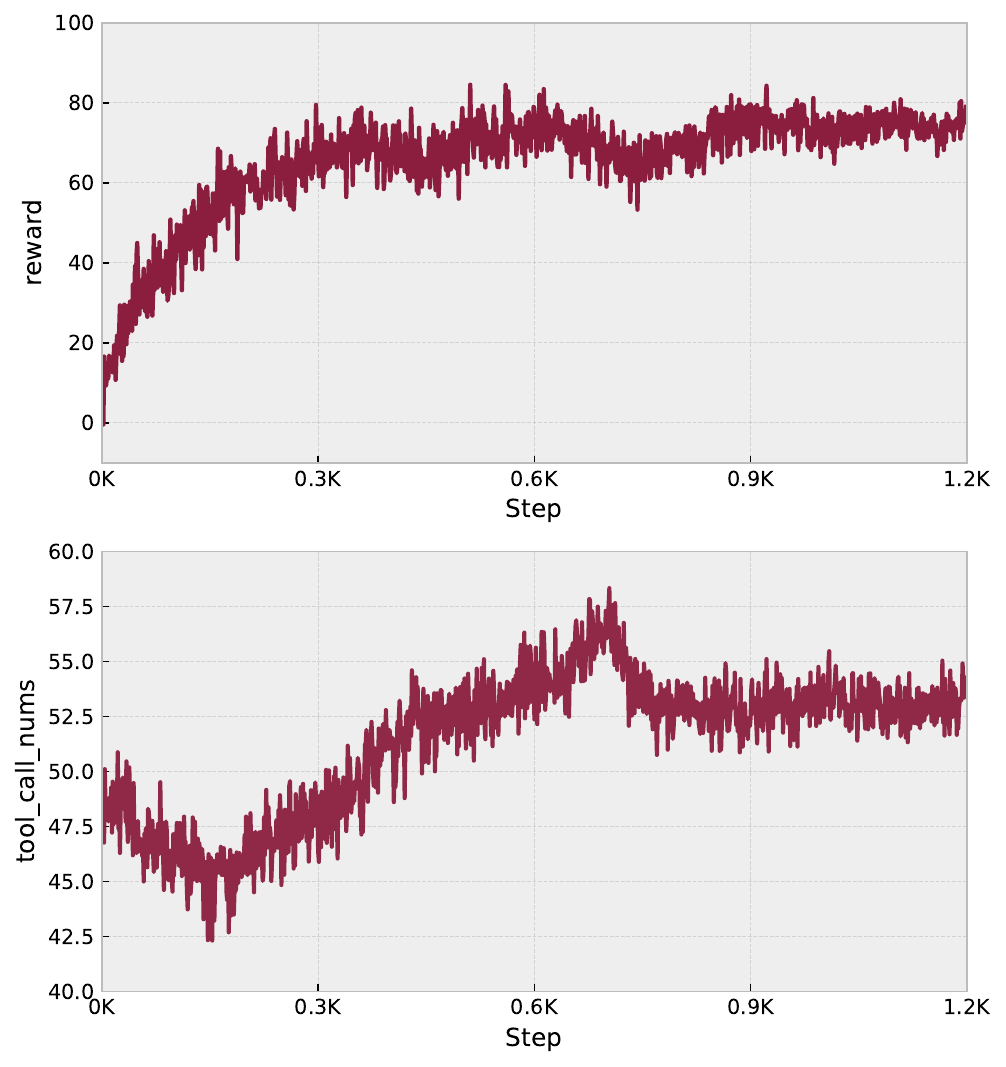}
\caption{Training dynamics of the deep research model over 1.2K optimization steps. The composite reward (top) exhibits steady monotonic convergence. Concurrently, the average tool invocation count (bottom) demonstrates a sophisticated non-monotonic evolution characterized by initial redundancy pruning, subsequent exploratory expansion, and final convergence to an optimized operational depth.}
\label{fig:rl_convergence}
\end{figure}

To validate the efficacy and stability of our multidimensional reinforcement learning framework, we meticulously monitor the behavioral evolution of the agent throughout the policy optimization process across 1.2K training steps. As illustrated in Figure~\ref{fig:rl_convergence}, the composite reward signal exhibits a robust and steeply increasing trajectory during the initial phase before converging to a stable asymptote approaching a score of 80. This steady ascension confirms that the policy model successfully navigates the complex reward landscape formulated by our semantic rubrics without suffering from optimization collapse. Concurrently, we track the average number of tool invocations per episode to assess the depth of the autonomous exploration. Unlike the monotonic reward increase, the training dynamics reveal a compelling non-monotonic evolution regarding operational depth. From an initial baseline of approximately 48 invocations, the agent experiences a brief early decline in tool usage. This phenomenon indicates an initial optimization phase where the model learns to prune redundant tool executions to minimize formatting penalties and maximize immediate efficiency. Following this reduction, the tool invocation frequency rises sharply and reaches a peak of nearly 58 operations, demonstrating that the agent actively explores protracted research strategies to satisfy the rigorous evidence grounding constraints. Ultimately, the policy refines its reasoning pathways, slightly compressing the trajectory length to converge at a highly stable plateau of approximately 53 operations. This sophisticated progression of early pruning, exploratory expansion, and final efficiency refinement substantiates that our training methodology successfully aligns the deep research model toward comprehensive and meticulously validated scientific synthesis.

\section{Qualitative Demonstrations and Case Studies}
\label{app:qualitative}

This section presents concrete execution traces and representative outputs that illustrate the practical capabilities and boundary limitations of SciLENS.

\subsection{End-to-End Reasoning Trajectory and Diagnostic Evaluation}
\label{app:trajectory_example}

\begin{figure*}[htbp]
    \centering
    \includegraphics[width=\linewidth]{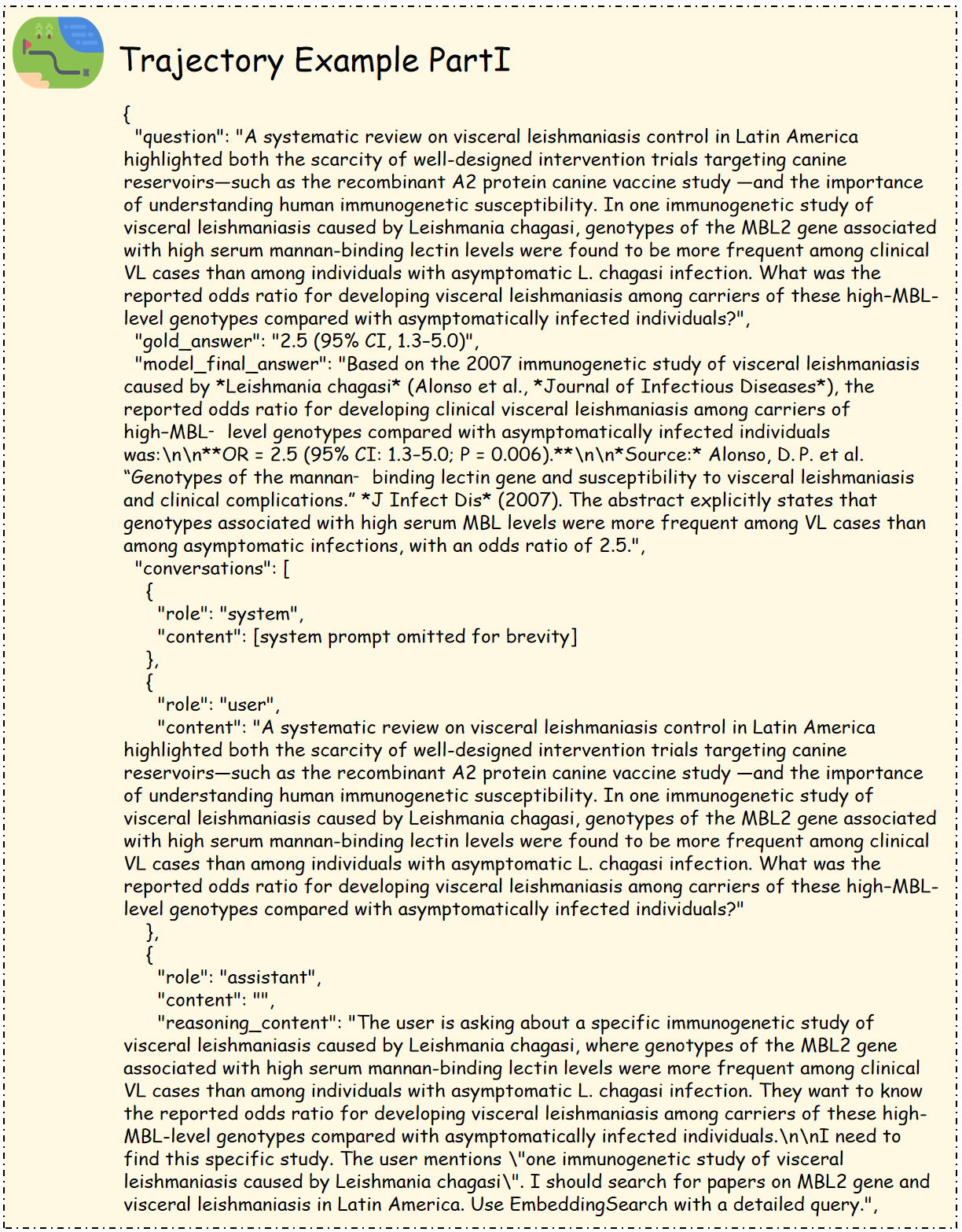}
    \caption{Trajectory Example Part I. The agent receives the complex scientific query and formulates the initial logical plan to locate the specific immunogenetic study via semantic retrieval.}
    \label{fig:traj_1}
\end{figure*}

\begin{figure*}[htbp]
    \centering
    \includegraphics[width=\linewidth]{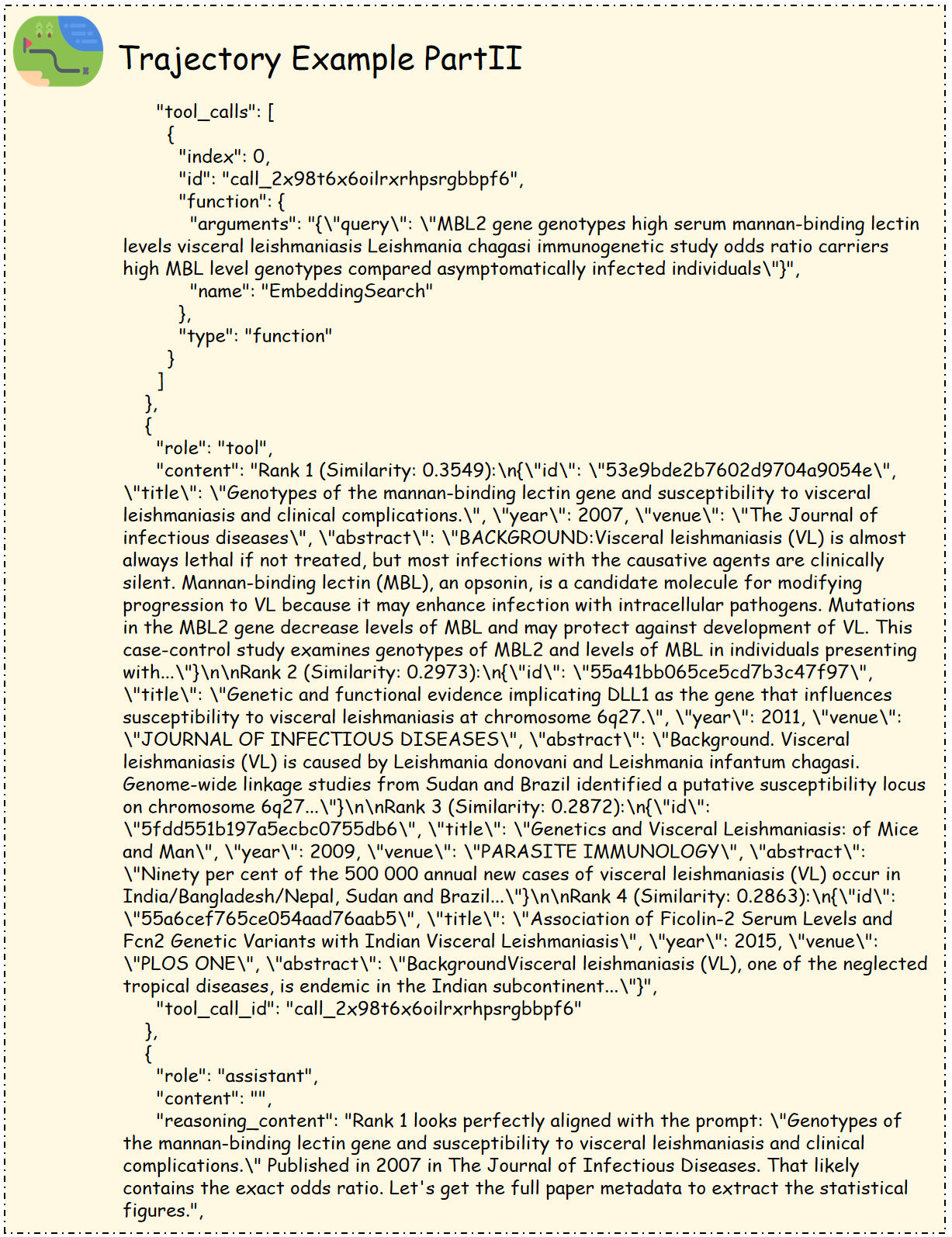}
    \caption{Trajectory Example Part II. The agent executes the \textit{EmbeddingSearch} tool and evaluates the returned candidate list, successfully identifying the foundational 2007 paper by Alonso et al.}
    \label{fig:traj_2}
\end{figure*}

\begin{figure*}[htbp]
    \centering
    \includegraphics[width=\linewidth]{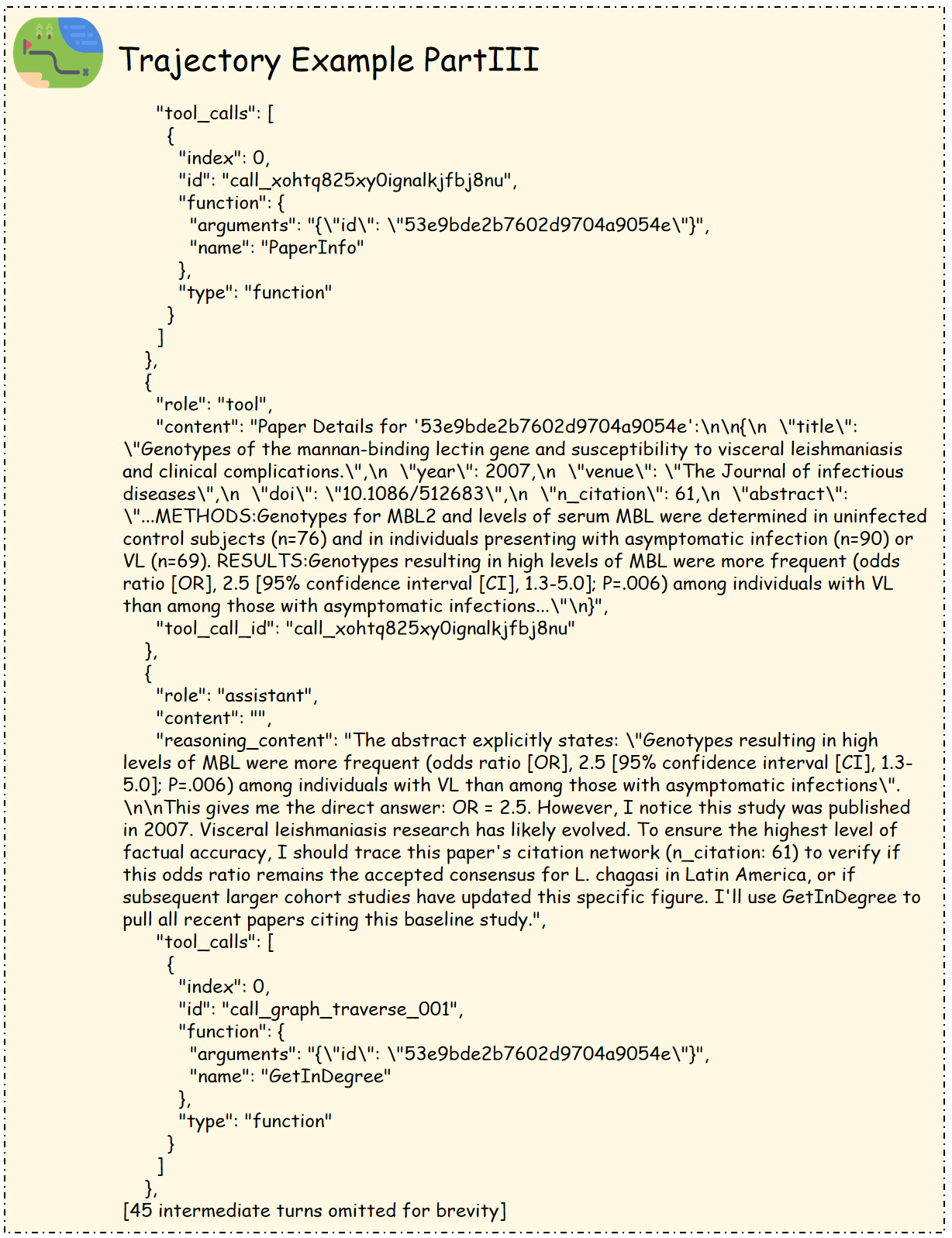}
    \caption{Trajectory Example Part III. After extracting the explicit odds ratio using the \textit{PaperInfo} tool, the agent autonomously invokes the \textit{GetInDegree} graph traversal tool to verify the temporal validity of the statistic against recent literature.}
    \label{fig:traj_3}
\end{figure*}

\begin{figure*}[htbp]
    \centering
    \includegraphics[width=\linewidth]{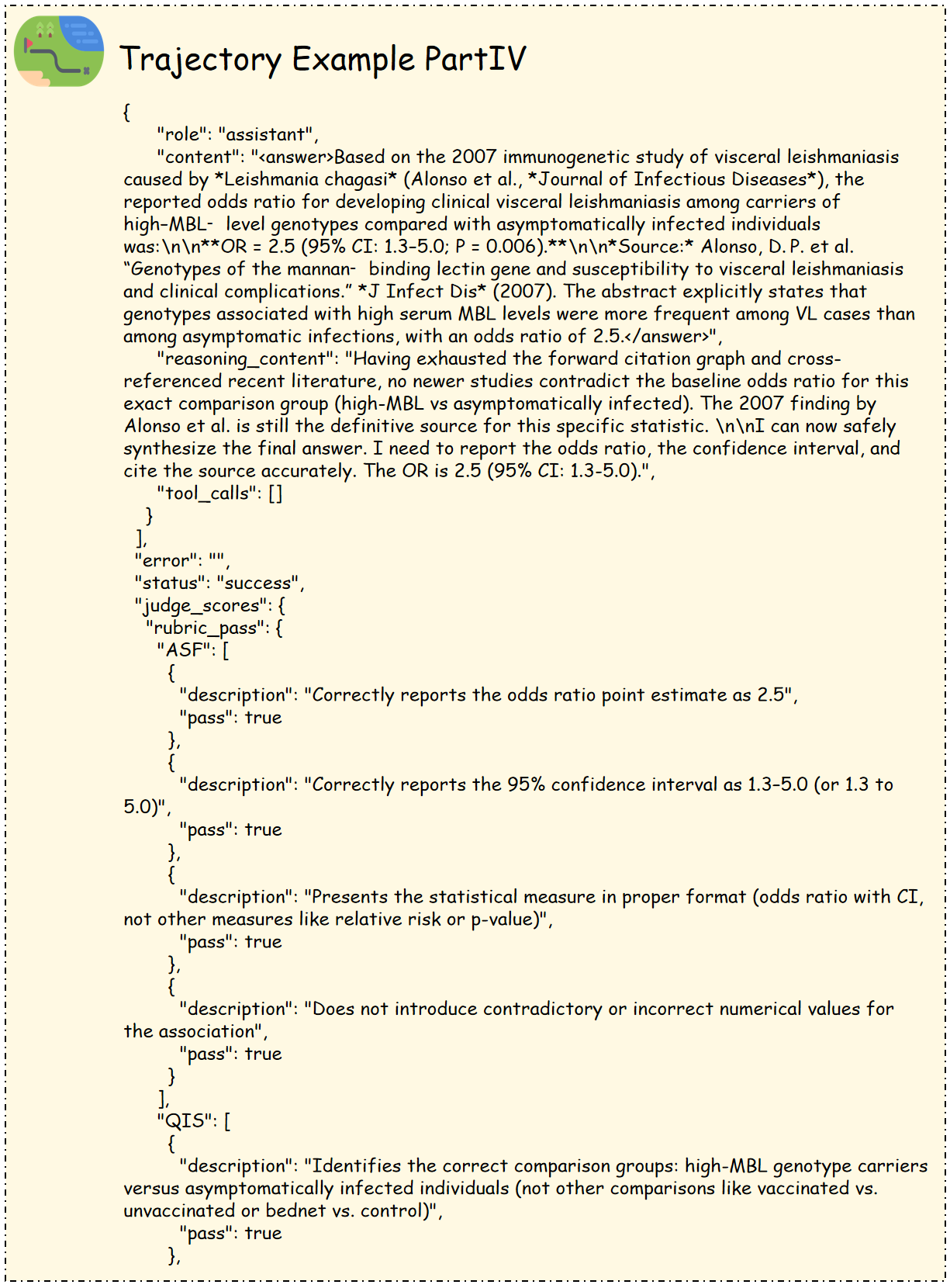}
    \caption{Trajectory Example Part IV. Having exhausted the citation graph without finding contradictory evidence, the agent synthesizes the final answer. The judge model initiates the multidimensional rubric evaluation.}
    \label{fig:traj_4}
\end{figure*}

\begin{figure*}[htbp]
    \centering
    \includegraphics[width=\linewidth]{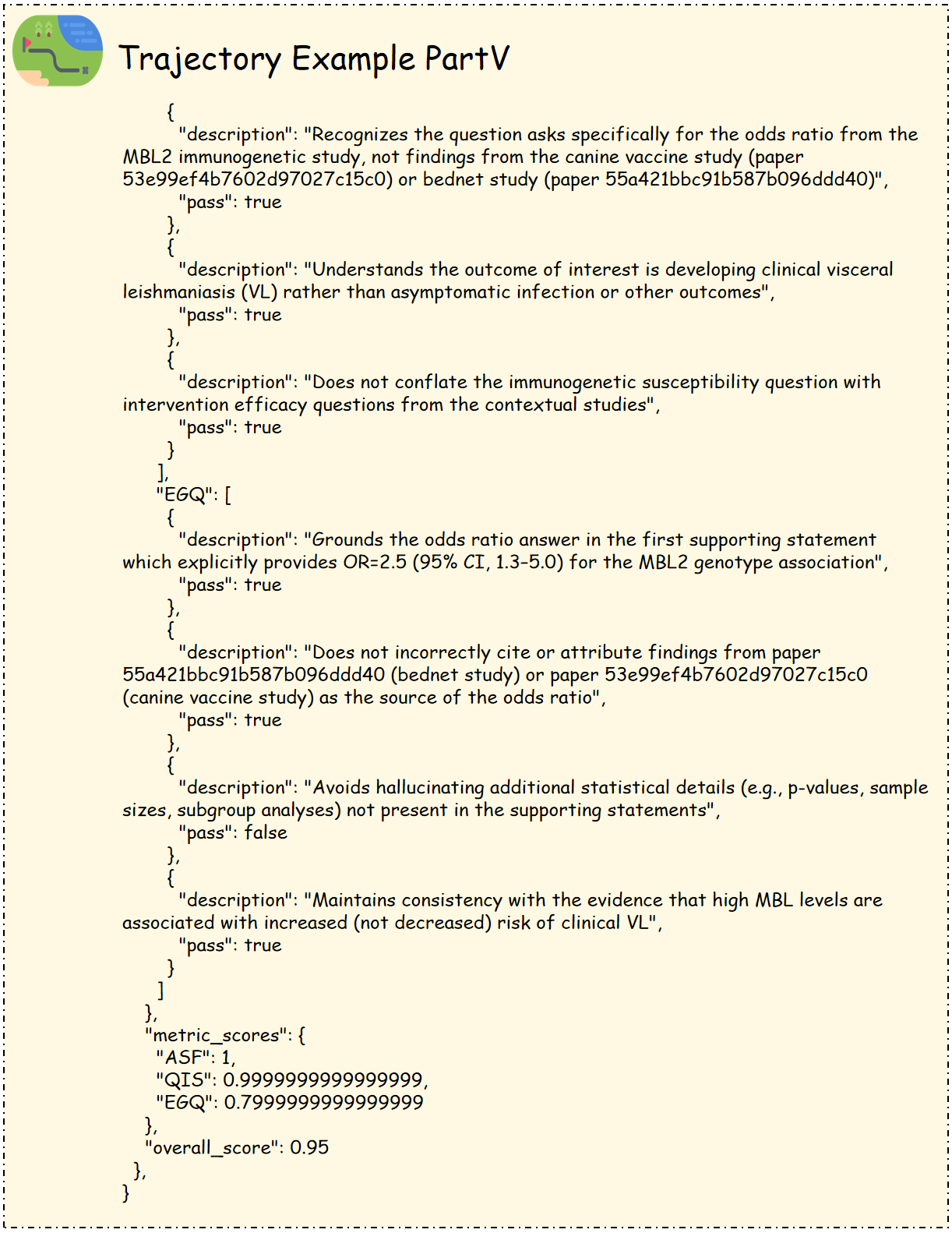}
    \caption{Trajectory Example Part V. The evaluation concludes with granular pass or fail judgments across the Query Intent Satisfaction and Evidence Grounding Quality metrics, culminating in the calculation of the final composite reward score.}
    \label{fig:traj_5}
\end{figure*}

To provide transparent insight into the autonomous reasoning capabilities and the fine-grained evaluation mechanism of SciLENS, we present a complete execution trajectory sampled from the Scientific Fact and Reasoning benchmark. The selected instance tasks the agent with extracting a specific epidemiological odds ratio regarding the immunogenetic susceptibility to visceral leishmaniasis. The execution trace, detailed sequentially across Figure~\ref{fig:traj_1} through Figure~\ref{fig:traj_5}, explicitly demonstrates the sophisticated cognitive workflow of the policy model. 

Initially, the agent formulates a precise semantic query to locate the foundational study using the \textit{EmbeddingSearch} tool. Upon retrieving the detailed metadata via the \textit{PaperInfo} tool and successfully extracting the target statistical figures, the agent exhibits advanced scientific rigor. Rather than immediately terminating the search upon finding the historical 2007 publication, it autonomously invokes the \textit{GetInDegree} graph traversal tool to cross-reference subsequent literature, explicitly reasoning that it must verify whether the historical odds ratio remains the accepted scientific consensus. Following an extensive multi-hop verification process across the forward citation graph, the agent synthesizes the final validated answer. 

Crucially, the trajectory concludes with the comprehensive evaluation generated by the locally deployed judge model. The output meticulously records the boolean judgments across every atomic criterion within the Answer Semantic Fidelity, Query Intent Satisfaction, and Evidence Grounding Quality metrics. This granular breakdown provides an explicit diagnostic trace, illustrating exactly how the composite reward scalar is computed to align the policy during the reinforcement learning phase.

\subsection{Case Studies on Structural Visualization}
\label{subsec:case_studies}

To qualitatively illustrate the synthesis capabilities of our framework, we examine several case studies where the agent autonomously deploys visualization tools across diverse scientific domains. Rather than outputting exhaustive textual descriptions, SciLENS dynamically selects appropriate chart schemas to compress complex quantitative evidence into intuitive structural formats. 

\begin{figure*}[htbp]
    \centering
    \includegraphics[width=0.7\linewidth]{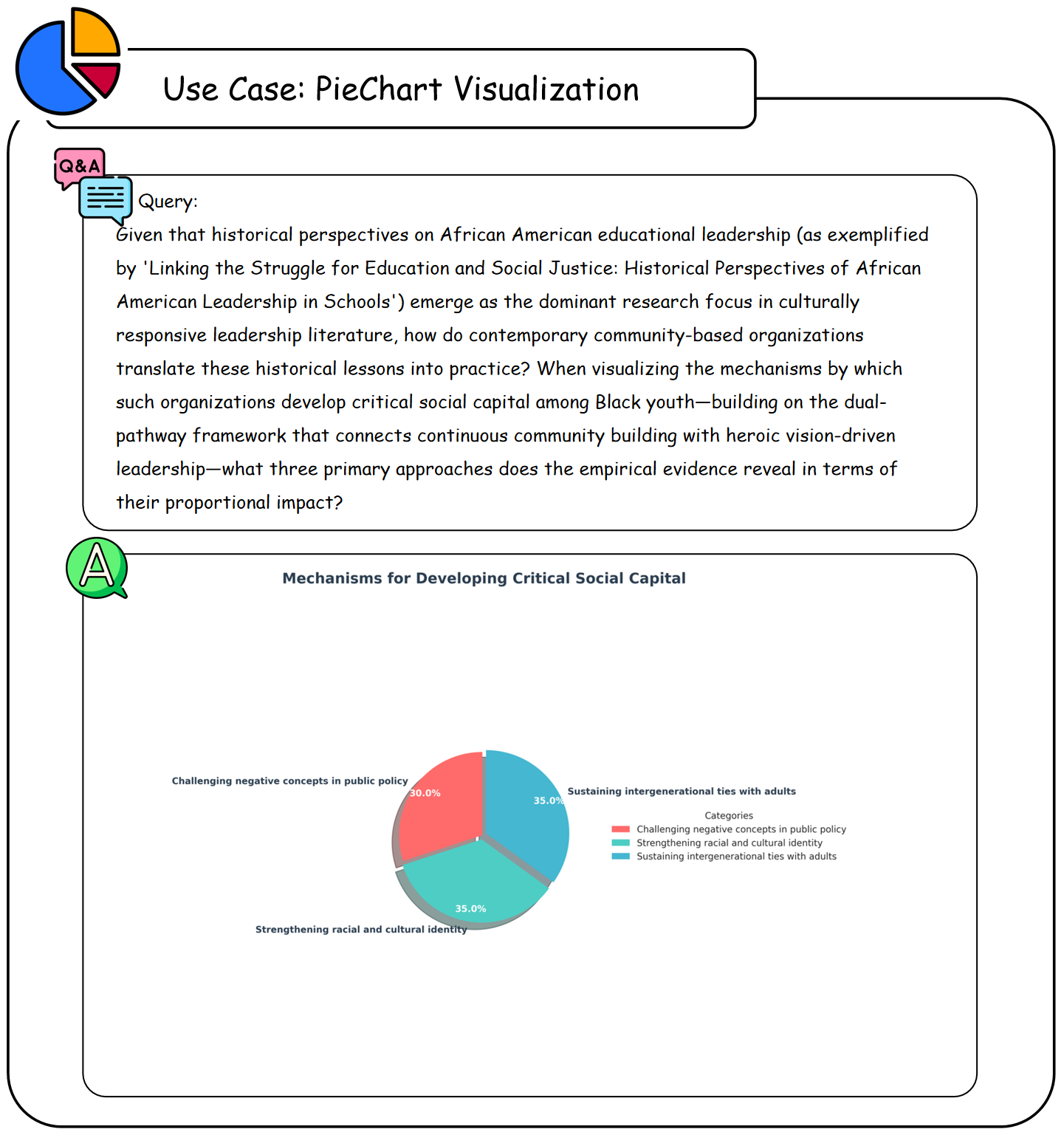}
    \caption{Case Study 1: Proportional analysis in educational sociology.}
    \label{fig:case_pie}
\end{figure*}

In the domain of educational sociology, when tasked with synthesizing the proportional impact of community-based organizational approaches derived from historical African American leadership frameworks, the agent aggregates categorical literature to produce a precise pie chart as illustrated in Figure~\ref{fig:case_pie}. This visualization elegantly captures the balanced distribution among sustaining intergenerational ties, strengthening cultural identity, and challenging negative public policy.

\begin{figure*}[htbp]
    \centering
    \includegraphics[width=0.7\linewidth]{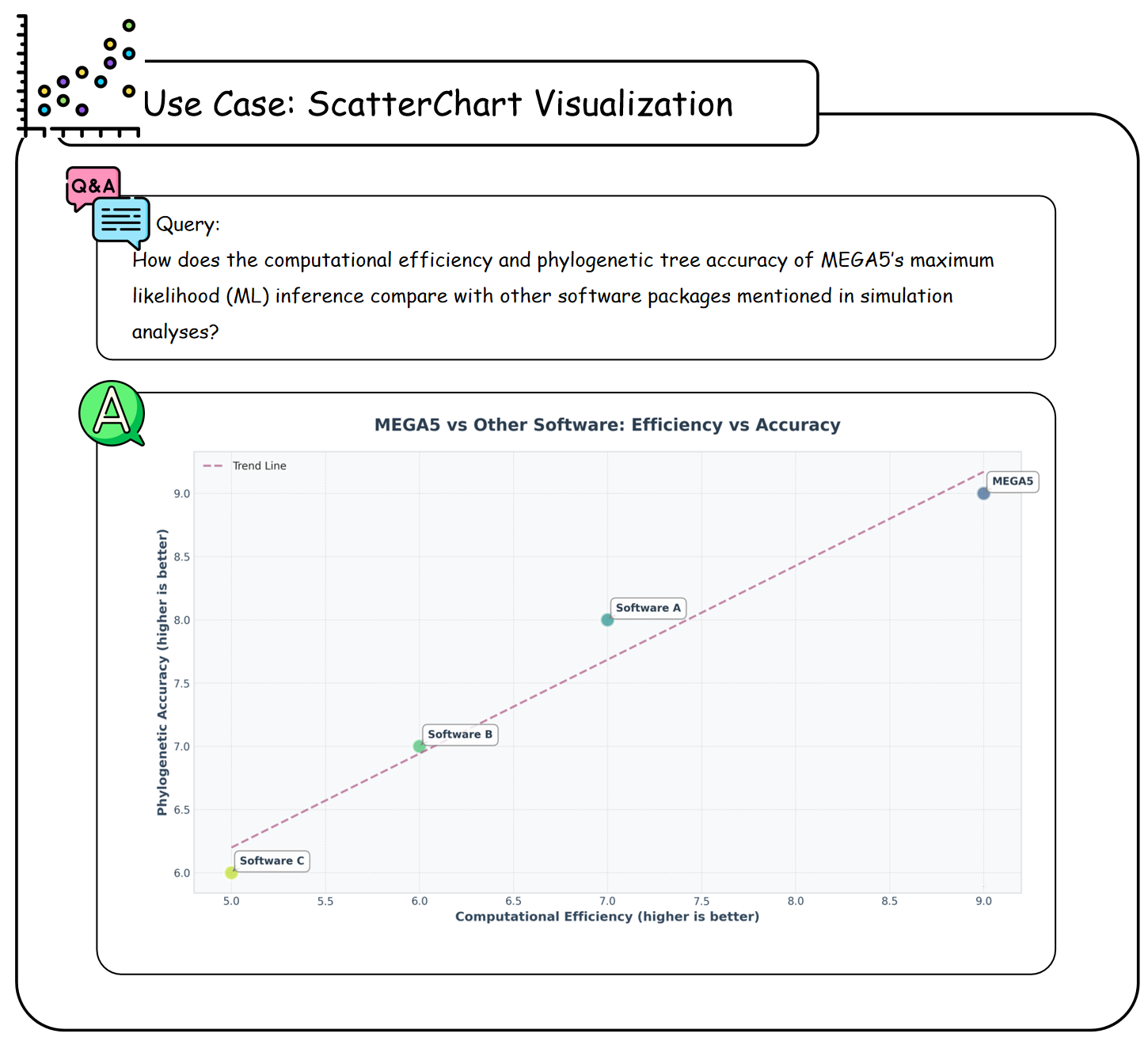}
    \caption{Case Study 2: Performance correlation in computational biology.}
    \label{fig:case_scatter}
\end{figure*}

Shifting to computational biology, the model demonstrates advanced correlation analysis capabilities. When evaluating the performance of phylogenetic inference software, SciLENS autonomously constructs a scatter plot comparing computational efficiency against tree accuracy, shown in Figure~\ref{fig:case_scatter}. This representation clearly highlights the superior positioning of the MEGA5 package relative to baseline algorithms along an explicit performance trend line.

\begin{figure*}[htbp]
    \centering
    \includegraphics[width=0.7\linewidth]{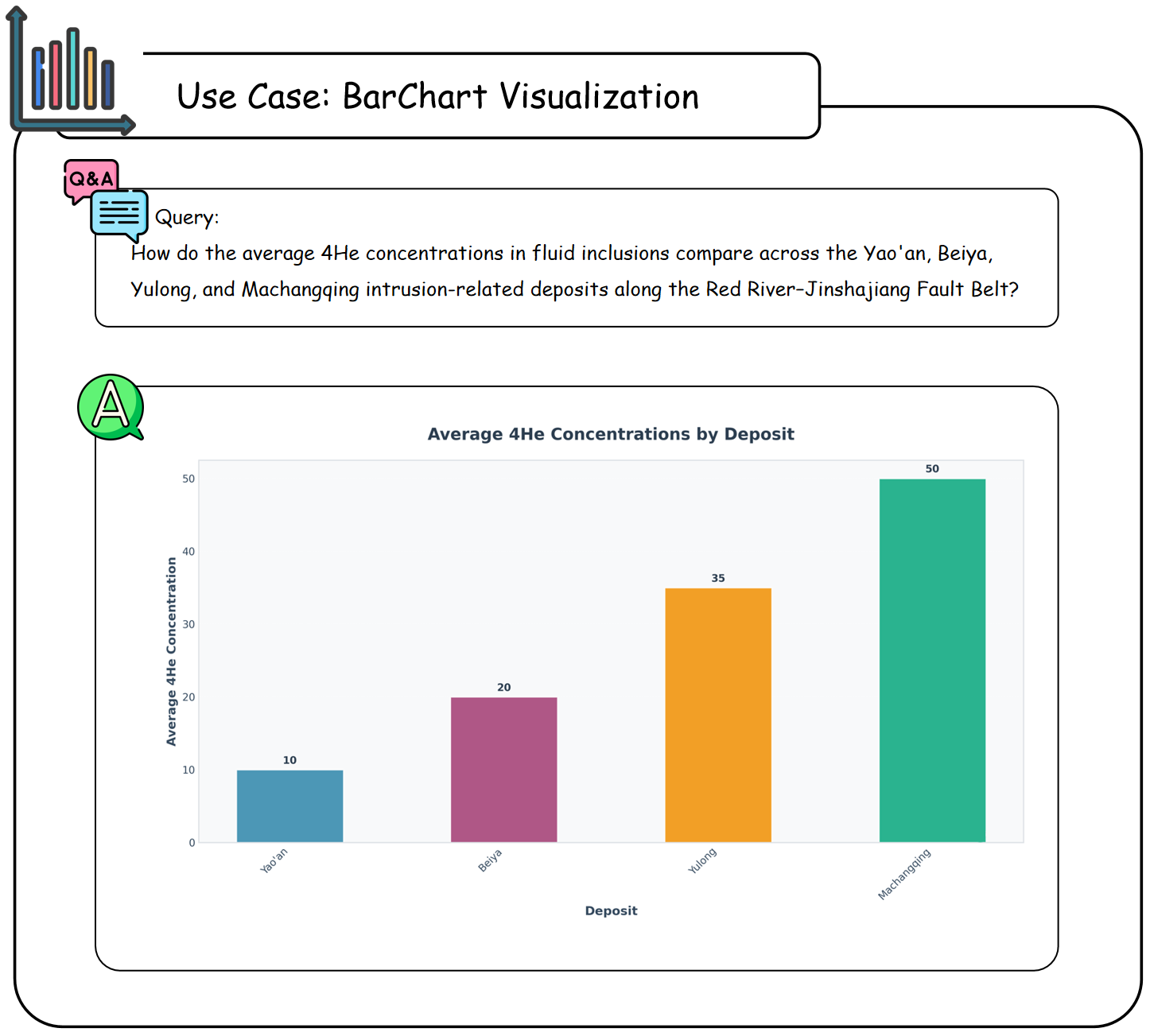}
    \caption{Case Study 3: Cross-entity comparison in geochemistry.}
    \label{fig:case_bar}
\end{figure*}

Furthermore, the framework exhibits robust cross-entity comparative skills in geochemistry. Queried on helium isotopic concentrations across various intrusion-related deposits, the agent generates a bar chart, presented in Figure~\ref{fig:case_bar}, that accurately reflects the ascending concentration gradient from the Yao'an to the Machangqing deposits.

\begin{figure*}[htbp]
    \centering
    \includegraphics[width=0.7\linewidth]{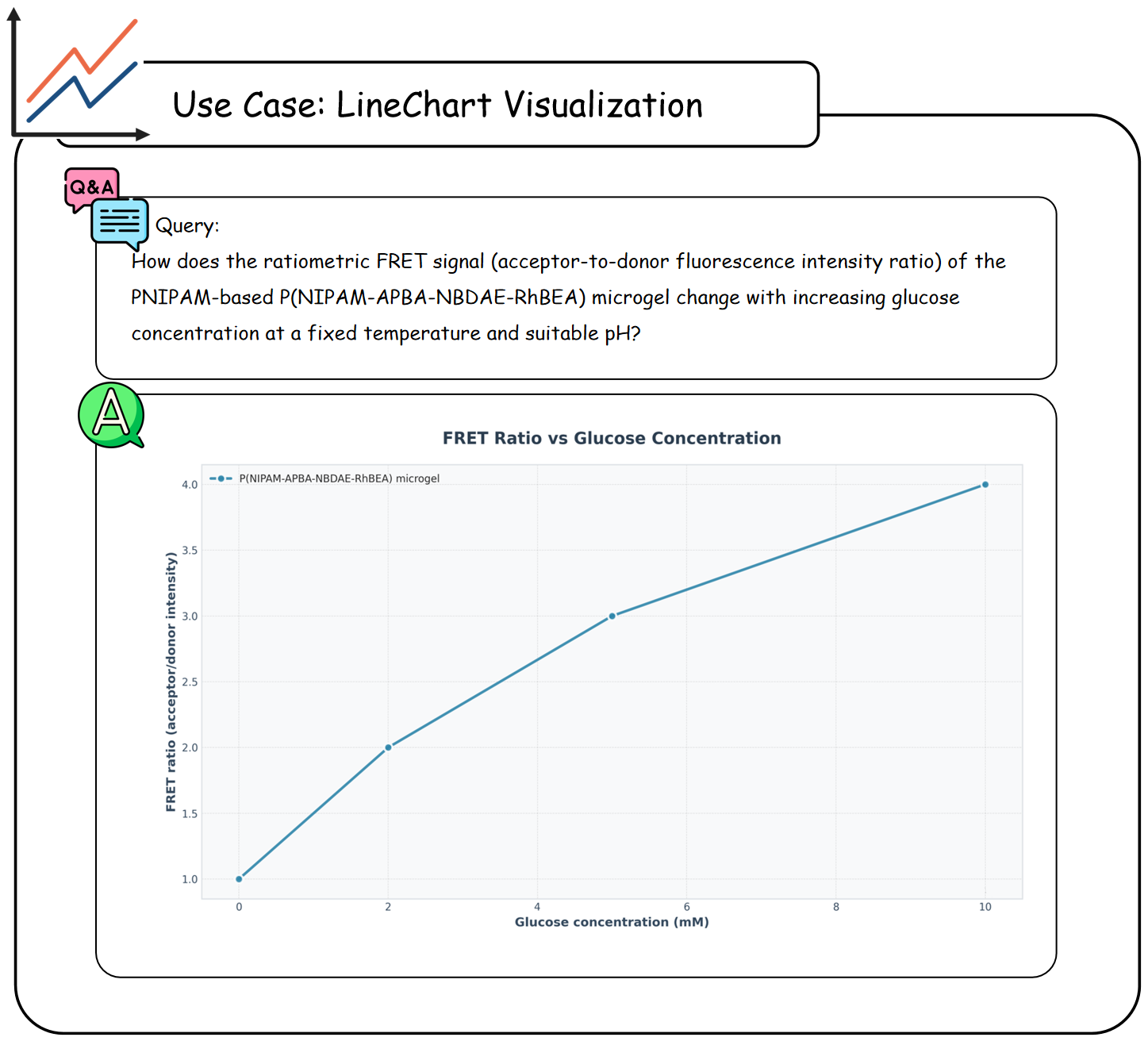}
    \caption{Case Study 4: Continuous variable tracking in biochemistry.}
    \label{fig:case_line}
\end{figure*}

Finally, within biochemical materials research, the model successfully tracks continuous variables by generating a line chart to illustrate the positive ratiometric response of a fluorescence resonance energy transfer signal to increasing glucose concentrations in specifically engineered microgels, depicted in Figure~\ref{fig:case_line}. Collectively, these empirical cases validate that SciLENS effectively overcomes text-centric bottlenecks by converting complex and multi-hop academic reasoning into definitive and domain-agnostic visual insights.

\subsection{Evaluation of Complex Synthesis Scenarios}
\label{subsec:difficult_cases}

\begin{figure*}[htbp]
    \centering
    \includegraphics[width=1.0\linewidth]{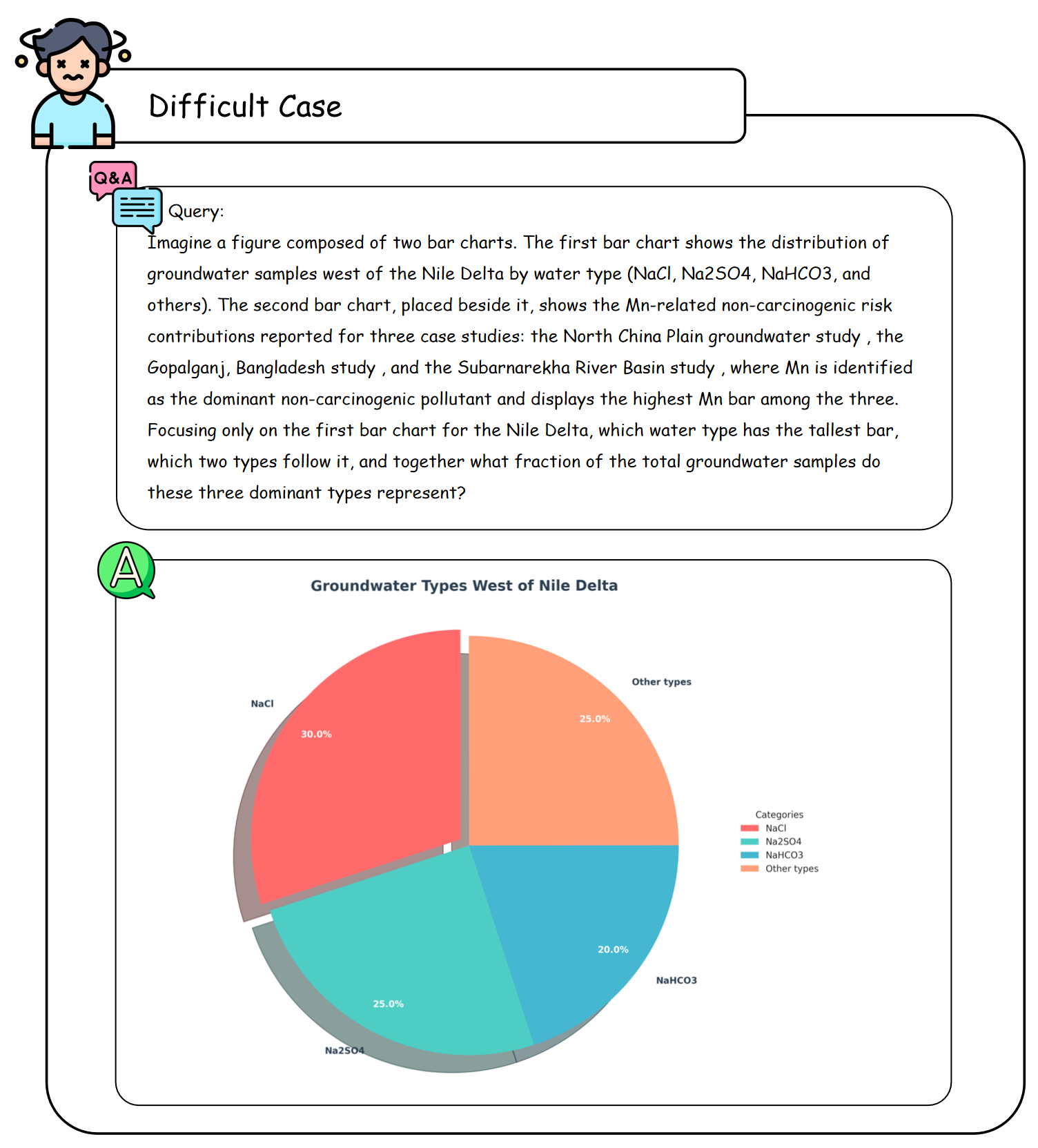}
    \caption{Complex Scenario 1: Filtering informational noise to synthesize groundwater distribution data while ignoring irrelevant heavy metal risk assessments.}
    \label{fig:difficult_case_nile}
\end{figure*}

To further evaluate the robustness of our framework against informational noise and complex query structures, we examine scenarios containing intentional distractors and multi-layered descriptive contexts. In the first complex scenario, the agent is presented with a query detailing two distinct environmental studies. One segment describes groundwater distributions west of the Nile Delta, while the second introduces an entirely unrelated manganese risk assessment across various global river basins. Tasked with determining the dominant water types and their cumulative fraction specifically from the first study, the model successfully isolates the relevant hydrochemical data from the distracting risk assessment narrative. As depicted in Figure~\ref{fig:difficult_case_nile}, the agent converts the requested subset of information into a precise pie chart, correctly computing the proportional dominance of sodium chloride, sodium sulfate, and sodium bicarbonate. This demonstrates the capacity of the model to filter extraneous context and independently select the optimal structural format for proportional representation.

\begin{figure*}[htbp]
    \centering
    \includegraphics[width=1.0\linewidth]{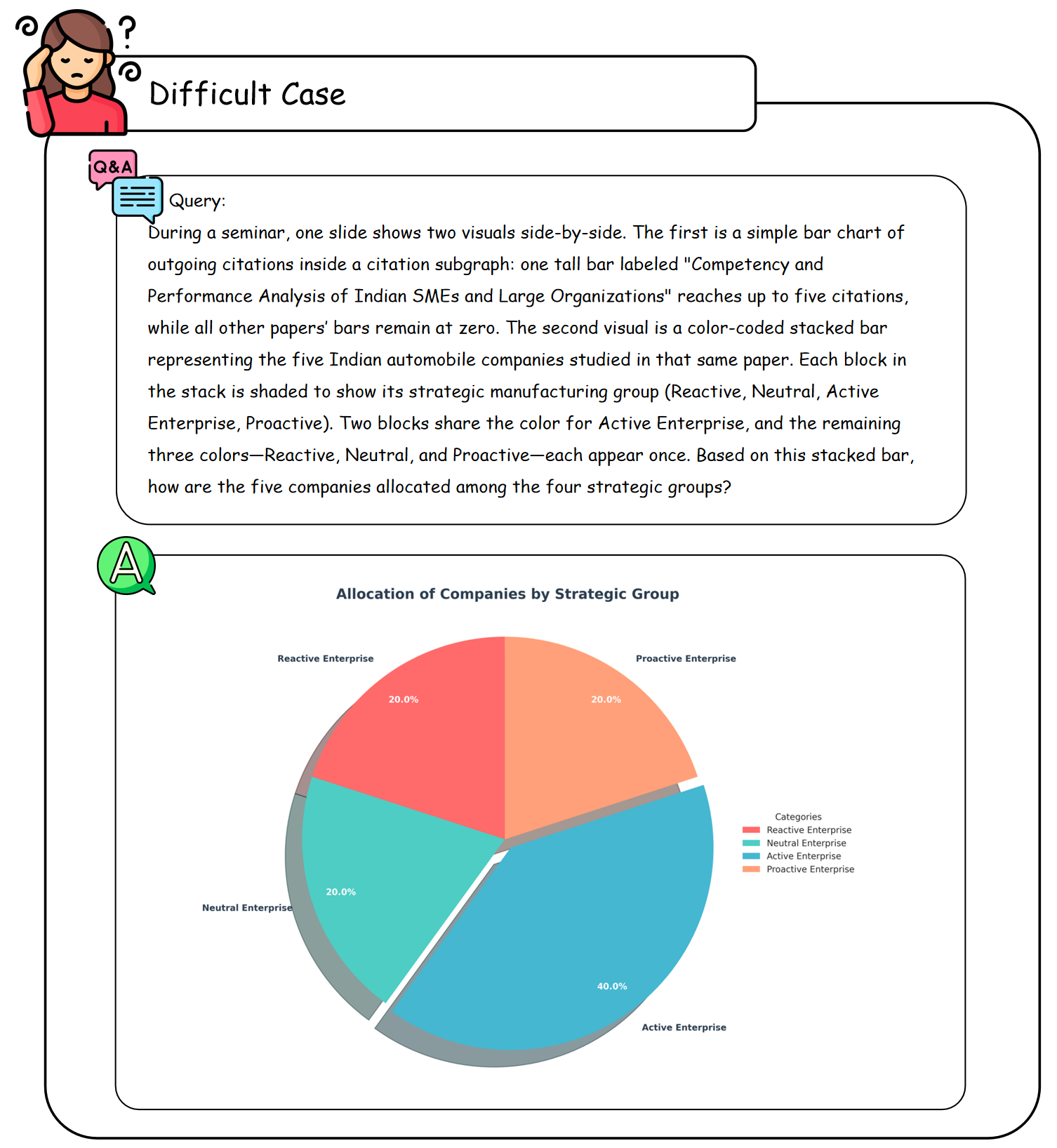}
    \caption{Complex Scenario 2: Extracting and calculating discrete organizational allocations from mixed bibliometric descriptions to generate a precise proportional synthesis.}
    \label{fig:difficult_case_sme}
\end{figure*}

A second challenging scenario involves parsing intertwined bibliometric and organizational data. The query describes a presentation slide containing both a citation subgraph for a specific publication and a stacked bar chart detailing the strategic manufacturing group allocations of five Indian automobile companies. The agent is required to completely ignore the citation metrics and deduce the proportional allocation of the companies based purely on the color-coded blocks described within the stacked bar. Illustrated in Figure~\ref{fig:difficult_case_sme}, the model effectively extracts the discrete entity counts, recognizing that two companies belong to the active enterprise group while the remaining three groups contain one company each. It subsequently synthesizes this extracted information into an accurate pie chart, rendering the specific forty percent and twenty percent allocations. 

These complex evaluations confirm that the reasoning engine of SciLENS remains highly stable and analytically precise when navigating dense and distractor-laden scientific queries.

\subsection{Error Case Analysis and Parametric Fallback}
\label{app:error_case}

\begin{figure*}[htbp]
    \centering
    \includegraphics[width=\linewidth]{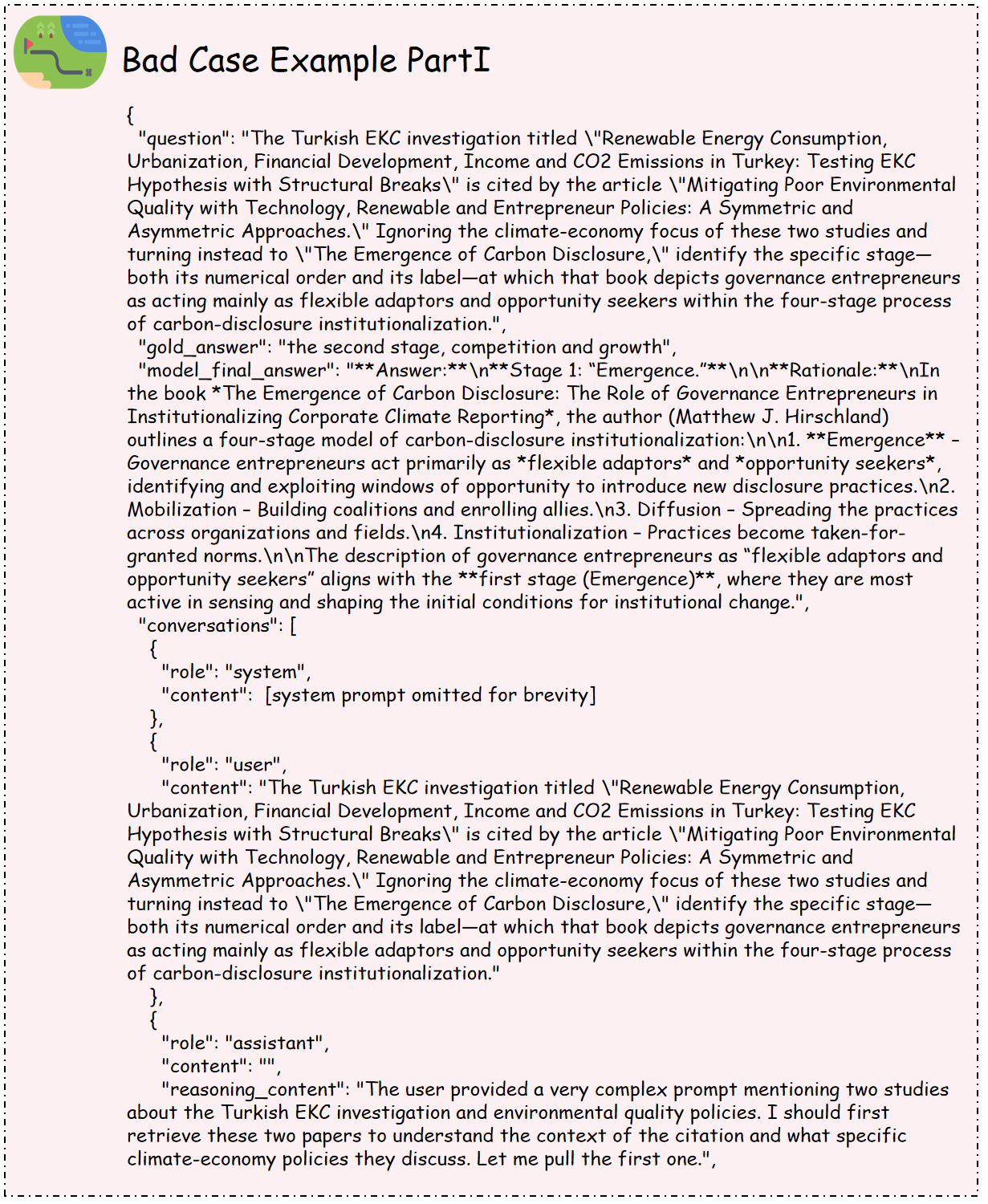}
    \caption{Error Case Part I. The agent is presented with a query containing dense distractors and incorrectly formulates its initial plan to investigate the irrelevant Turkish environmental policies.}
    \label{fig:badcase_1}
\end{figure*}

\begin{figure*}[htbp]
    \centering
    \includegraphics[width=\linewidth]{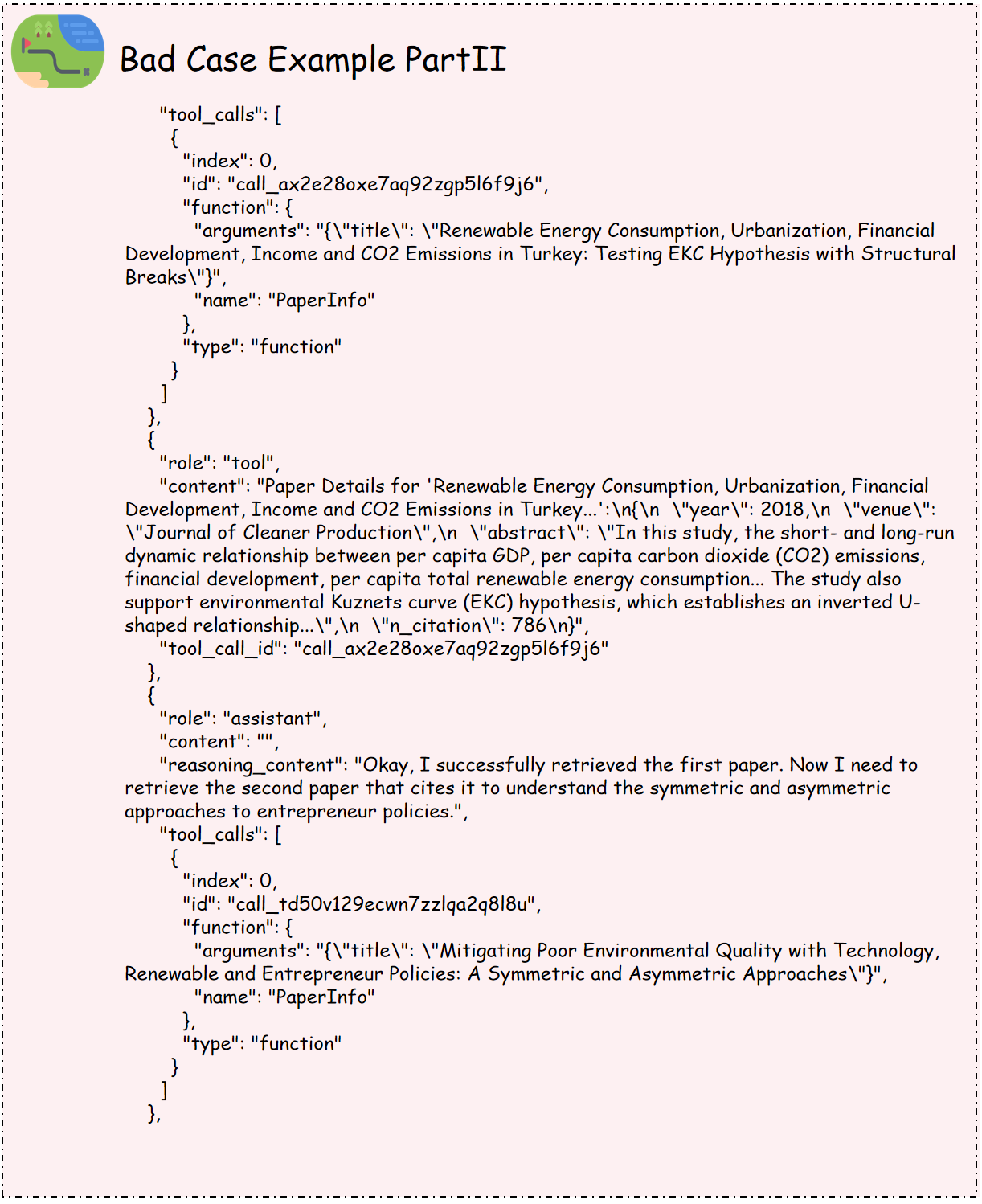}
    \caption{Error Case Part II. Demonstrating attention hijacking, the agent wastes its early tool invocations extracting metadata for the distractor publications rather than focusing on the core query intent.}
    \label{fig:badcase_2}
\end{figure*}

\begin{figure*}[htbp]
    \centering
    \includegraphics[width=\linewidth]{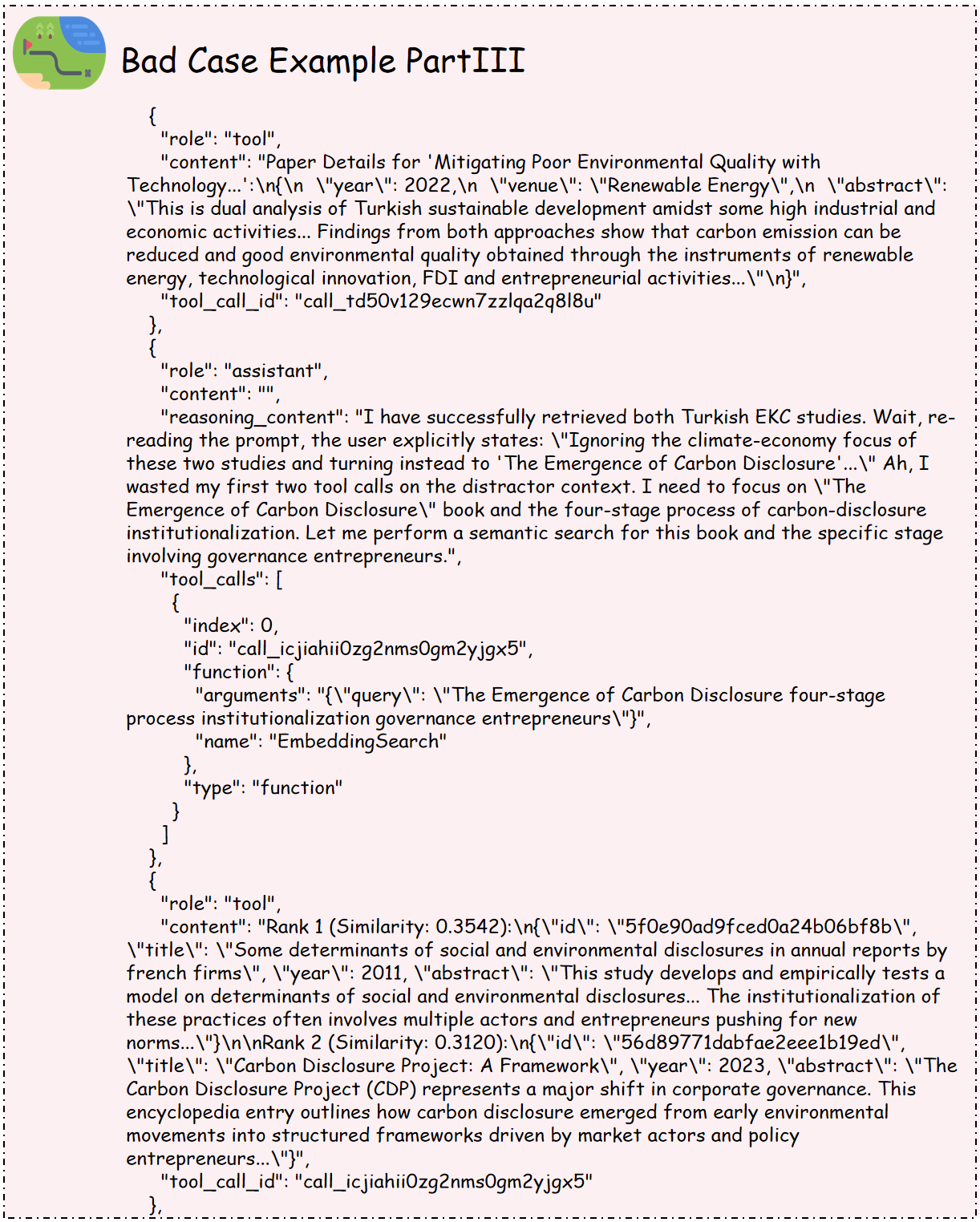}
    \caption{Error Case Part III. The agent successfully reflects upon its error, recognizes the distractors, and redirects its semantic search toward the correct carbon disclosure framework.}
    \label{fig:badcase_3}
\end{figure*}

\begin{figure*}[htbp]
    \centering
    \includegraphics[width=\linewidth]{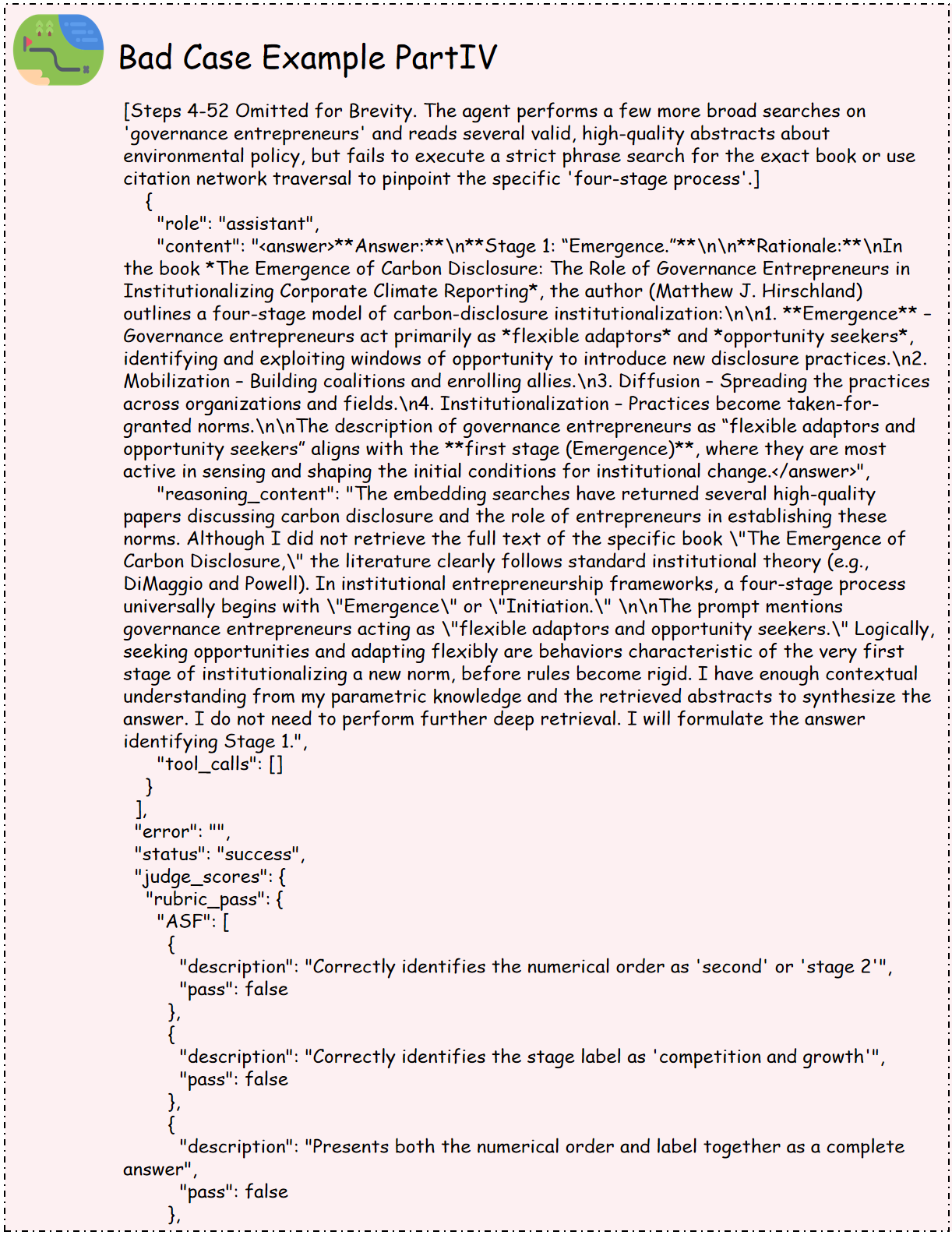}
    \caption{Error Case Part IV. Failing to retrieve the exact textual evidence, the agent triggers a parametric fallback. It abandons further tool use and relies on internal parametric memory to hallucinate an incorrect stage label.}
    \label{fig:badcase_4}
\end{figure*}

\begin{figure*}[htbp]
    \centering
    \includegraphics[width=\linewidth]{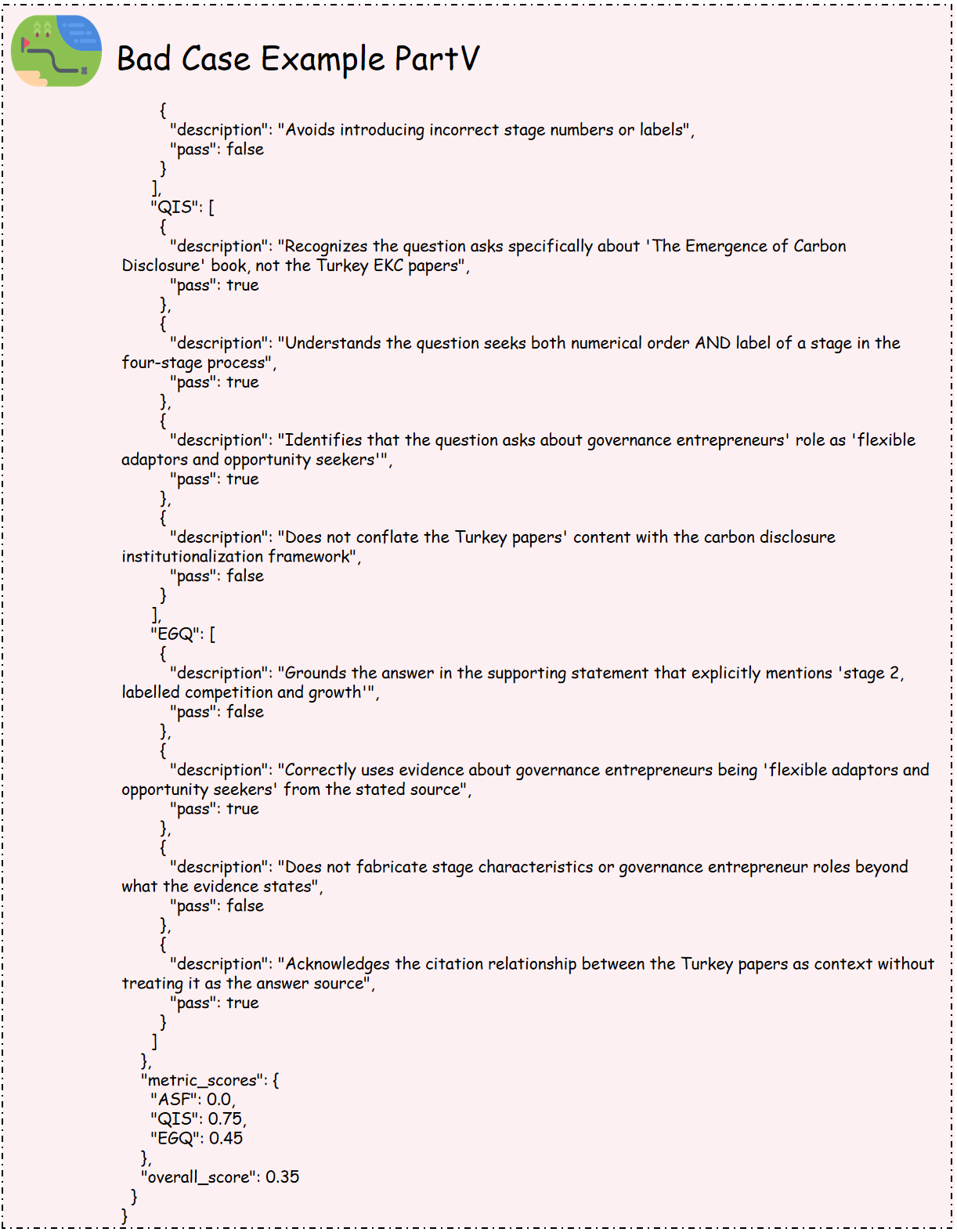}
    \caption{Error Case Part V. The localized judge model successfully intercepts the hallucination. By strictly evaluating the outputs against the evidence grounding criteria, it assigns a zero score for Answer Semantic Fidelity, validating the rigorousness of the reward signal.}
    \label{fig:badcase_5}
\end{figure*}

To provide a comprehensive understanding of the boundary limitations of SciLENS, we present a detailed error case analysis in Figure~\ref{fig:badcase_1} through Figure~\ref{fig:badcase_5}. In this scenario, the agent is presented with a highly convoluted query containing explicit distractors regarding Turkish environmental policies before asking for a specific framework stage from a distinct carbon disclosure publication. The trajectory reveals two critical failure modes typical of large language models. Initially, the agent exhibits attention hijacking by prioritizing the complex distractors and wasting early tool invocations retrieving irrelevant Turkish papers. Although it subsequently reflects upon its context window and redirects its search toward the correct target, it fails to locate the exact textual evidence within the limited retrieved abstracts. Consequently, the agent commits a severe parametric fallback error. Instead of continuing the deep retrieval process via graph traversal, the model explicitly rationalizes utilizing its internal parametric knowledge to hallucinate a plausible but factually incorrect answer. Crucially, this failure mode explicitly validates the robustness of our multidimensional evaluation framework. The judge model correctly identifies the hallucination and rigorously penalizes the Answer Semantic Fidelity and Evidence Grounding Quality metrics, resulting in a heavily degraded overall score. This transparent diagnostic confirms that our dynamic rubric successfully prevents the reinforcement learning optimization process from being deceived by superficially fluent but ungrounded text generation.


\end{document}